\documentclass{iopart}

\usepackage{iopams}
\usepackage{natbib}
\bibpunct{(}{)}{;}{a}{}{,} 
\usepackage{graphicx}
\usepackage{xspace}

\def\pr{\mathrm{Pr}}

\lineup 
\def\lenstool{\textsc{lenstool}\xspace}

\begin{document} 

\title[Bayesian strong lensing modelling of galaxy clusters]%
{A Bayesian approach to strong lensing modelling of galaxy clusters}

\author{ 
E.~Jullo$^{1,2}$, 
J-P.~Kneib$^{2}$, 
M.~Limousin$^{3}$, 
\'A.~El\'iasd\'ottir$^{3}$,
P.~J.~Marshall$^{4}$ and
T.~Verdugo$^{5}$
} 

\address{$^1$ European Southern Observatory, Alonso de Cordova, Santiago, Chile} 
\address{$^2$ OAMP, Laboratoire d'Astrophysique de Marseille - UMR 6110 - Traverse du siphon, 13012 Marseille, France} 
\address{$^3$ Dark Cosmology Centre, Niels Bohr Institute, University of Copenhagen, Juliane Maries Vej 30, 2100 Copenhagen, Denmark} 
\address{$^4$ Physics Department, University of California, Santa Barbara, CA 93106-9530, USA} 
\address{$^5$ Instituto de Astronomia, UNAM, AP 70-264, 04510 Mexico DF}
\ead{ejullo@eso.org} 

\submitto{``Gravitational Lensing'' Focus Issue of the New Journal of Physics (invited)}

\begin{abstract}

 In this paper, we describe a procedure for modelling strong lensing
 galaxy clusters with parametric methods, and to rank models
 quantitatively using the Bayesian evidence.  We use a
 publicly-available Markov Chain Monte-Carlo (MCMC) sampler
 (``BayeSys''), allowing us to avoid local minima in the likelihood
 functions.

 To illustrate the power of the MCMC technique, we simulate
 three clusters of galaxies, each composed of a cluster-scale halo and a set
 of perturbing galaxy-scale subhalos.  We ray-trace three
 light beams through each model to produce a catalogue of multiple
 images, and then use the MCMC sampler to recover the model parameters in
 the three different lensing configurations.

 We find that, for typical HST-quality imaging data, 
 the total mass in the Einstein radius is recovered with
 $\sim$ 1\% to 5\% error according to the considered lensing configuration.
 However, we find that the mass of the galaxies is strongly degenerate
 with the cluster mass when no multiple images appear in the cluster
 centre.  The mass of the galaxies is generally recovered with a 20\%
 error, due largely to the poorly constrained cut-off radius.

 Finally, we describe how to rank models quantitatively using the
 Bayesian evidence.  We confirm the ability of strong lensing to
 constrain the mass profile in the central region of galaxy
 clusters in this way. Ultimately, such a method applied to strong lensing
 clusters with a very large number of multiple images may provide
 unique geometrical constraints on cosmology.

 The implementation of the MCMC sampler used in this paper has been
 done within the framework of the \lenstool software package, 
 which is publicly available.\footnote{http://www.oamp.fr/cosmology/lenstool/}

\end{abstract}


\maketitle

\section{Introduction}

  Strong gravitational lensing is produced when a distant object (such
  as a galaxy or a quasar) is serendipitously aligned with a critical
  foreground mass concentration. Such a phenomenon was first observed
  by \citet{walsh1979} who discovered a double quasar strongly
  lensed by a distant galaxy. In the 1980's, with the advent of CCD
  imaging and its application to astronomy, giant gravitational arcs
  in galaxy cluster cores were discovered by two independent teams
  \citep{lynds1986,soucail1987}.  The lensing explanation proposed by
  \citet{paczynski1987} was soon confirmed by \citet{soucail1988}, who
  measured the redshift for the giant arc in Abell~370 as being
  roughly twice that of the cluster redshift. Together with the
  multiply-imaged quasars, giant arcs in galaxy clusters turned strong
  gravitational lensing from a theoretical curiosity into a powerful
  tool to probe the mass distributions of galaxies and galaxy cluster
  cores.
 Although rare in current surveys, strong lensing events are expected
 to number as many as a few hundred thousand over the whole sky
 \citep{cabanac2007}.

 In order to fully exploit strong gravitational lensing events, one generally
 needs high resolution imaging coupled to deep spectroscopy to measure
 the redshift of both the lensing object and the lensed sources. By combining,
 \emph{Hubble Space Telescope} (HST) images with ground-based spectroscopy
 on 8-10m telescopes, strong lensing analysis has proved to be very successful
 at constraining the mass distribution of
 galaxies \citep[e.g.\ ][]{munoz1998,koopmans2006}  and galaxy cluster cores
 \citep[e.g.\ ][]{kneib1996, abdelsalam1998, smith2005, halkola2006}.

 Nowadays, one particularly interesting application of strong lensing
 is to constrain the dark matter (DM) distribution in cluster cores,
 and contrast it with predictions of numerical simulations. For
 example, we would like to measure accurately the inner slope and the
 concentration parameter of the DM density profile, to probe DM
 properties and its link with the baryonic component \citep[][and
 references therein]{sand2007}.  Indeed, numerical simulations seem to
 advocate a cuspy DM slope that could be described by an NFW
 \citep{navarro1997} or a S\'ersic \citep{sersic1968,merritt2005}
 profile.  Observations are not yet giving definitive answers relative
 to the value of the inner slope \citep{gavazzi2003, sand2004,
 sand2007} or the concentration \citep{kneib2003,gavazzi2003}, but
 progress is being made steadily. 

 For example, in Abell~1689, after much disagreement over its
 concentration \citep{clowe2001, king2002a, bardeau2005,
 broadhurst2005,halkola2006}, \citet{limousin2007b} came to a consensus
 value of $c_{vir} \sim 6 - 8$ after careful and detailed modelling of
 the previously-analysed data combined with new multiple image
 identifications, redshifts and weak lensing source galaxy colours.  %
 \citet{comerford2007} discuss the issues related to the determination
 of the concentration parameter with different techniques, and compare
 its measurement in a large compilation of galaxy clusters with the
 distribution of $c_{vir}$ in numerical simulations.

 Numerical studies have shown that the concentration parameter of the
 NFW potential is quite sensitive to complex structures along the line of
 sight \citep{king2007} or triaxiality of the dark matter halos
 \citep{corless2006}. Improved datasets, but also more
 advanced techniques are needed to accurately model the mass distribution of
 gravitational lenses such as these. This movement towards more complex
 models has generated two competitive methodologies for lens modelling.

 So-called ``non-parametric'' methods, where the mass distribution or
 lens potential is reconstructed as a map defined on a grid of pixels,
 have been developed to constrain the mass distribution of (admittedly
 well-constrained) galaxy-scale lenses \citep{saha1997,
 abdelsalam1998}, initially for the purpose of probing the large
 diversity of possible mass models with a view to investigating in
 particular the modelling degeneracy present in the measurement of the
 Hubble constant.  Since 1997, non-parametric modelling has been
 intensively tested and greatly improved to overcome the lack of
 constraints very common in strong lensing \citep[e.g.\
 ][]{koopmans2005, diego2005, kochanek2006conf}.  However, the
 flexibility of these methods arising from their very large number of
 parameters has to be controlled to avoid over-fitting the data.
 Recent work on regularisation techniques \cite{bradac2005,suyu2006}
 has improved the situation in this regard somewhat. However, physical
 understanding often comes from the measurement of quantities such as
 total mass, profile slope, and so on, which still have to be
 extracted from the flexible reconstructed maps.

 ``Parametric'', or rather, simply-parameterised models therefore have
 two advantages: the assumption of a physical model leads to
 inferences that are directly related to physical quantities, while
 the model fits the data with relatively few free parameters compared
 to a ``non-parametric'' model.

 Effectively the regularisation of the mass distribution is achieved
 through the physical model itself.  The predicted surface density
 maps are smooth (by design), a situation perhaps valid only for quiet
 systems where the galaxy dynamics are well understood. The modelling
 of merging and perturbed systems is clearly the next challenging step
 for parametric methods.

 Another important issue in both parametric and non-parametric methods
 is the way the parameter space is explored. In this paper, we have
 used the parametric gravitational lensing package \lenstool to
 perform the lens modelling.  Given a parametrization describing the
 lens, this software explores the parameter space around the best-fit
 region, reproducing the location of the observed multiple images
 within the supplied uncertainties.  The first versions of the
 software \citep{kneib1993,smith2005} were based on a downhill
 $\chi^2$ minimization. However, this technique is very sensitive to
 local minima in the likelihood distribution; as a result, the
 modelling of complex systems would rapidly become too involving and
 inefficient.

 In order to face the current and future observational data, we have
 thus implemented a new optimization method based on a Bayesian 
 Markov Chain Monte Carlo (MCMC) approach. We will investigate here the
 merits of this new method on simulated strong lensing
 clusters.

 In the first part of the paper, we explain how to model a cluster of
 galaxies, and how to identify systems of multiple images. Then, we
 describe the implementation of the MCMC package
 \textsc{bayesys} \citep{skilling2004} in the \lenstool
 software.  In the second part, we analyse the performance of the
 Bayesian MCMC sampler by studying the degeneracies between the
 parameters of the Peudo-Isothermal Elliptical Mass Distribution
 \citep[PIEMD,][]{kk1993}, the pseudo-elliptical Navarro, Frenk \&
 White \citep[NFW, e.g.\ ][]{navarro1997,golse2002a} 
 and the pseudo-elliptical S\'ersic
 potentials. In the last section, we use the Bayesian evidence to rank
 the models that best reproduce systems of multiple images simulated
 from galaxy clusters with flat inner mass profiles.  Finally, we
 discuss the limitations of the strong lensing modelling.

 Note that the \lenstool\footnote{This software is publicly
 available at: http://www.oamp.fr/cosmology/lenstool} Bayesian MCMC
 implementation has already been used to model Abell~1689
 \citep{limousin2007b}, Abell~68 \citep{richard2007}, MS~2053
 \citep{verdugo2007} and Abell~2390 \citep{jullo2007}. All our results
 are scaled to the flat, low matter density \textsc{$\Lambda$cdm}
 cosmology with $\Omega_{\rm{M}} = 0.3, \ \Omega_\Lambda = 0.7$. When
 necessary, we scale the masses and distances according to a Hubble
 constant of \textsc{H}$_0 = 70$ km\,s$^{-1}$ Mpc$^{-1}$.

\section{Definitions and Methodology}

\subsection{Definition}

 The gravitational lensing transformation is a mapping from the source
 plane to the image plane \citep{schneider1992}:

\begin{equation}
\label{eq:lensequation}
\beta = \theta - \nabla \varphi(\theta)\;,
\end{equation}

 \noindent where $\theta$ and $\beta$ are the image and source
 positions respectively and $\varphi(\theta)$ is the lens potential
 computed at the image position. Depending on the strength of the
 gradient of the lens potential, one can easily see that for a given
 source position $\beta$, multiple images (at different $\theta$) can
 solve the lensing equation. When this is happening it corresponds to
 the strong lensing regime.

 The lens potential is the product of angular diameter distances
 ratio: $D_{LS}/D_{OS}$ (Lens-Source distance over Observer-Source
 distance) and the projected Newtonian potential $\phi(\theta)$ at the
 image position:

\begin{equation}
\label{eq:lensequation2}
\varphi(\theta)={2\over c^2}{D_{LS}\over D_{OS}}\phi(\theta)\;.
\end{equation}

 Hence, once the distance of the lens and the source are known, solving
 the lensing equation for different multiple images, allows to
 directly constrain the Newtonian potential, or equivalently the mass
 distribution of the lens.

%
%
\subsection{Modelling the different cluster mass components}

 Observations of clusters of galaxies reveal two components:
 cluster-scale halos (which includes both DM and the baryonic intra
 cluster gas) and galaxy-scale halos (made of stars and DM).
 Similarly, N-body simulations of clusters show that the mass
 distribution of subhalos inside a cluster halo follows a Schechter
 function \citep[e.g.\ ][]{shaw2006}.

 Thus, cluster gravitational potential can be decomposed in
 the following manner:

\begin{equation} 
\label{eq:decomposition}
\phi_{tot} = \sum_i \phi_{c_i} + \sum_j \phi_{p_j}\;, 
\end{equation}

 \noindent where we distinguish the cluster-scale smooth and large
 potentials $\phi_{c_i}$, and the subhalo potentials $\phi_{p_j}$
 providing small perturbations \citep{natarajan1997}. In the
 following, we consider a subhalo as a clump of matter containing a
 galaxy: we assume that there are no dark galaxies in clusters.  This
 decomposition has been successful in reproducing the observed systems
 of multiple images and in constraining the size of the subhalos in
 clusters \citep[e.g.\ ][]{smith2005,natarajan2006}.  We now describe
 in more detail how we model the cluster-scale halos and galaxy-scale
 subhalos.

\subsubsection{Smooth cluster-scale halos}

 The smooth cluster-scale halos represent both the DM and the
 intra-cluster gas. With enough constraints, each of these two
 component could in principle be modelled separately, but in this work
 they are modelled together as a single mass component.  The number of
 such halos is not easy to evaluate; generally one starts with a
 single halo -- except when X-Ray observations or the distribution of
 the galaxies clearly show a multi-modal distribution -- and increases
 the complexity of the model from there.

 In the case of a multi-modal distribution or a clearly bad fit to the
 data with a single halo, additional halos can be included to the
 model until a good fit is reached. In the \lenstool literature to
 date no more than two cluster-scale halos have been needed to achieve
 a good model (e.g.\ Abell~2218, Abell~1689), but this may change in
 the near future with the expected improvement of the strong lensing
 data (in particular with more spectroscopic redshifts) or when
 properly taking into account external constraints.

 Each halo in a model (both the cluster-scale and the galaxy-scale
 described below) is parametrized by a position on the sky
 ($x_c,y_c$), a projected ellipticity of the mass distribution
 ($\epsilon_\Sigma$) (see also ~\ref{sec:pseudoser} for the
 pseudo-elliptical developments of the S\'ersic potential), a position
 angle (PA), and a set of parameters specific to the choice of
 potential profile used to describe the halo. In this paper, we
 consider either the SIE, NFW, PIEMD, or S\'ersic profiles, described
 by either 1, 2, 3, or 3 parameters respectively (see
 Table~\ref{potentials} for the analytic description of each
 potential. See also \citet{limousin2005} for the surface density
 definitions of the PIEMD and NFW potentials). 

\def\inttab{\rule{0pt}{5ex} \indent}
\begin{table*}
\caption{\label{potentials} \textsc{lenstool} most used potentials.}
\begin{indented}
\item[]\begin{tabular*}{\linewidth}[c]{lclp{6cm}}
\br
\multicolumn{3}{l}{SIE}\\
\mr
\inttab $\epsilon_\varphi$ & = & $ \epsilon_\Sigma / 3 $ & $(\epsilon_\Sigma < 0.4)^a  $ \\
\inttab $\rho$ & = & $ \rho_0 / \tilde{R}  $ \\
\inttab $\rho_0$ & = & $ \displaystyle \frac{\sigma^2}{2\pi G} $ \\
\mr
\multicolumn{3}{l}{PIEMD}\\
\mr
\inttab $\epsilon_\varphi$ & = & $ \displaystyle \frac{1-\sqrt{1-\epsilon_\Sigma^2}}{\epsilon_\Sigma} $ \\
\inttab $\rho$ & = & $ \displaystyle \frac{\rho_0}{(1+\frac{\tilde{R}^2}{r_c^2})(1+\frac{\tilde{R}^2}{r_{cut}^2})}  $ \\
\inttab $\rho_0$ & = &  $ \displaystyle \frac{\sigma_\infty^2}{2\pi G r_c^2} $ & $ (\sigma_0 \simeq \sigma_\infty/1.46)^b  $ \\
\mr
\multicolumn{3}{l}{NFW}\\
\mr
\inttab $\epsilon_\varphi$ & = & $\epsilon_\Sigma / 2.27$ & $(\epsilon_\varphi < 0.25)^c$ \\
\inttab $\rho$ & = & $\displaystyle \frac{\delta_c \rho_c}{\frac{\tilde{R}}{r_s}(1+\frac{\tilde{R}}{r_s})^2} $ \\
\inttab $\delta_c$ & = & $\frac{200}{3} \frac{c^3}{\ln(1+c) - c/(1+c)} $ & $r_s = \frac{r_{vir}}{c}$ \\
\mr
\multicolumn{3}{l}{S\'ersic}\\
\mr
\inttab $\epsilon_\varphi$ & = & $\epsilon_\Sigma / 3.55$ & $(\epsilon_\varphi < 0.25)$ \\
\inttab $\ln \left( \frac{\Sigma}{\Sigma_e} \right)$ & = & $-b_n \left[ \left( \frac{\tilde{R}}{R_e} \right)^\frac{1}{n}-1 \right] $ \\
\inttab $b_n$ & $\simeq$ &  $2 n - \frac{1}{3} + \frac{4}{405 n} + \frac{46}{25515 n^2}^d $ \\ 
\br
\multicolumn{3}{l}{$^a$ \citet{kneib1993}} \\
\multicolumn{3}{l}{$^b$ \citet{golsephd}} \\
\multicolumn{3}{l}{$^c$ \citet{golse2002b}} \\
\multicolumn{3}{l}{$^d$ \citet{ciotti1999}} 
\end{tabular*}
\end{indented}
\end{table*}

\begin{figure}
\begin{indented}
\item[] \includegraphics[width=\linewidth]{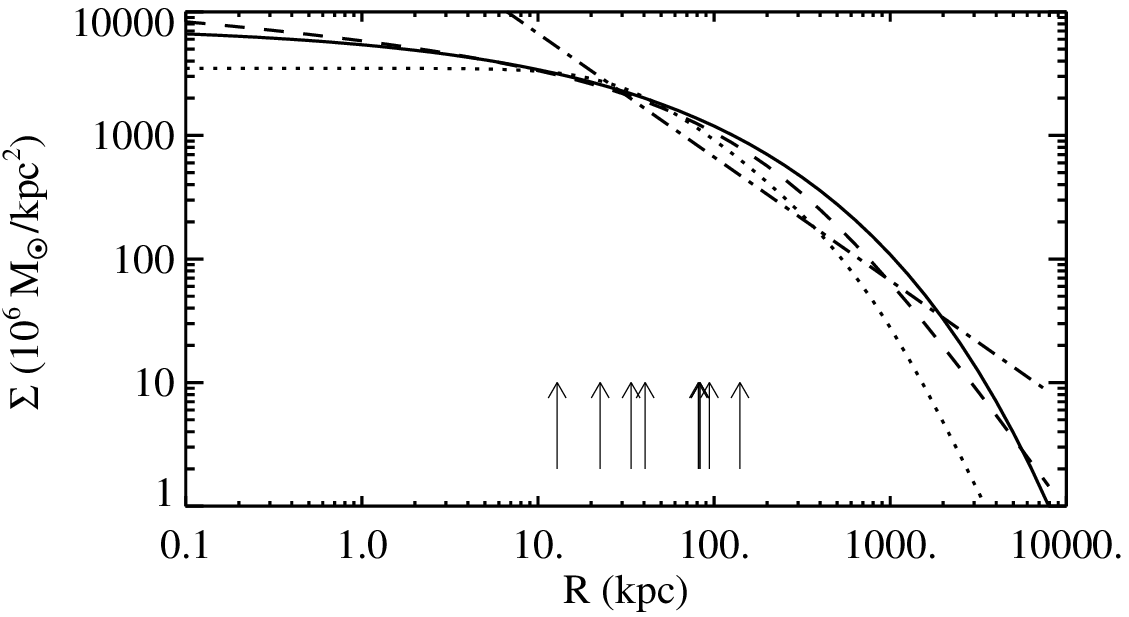}
\caption{ \label{fig:sdens} 
Surface density comparison between the S\'ersic (solid line), the NFW
(dashed line), the PIEMD (dotted line) and the SIS profiles
(dot-dashed line) The surface densities correspond to the fit
performed in section~\ref{sec:evid} and extended to very small and
large radii.  The arrows mark the multiple images positions used as
constraints. }

\end{indented}
\end{figure}

 In Figure~\ref{fig:sdens}, we compare the surface density of the SIS,
 the S\'ersic, the NFW and the PIEMD profiles both in the very central
 and in the very outer regions. These regions are accessible either to
 strong or weak lensing. These profiles are the best fit to the set of
 plotted multiple images.  We clearly note the flat core of the PIEMD
 profile up to 10 kpc and in contrast the monotonically increasing
 slope of the NFW and the S\'ersic profiles. The SIS profile slope is
 constant and hardly follows the other profiles. 

 Given the data (e.g.\ strong lensing or dynamics data), the cluster
 brightest galaxy -- also called the cD galaxy in the following -- can
 either be included in the cluster-scale halo or modelled separately.
 However, \citet{smith2005} showed that the centre of mass of the
 cluster-scale halo can be different from the cD galaxy centre.
 Therefore, it is generally justified to model the cD galaxy as an
 additional subhalo.

\subsubsection{Galaxy-scale components}

 \citet{kneib1996} first demonstrated that the inclusion of
 galaxy-scale subhalos was necessary to reproduce the observed systems
 of multiple images, particularly those appearing near cluster
 galaxies. These galaxy-scale subhalos or perturbers can be probed in a
 direct way using weak galaxy-galaxy lensing techniques
 \citep{natarajan1997,natarajan2002b}, however in this paper we will
 concentrate only on the strong lensing aspects.

 The number of subhalos to include in a model needs to be quantified.
 To date, a conservative attitude has been adopted: all the massive
 cluster member galaxies with cluster-centric radii out to
 approximately two times the limits of the strong lensing region are
 included. This is generally achieved by selecting galaxies within the
 cluster red sequence and selecting them brighter than a given
 luminosity limit. Moreover, the subhalos shape (ellipticity and
 orientation) is usually taken to be the same as its galaxy .

 Recently, \citet{wambsganss2005} and \citet{king2007} have raised the
 issue of multiple halos/subhalos along the line of sight that
 increase the projected surface density and thus affect the lensing
 strength.  While not large, this effect is a systematic, and so
 lensing models must consider the possibility of such gravitational
 perturbations.  In practice, the mass distribution along the line of
 sight can be understood from spectroscopic and photometric
 measurements in the field of view.

 Here, we propose a set of criteria for including perturbing subhalos
 in a model. The basic idea is to measure their strong lensing
 deviation angle and compare it to the spatial resolution $\delta$ of
 the lensing observations ($\delta \sim 0.1''$ for HST). A subhalo is
 included in the model if it can increase significantly the deflection
 angle at its associated galaxy position.  For a cluster member galaxy
 if its Einstein radius $R_{Einstein}>\delta/\mu$ (where $\mu$ is the
 magnification of the cluster-scale halo at the position of the
 galaxy) then it is included, otherwise its lensing contribution is
 not important and it is disregarded. For galaxies not part of the
 cluster, if $R_{Einstein}>\delta/\mu$ and the associated galaxy is in
 projection out of the strong lensing region, we include it in the
 model at the cluster redshift by rescaling its mass so that the
 global lensing effect is preserved.  Finally, if the galaxy is in the
 strong lensing region and its lensing effect is detectable then the
 associated subhalo must be included with a proper multi-plane lensing
 technique (we will not discuss such a case here as it is beyond the
 scope of this paper).

 Accounting for all the subhalos in a galaxy cluster as individually
 optimisable potentials would lead to an under-constrained problem.
 Assumptions must be made in order to make the number of parameters
 commensurate with the number of constraints.  \citet{koopmans2006}
 have shown that a strong correlation exists between the light and the
 mass profiles of elliptical galaxies in the field. Consequently, in a
 first approximation, the subhalos position, ellipticity and
 orientation are matched to their luminous counterpart.

 As we will show in the second part of this paper, apart
 from a few subhalos perturbing multiple images close to them, the 
 vast majority of subhalos act merely to increase the total mass enclosed 
 in the Einstein radius. Strong lensing provides few constraints on the
 mass profile parameters of most individual subhalos.

 We therefore reduce the number of subhalo parameters by asserting
 exact scaling relations between the subhalo masses and their
 associated galaxy luminosities.  Following the work of
 \cite{brainerd1996}, we model cluster subhalos with PIEMD
 potentials. The mass profile parameters in this model are the core
 radius ($r_{core}$), cut-off radius ($r_{cut}$), and velocity
 dispersion ($\sigma_0$), which we take to scale with the galaxy
 luminosity $L$ in the following way:

\begin{equation} 
\label{eqscaling}
\left\{ \begin{array}{l}
\sigma_0  =   \sigma_0^\star (\frac{L}{L^\star} )^{1/4}\;,  \\ 
r_{core}  =  r_{core}^\star (\frac{L}{L^\star} )^{1/2}\;, \\ 
r_{cut}  =   r_{cut}^\star (\frac{L}{L^\star} )^\alpha\;.  \\
\end{array} 
\right.
\end{equation}

 The total mass of a subhalo scales then as:  

\begin{equation}
 M = (\pi/G)(\sigma_0^\star)^2 r_{cut}^\star (L/L^\star)^{1/2+\alpha}\;,
\end{equation}

 \noindent where $L^\star$ is the typical luminosity of a galaxy at
 the cluster redshift, and $r_{cut}^\star$, $r_{core}^\star$ and
 $\sigma_0^\star$ are its PIEMD parameters. When $r_{core}^\star$
 vanishes, the potential becomes a singular isothermal potential
 truncated at the cut-off radius. This is generally the type of
 potential used in weak galaxy-galaxy lensing studies to measure the
 tidal radius of galaxy-scale subhalos in clusters or in the field
 (see \citet{limousin2005,limousin2006a}).

 In these scaling relations, the velocity dispersion scales with the
 total luminosity in agreement with the Tully-Fisher and the
 Faber-Jackson relations for spiral and elliptical galaxies
 respectively.  The $r_{cut}$ relation is more hypothetical. If
 $\alpha = 0.5$, it assumes a constant mass-to-light ratio independent
 of the galaxy luminosity. If $\alpha = 0.8$, the mass-to-light ratio
 scales with $L^{0.3}$ similar to the scaling of the fundamental plane
 \citep{natarajan1997,jorgensen1996,halkola2006}.

%
%
\subsection{Constraints} 

\subsubsection{Multiple images}

 In the strong lensing regime, the light coming from a background
 galaxy (the source) passes through a high density region and is
 lensed into multiple images. The position, shape and flux of each
 multiple image depends on the properties of the lens and the redshift
 of the source. The precise measurement of the source redshift, and of the
 image properties (such as position, ellipticity and orientation)
 provides strong constraints on the lens model. 

 In general, image properties can be inferred from their light
 distributions. Indeed, the first order moment provides the image
 position, and the PSF-corrected second order moment gives the
 ellipticity and the position angle of the image. Note however, that
 the ellipticity of a curved arc is somewhat ill-defined, so this
 information can only be used if the images are relatively compact. In
 this paper, we only consider the multiple image's position as a
 constraint, and we discuss the associated likelihood in the next
 section.

 Sometimes, the background galaxy presents several bright regions that
 can be individually identified in each multiple image.  Matching
 these bright regions in each image brings even tighter constraints to
 the lensing model. 

 The images flux can also be considered as a constraint. However, the
 amplification can vary strongly across highly extended images, and
 properly computing the amplification to measure the total flux in
 each image is usually not straightforward.

 Finally, the redshift of the source is a strong constraint on the
 lens model.  A spectroscopic determination is best, but a photometric
 redshift (e.g. \citet{ilbert2006}) can be sufficient if accurate
 enough (e.g. $\sigma_z<0.05$ introduces a 2\% error on the $D_{LS}/D_{OS}$
 ratio for a lens and a source at redshifts $z_L=0.2$ and $z_S=1$
 respectively) and with no multiple peak in its probability
 distribution (no catastrophic redshift).

 For well-defined photometric redshifts, \lenstool provides a way of
 introducing accurately the redshift likelihood as a prior for the
 model.

 Including an uncertain source redshift as a free parameter to be
 inferred from the data gives the model more freedom, albeit at some
 extra computational cost.  However, due to the other available
 constraints, it may lead to a more accurate redshift for that image
 system. This procedure may also raise questions about a photometric
 or spectroscopic measured redshift if the model favours a different
 range of values.

 The correct identification of multiple images is probably the most
 complex task in strong lensing modelling. 

 Initially we consider (as a guide) only generic geometrical
 lensing configurations -- cusp, fold and saddle \citep{blandford1986} --
 for single cluster-scale halo. 
 Having found a basic model that satisfies the most obvious or most
 straightforward multiple image system, the perturbations due
 to galaxy-scale subhalos can be taken into account. Generally,
 subhalos do not create strong lensing events by themselves, but
 affect the multiple images produced by the cluster-scale
 halo. They can deflect their position or
 occasionally further divide a multiple image.

 Comparing the colours of multiple images is another straightforward
 technique.  As lensing is achromatic, multiple images must have
 similar colours unless the images' fluxes are strongly contaminated
 with or reddened by nearby galaxies.

 It is important to realize that the identification process of
 multiple images is both iterative and strongly linked to the determination
 of the mass profile, starting from the most obvious systems close to
 the cluster centre and progressively adding perturbations and new
 systems. New multiple images can be predicted before they are
 observationally confirmed.

\subsubsection{Other lensing constraints}

\paragraph{Single images}

 Single images with known redshift lying close to the strong
 lensing region (typically when $R_{Einstein}< r < 2R_{Einstein}$) can
 also be included in the lens model. Indeed, they can help in
 constraining the parts of the model where no multiple image system is
 detected. Such constraints have been neglected up to now.  We propose
 here an efficient way to include them in the $\chi^2$ determination.

 In essence, we add a penalizing term to the likelihood if an observed
 single image is predicted to be multiple, and if at least one of the
 counter-images could effectively be detected in the observed
 data image. The penalizing term is a function of $n_k$, the number of
 predicted images above the detection limit (defined to be 
 3 times the sky noise flux in the object detection aperture).

 The penalizing term is implemented in the following way:
\begin{equation}
\chi_{single}^2 = \sum_{j=1}^{n_k} \frac{[x_{single}
                                - x^j(\btheta)
                                  ]^2}{\sigma_{single}^2}.
\end{equation}
 Here, $x_{single}$ is the position of the observed single image and
 $x^j(\btheta)$ is the position of a detectable image predicted by the
 current model, whose parameters are $\btheta$ and $\sigma_{single}$
 is the position error of the observed single image.

 This implementation provides a smooth way of converging to the best
 $\chi_{single}^2$.  Once $\chi_{single}^2 = 0$ (as it must be if truly single),
  the single image is
 no more a constraint.  Consequently, this definition only imposes an
 upper limit on the enclosed mass at the single image position.
 The truly singly-imaged systems do not add  to the overall number of
 degrees of freedom, nor to the final global chi-squared value.
 However, they do accelerate the convergence on the best fitting
 parameter region.

 This penalizing term must be used with some care; in particular,
 instances where $\chi_{single}^2>0$ have to be flagged and investigated, as
 they indicate either a failure of the model or that the single image
 identification was incorrect. Indeed, 
 this is one way in which new multiple images may
 be found.

\paragraph{Location of critical lines}

\begin{figure}
\begin{indented}
\item[]\begin{tabular}{@{}c}
\includegraphics[width=\linewidth]{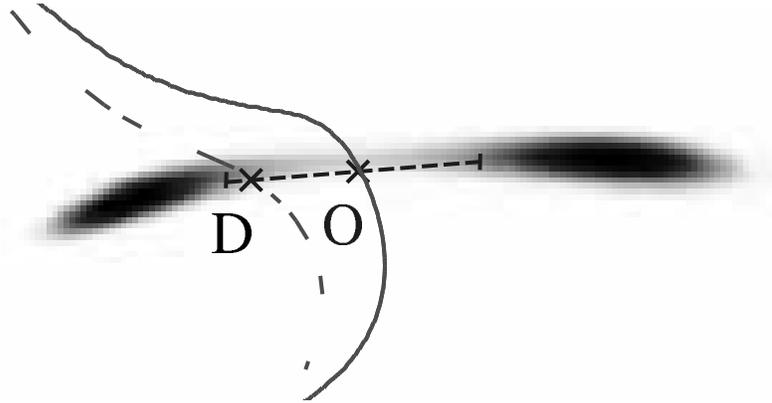} \\
\end{tabular} 
\end{indented}
\caption{\label{fig:cl} Merging of two multiple images and
determination of the distance between the true critical line
(\textit{solid line}, showing the surface brightness saddle point) 
and a predicted
critical line (\textit{dashed line}). The dashed segment represents
the prior that would be set on the critical line location.}
\end{figure}

 In the case of fold images, the position of the critical line passing
 in between the 2 images can sometimes be observed as a saddle point in the
 surface brightness of the images.  We can use this information to put
 a constraint on the lens model by minimizing the distance between the
 position where the image isophotes cross and the critical line
 predicted by a model, as shown in Fig.~\ref{fig:cl}.

 The prior segment for the critical line position can be defined by a centroid
 $\mathbf{O}$, a position angle and a Gaussian error size on the position
 $\sigma_{cl}$, hence, the corresponding $\chi^2$ can be given as:

\begin{equation}
\label{eq:cl}
\chi^2_{cl} = \frac{||\mathbf{O} - \mathbf{D}||^2}{\sigma_{cl}^2}\;,
\end{equation}

 \noindent where $\mathbf{D}$ is the intersection of the predicted
 critical line and the defined prior segment.

 This constraint merely reinforces the weight of the considered system
 of multiple images in the model. 

 By focusing on the crossing isophote, it makes
 of use of more of the imaging information than just the centroids of the
 multiple images. As such,
 it is a low-cost constraint in terms
 of computation time and definitely accelerates the  convergence on
 the best fit region.  Of course, since constraints must be
 independent observations, this constraint must be observable and not
 computed from the image positions.

 At the end of the optimisation, we check that
 $\chi^2_{cl}<1$.  If this is not satisfied, then 
 either the critical constraint was wrongly
 identified or the model has not yet fully converged.

\paragraph{Weak shear signal}

 Outside the strong lensing region, the weak shear signal can be used
 to constrain the model on larger angular scales. Considering a catalogue of
 background galaxies with PSF-corrected shape measurements, one can
 minimize the difference between the ellipticity of each galaxy and
 the reduced shear predicted by a mock model at the galaxy
 location~\citep[see e.g.\ ][and references therein]{marshall2002}.
 We will discuss the weak lensing implementation in a forthcoming
 paper

\subsection{The multiple images' likelihood}

\label{likelihood}
 We assume that the noises associated with the measurement of the
 images position are Gaussian and uncorrelated from one image to
 another. The noise covariance matrix for all the considered systems
 of multiple images is therefore diagonal. Hence, the usual definition
 of the likelihood function applies and becomes, in this case,  

\begin{equation}
\mathcal{L} = \pr(D|x(\btheta)) 
            = \prod_{i=1}^N \frac{1}{\prod_j \sigma_{ij} \sqrt{2\pi}}\exp^{-\frac{\chi^2_i}{2}}\;,
\end{equation}

 \noindent where $N$ is the number of sources, and $n_i$ is the number
 of multiple images for source $i$. The contribution to the overall
 $\chi^2$ from multiple image system $i$ is 

\begin{equation}
\chi^2_i = \sum_{j=1}^{n_i} 
	\frac{[x^j_\mathrm{obs} - x^j(\btheta)]^2}{\sigma_{ij}^2}\;,
\end{equation}

 \noindent where $x^j(\btheta)$ is the position of image $j$ predicted
 by the current model, whose parameters are $\btheta$ and $\sigma_{ij}$
 is the error on the position of image $j$. 

 The accurate determination of $\sigma_{ij}$ depends on the image S/N
 ratio. For extended images, a pixellated approach is the only
 accurate method which takes the S/N ratio of each pixel into account
 \citep{dye2005,suyu2006}.  However, this method is very time
 consuming. Therefore, in a first approximation, the image position
 error can be determined by fitting a 2D Gaussian profile to the image
 surface brightness.  In this case, the fit error contains implicitly
 the S/N ratio of each pixel. However, this assumes that the
 background galaxy is compact and its surface brightness profile is
 smooth so that the brightest point in the source plane match the
 brightest point in the image plane. In this paper, for simplicity,
 the image positions are determined by inverting the lens equation for
 a given source position. Therefore, the images are point-like. We
 assign them identical $\sigma_{ij}$ so that they have the same weight
 in the likelihood computation. Of course, this procedure is valid
 only in simulations where the source positions are known a priori and
 could not be applied to real cases.

 A major issue of the $\chi^2$ computation is of how to match the
 predicted and observed images one by one. Many techniques have been
 proposed so far to find the roots of the lens equation \citep[see
 e.g.][]{dominik1995}.  Unfortunately, the matching of the predicted
 to the observed images one by one becomes problematic when their
 respective positions do not match closely.  This always happens
 during the first steps of the optimisation. We have found no
 algorithm that performs this matching automatically.

 In contrast, the algorithm implemented in \lenstool is a simplex
 method \citep{press1986} of image transport \citep{schneider1992}. By
 definition, the observed image is coupled to the predicted image all
 along the iterative refinement of the predicted position.  The
 $\chi^2$ is therefore easy to compute.  However, in models producing
 different configurations of multiple images (e.g. a radial system
 instead of a tangential system), the method fails and that particular
 model is then rejected.  This usually happens when the model is not
 yet well determined, and it can slow the convergence of the model
 significantly.

 To get around this complexity, we can compute the $\chi^2$ in the
 source plane (by computing difference of the source position for a
 given parameter sample $\btheta$) instead of the image plane. The
 source plane $\chi^2$ is written as

\begin{equation}
\chi^2_{S_i} = \sum_{j=1}^{n_i} 
	\frac{[x^j_\mathrm{S}(\btheta)-<x^j_\mathrm{S}(\btheta)>]^2}
               {\mu_j^{-2}\sigma_{ij}^2}\;,
\end{equation}

 \noindent where $x^j_\mathrm{S}(\btheta)$ is the source position of
 the observed image $j$, $<x^j_\mathrm{S}(\btheta)>$ is the barycenter
 position of all the $n_i$ source positions, and $\mu_j$ is the
 magnification for image $j$. Written in this way, there is no need to
 solve the lensing equation and so calculation of the $\chi^2$ is very
 fast.

 The MCMC method we have implemented in \lenstool supports both the
 source and the image plane $\chi^2$ methods.  However, with the image
 plane method many models have to be tested and eventually rejected
 before the Bayesian sampler (see below) focuses on the best fit
 region.  This unnecessarily increases the computation time. In this
 paper, we first ``size up'' the best fit region with the source plane
 method, and then refine the models with the image plane method.

\begin{figure}
\begin{indented}
\item[]\includegraphics[width=\linewidth]{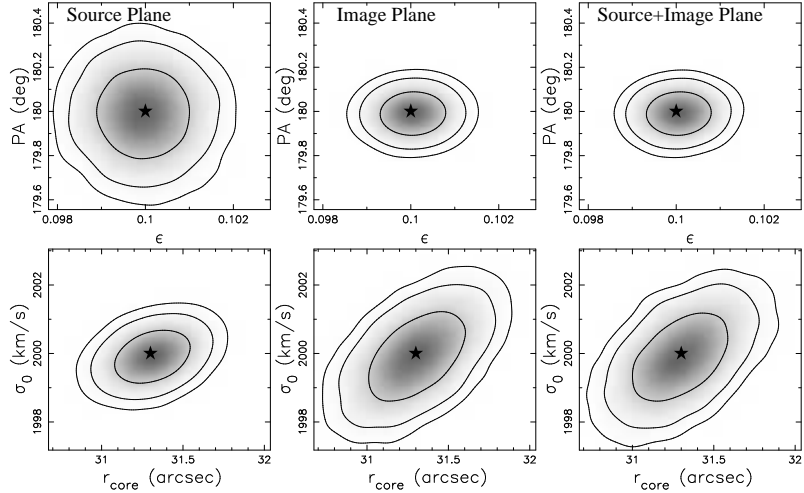}
\end{indented}
\caption{ \label{fig:srcimg} 
 2D marginalized posterior PDF of a simulated cluster of galaxies. The
 left, middle and right columns are respectively obtained by computing
 the likelihood  with the source plane method, with the image plane
 method and successively with the source plane and the image plane
 methods. In terms of computation time, the combined method source
 plane -- image plane is about 8 times faster than the image plane
 method alone. } 

\end{figure}

 Figure~\ref{fig:srcimg} shows that the posterior PDF are similar when
 computed with the image plane method alone or with the successive
 source plane+image plane method. However, this latter 
 method is about 8 times faster than the image method alone.

%
%
\section{A Bayesian Markov Chain Monte Carlo method}

\label{mcmc}

 We have implemented the Bayesian MCMC package \textsc{BayeSys}
 \citep{skilling2004} to perform the lens model fitting. By model, we
 mean a multiple-component (and hence multi-scale) mass distribution
 as described above, with a set of priors for its parameters.

 Theoretically, the Bayesian approach is better suited than regression
 techniques in situations where the data by themselves do not
 sufficiently constrain the model. In this case, prior knowledge about
 the parameter Probability Density Function (PDF) helps to reduce the
 model's degeneracies. The Bayesian approach is well-suited to strong
 lens modelling, given the few constraints generally available to
 optimize a model.

 The Bayesian approach provides two levels of inference: parameter
 space exploration, and model comparison. The first level can be
 achieved using the unnormalised posterior PDF (equal to the product
 of the likelihood and the prior); the second requires the calculation
 of the normalisation of the posterior, known as the evidence.  All
 these quantities are related by Bayes Theorem,

\begin{equation}
\label{bayesTheorem} 
\pr(\btheta|D,M) = \frac{\pr(D|\btheta,M)\pr(\btheta|M)}{\pr(D|M)}\;, 
\end{equation} 

 \noindent where $\pr(\btheta|D,M)$ is the posterior PDF,
 $\pr(D|\btheta,M)$ is the likelihood of getting the observed data $D$
 given the parameters $\btheta$ of the model $M$, $\pr(\btheta|M)$ is
 the prior PDF for the parameters, and $\pr(D|M)$ is the evidence. 

 The posterior PDF will be the highest for the set of parameters
 $\btheta$ which gives the best fit and is consistent with the prior
 PDF, regardless of the complexity of the model $M$.  Meanwhile, the
 evidence $\pr(D|M)$  is the probability of getting the data $D$ given
 the assumed model $M$. It measures the complexity of model $M$, and,
 when used as in model selection, it acts as Occam's razor: ``All
 things being equal, the simplest solution tends to be the best one.''
 Here, the simplest solution tends to be the model with the smallest
 number of parameters and with the prior PDF the closest to the
 posterior PDF.  In contrast, the commonly-used reduced $\chi^2$
 analysis is only a rough approximation to the evidence analysis,
 although it does provide an absolute estimator of goodness-of-fit
 (provided the error estimates on the data are accurate).

 In information theory, the evidence combines the likelihood and the 
 information $I$, or negative entropy:

\begin{equation} 
\label{eq:information}
I = \int \pr(\btheta|D,M)\log (\pr(\btheta|D,M)/\pr(\btheta|M))
\mathrm{d}\btheta\;.
\end{equation} 

 \noindent where the sum is performed over the whole parameter space
 and $\pr(\btheta|D,M)$ is the posterior PDF and $\pr(\btheta|M)$ is
 the prior PDF.

 The negative entropy measures the information we have obtained in
 computing the posterior PDF from the input prior PDF.  It represents
 a ``distance'' between the prior PDF and the posterior PDF. It can
 also be understood as the volume of the prior PDF over the posterior
 PDF, which can be very large for high signal to noise data. [In this
 case the task of parameter space exploration is like searching for a
 ``a needle in a haystack,'' and the entropy measures the ratio of the
 needle's volume (the posterior PDF) to the haystack's volume (the
 prior PDF)]. 

 In general, the information is much bigger than unity because the
 ``distance'' between the prior PDF and the posterior PDF is large.
 For this reason, we use annealed Markov Chains to converge
 progressively from the prior PDF to the posterior PDF.

 Technically, we run 10 interlinked Markov chains at the same time to
 prevent any Markov chain from falling in a local minimum. The MCMC
 convergence to the posterior PDF is performed with a variant of the
 ``thermodynamic integration'' technique \citep{oruanaidh1996} called
 \textit{selective annealing}.  

 ``Selective'' stands for the following process.  At each step, 10 new
 samples (one per Markov chain) are drawn randomly from the current
 posterior PDF (which corresponds to the prior PDF at the beginning).
 These samples are weighted according to their likelihood raised to
 the power of $\delta_\lambda$ (see below) and selected with a variant
 of the Metropolis-Hasting algorithm
 \citep{metropolis1953,hastings1970}.  Roughly, the samples with the
 worst likelihood are deleted and the ones with the best likelihood
 are duplicated so that we always keep $10$ Markov chains running at
 the same time.  Then, \textsc{bayesys} provides 8 exploration
 algorithms to randomly move the new samples in the parameter space
 and keep the 10 Markov chains uncorrelated \citep[see][for more
 details]{skilling2004}. This new set of randomly mixed samples is
 appended to the Markov chains and used as a new seed for the next
 step.

 The \textsc{bayesys} production of new samples is fast but the
 likelihood computation by \lenstool is slow. For each observed image,
 we must compute the gradient of every potential and sum them to
 compute the deviation angle and determine the source position.
 Therefore, the optimization process takes longer with more images
 and/or more potentials.  However, if the $r_{cut}^\star$ or
 $\sigma_0^\star$ parameters are fixed, the luminosity-scaled subhalo
 gradients can be computed just once (at the first iteration), thus
 reducing drastically the computation time.

 The ``annealing'' term of the ``selective annealing'' technique
 controls the convergence speed.  The slower and smoother the
 convergence, the more accurate is the evidence and the
 better-characterised is the posterior.  The annealing process is best
 seen by re-writing Bayes theorem:

\begin{equation} 
\pr(\btheta|D,M) = \frac{\pr(D|\btheta,M)^\lambda
\pr(\btheta|M)}{\pr(D|M)}\;. 
\end{equation}

 Here, $\lambda$ is the cooling factor for the annealing.  During a
 so-called ``burn-in'' phase, the likelihood influence is raised
 progressively from $\lambda = 0$ to $\lambda = 1$ by step of  
 $\delta_\lambda \sim \mathrm{Rate}/(\log \mathcal{L}_{max} - \log
 \bar{\mathcal{L}})$ where $\bar{\mathcal{L}}$ is the mean likelihood
 value of the 10 samples and Rate is an arbitrary constant set by the
 user.  At the beginning of the optimization, $\delta_\lambda$ is
 small because the likelihood dispersion of the 10 samples is large.

 As seen above, the samples are weighted and selected according to
 their likelihood raised to the power of $\delta_\lambda$. Thus,
 whatever the likelihoods are widely separated, $\delta_\lambda$
 decreases and the convergence automatically slows in proportion to
 compensate.

 In the small-convergence speed limit, the relative information between the
 beginning and end of a MCMC step is approximately constant and equals
 to Rate$^2$ \citep{skilling2004}. 

\begin{figure} 
\begin{indented} 
\item[]\begin{tabular}{@{}cc} 
\includegraphics[width=0.48\linewidth]{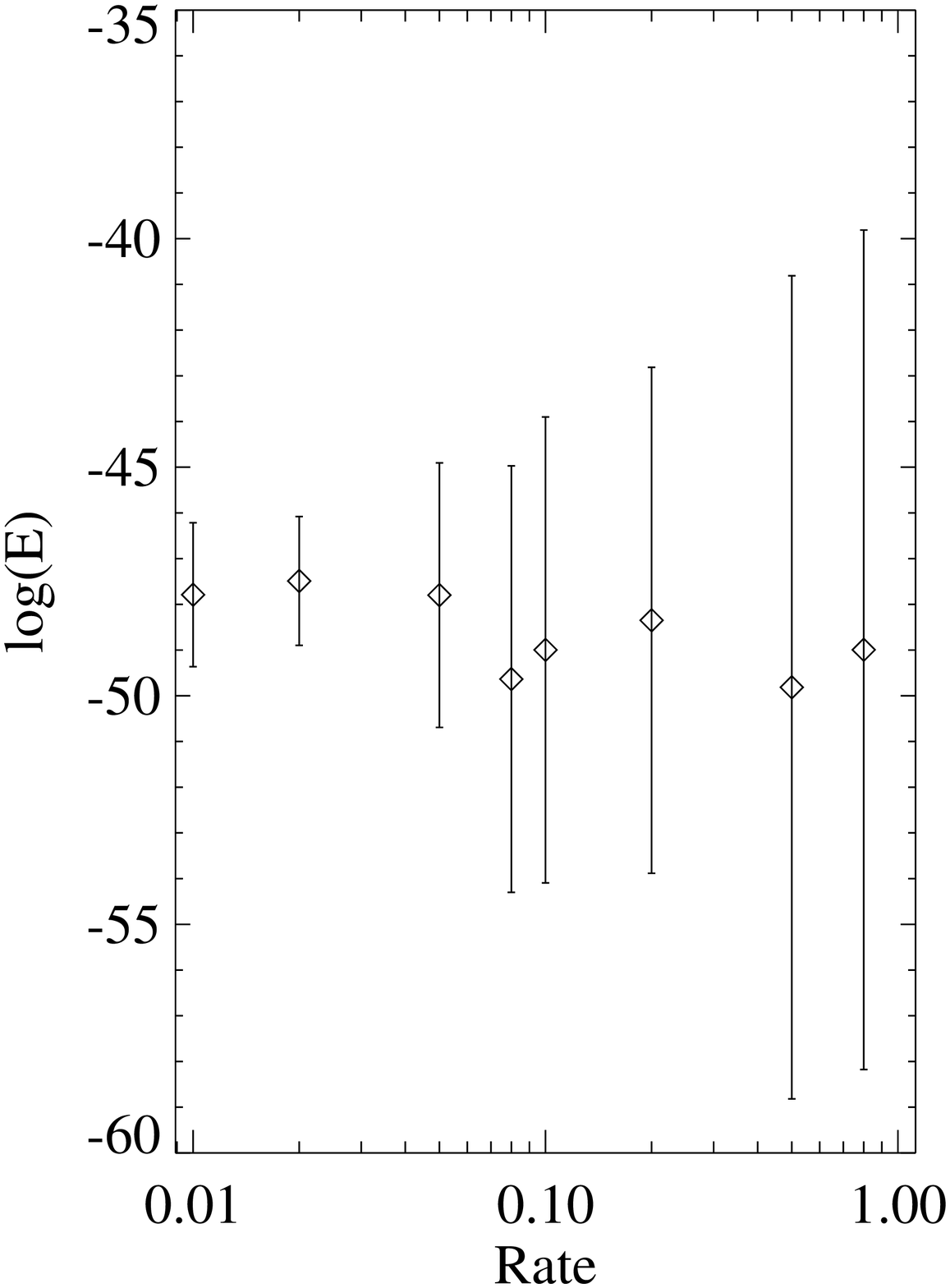} & 
\includegraphics[width=0.48\linewidth]{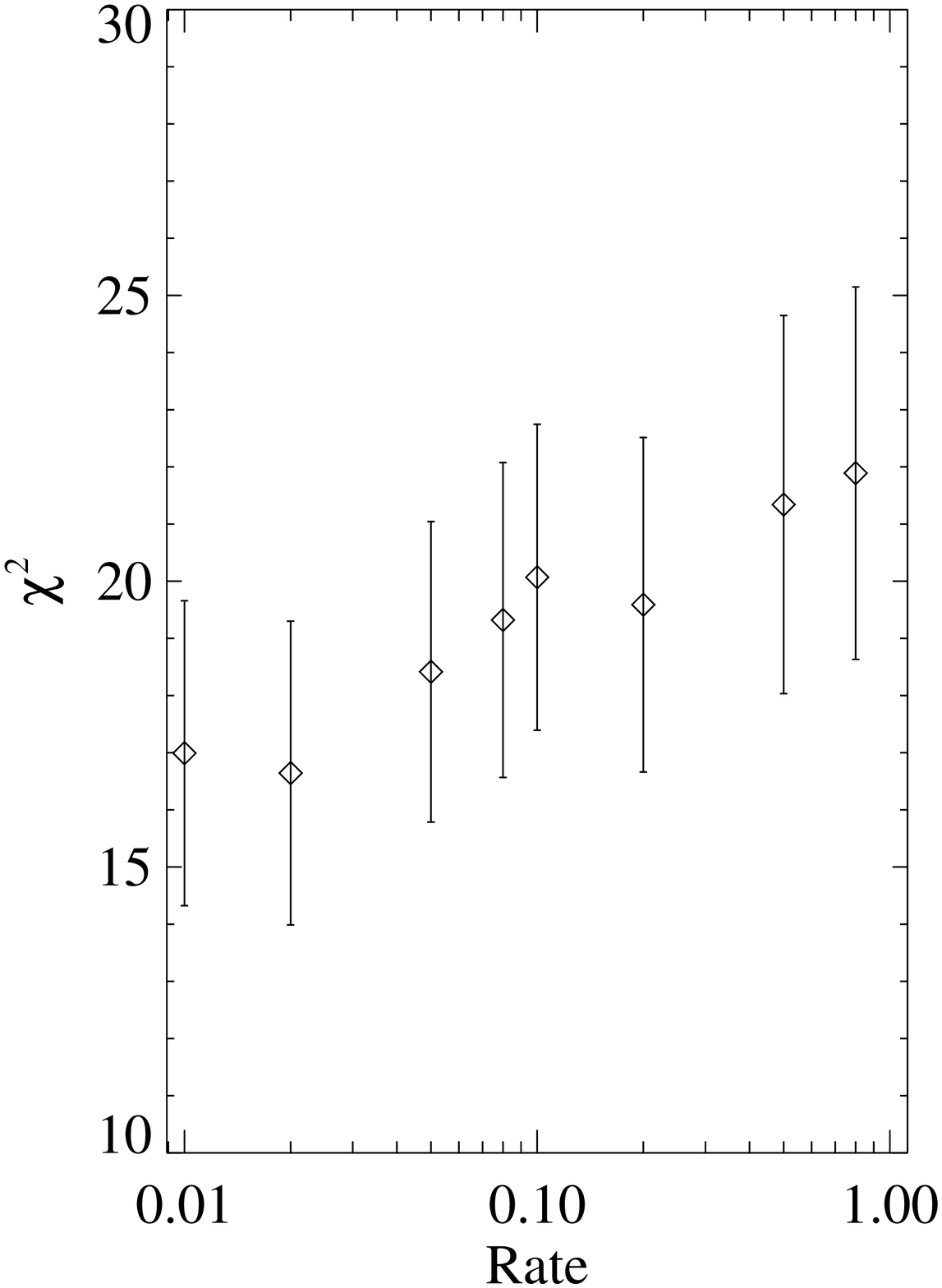}  
\end{tabular} 
\end{indented} 
\caption{\label{evidStudy} Evidence and $\chi^2$ evolution in function
of the convergence speed parameter ``Rate.'' } 
\end{figure}

 By decreasing Rate, the user decreases the information rate per MCMC
 step and thus the evidence error (see left panel of
 Figure~\ref{evidStudy}) but at the price of slower convergence. 

 The right panel of Figure~\ref{evidStudy} shows that, within the
 error bars, the median $\chi^2$ is stable when Rate decreases. A
 lower Rate implies a slower convergence speed. The chains will
 contain more samples and hence better explore the parameter space
 towards the best fit region.  This explains the slight decrease of
 the median $\chi^2$ when Rate decreases.  Alternatively, the spread
 of $\chi^2$ is similar for all Rate values, indicating that the
 convergence speed does not affect the parameter space exploration
 around the median $\chi^2$.

 From our experience, we have found that a value between 0.1 and 0.5
 gives evidence values that are accurate enough for our purposes,
 while returning the posterior PDF in a reasonable amount of
 computation time. From Figure~\ref{evidStudy}, we can see that the
 uncertainty on the logarithm of the evidence is approximately 4
 units: this corresponds to an odds ratio of 50 to 1, a sufficiently
 convincing value.  In the rest of this paper, we will use a Rate of
 0.1 unless otherwise specified.

%
%
\subsubsection{MCMC output}

 Contrary to maximum likelihood methods (like the downhill method used
 by \citeauthor{kneib1993}~\citeyear{kneib1993}), the Bayesian MCMC
 sampler does not look for the best sample of parameters. Instead, it
 samples the posterior PDF, drawing more samples where the posterior
 PDF is higher.

 The more samples we collect after the burn-in phase, the better the
 resolution of the posterior PDF. This is of particular interest given
 that we use one- and two-dimensional histograms to represent the
 marginalized posterior PDFs $\pr(\theta_i|M)$ and
 $\pr(\theta_i,\theta_j|M)$. The number of histogram bins is limited
 by the number of samples. To determine the bin sizes, we use the
 Freedman \& Diaconis rule \citep{freedman1981}. They have shown that
 in order to get the best fit between a PDF and the corresponding
 histogram, the bin size should be:

\begin{equation} 
\mathrm{Bin\ size}=2\mathrm{IQR}(\theta_i)N^{-1/3}\;, 
\end{equation}

 \noindent where IQR is the interquartile range of the $\theta_i$
 samples and $N$ is the number of samples.

 The produced 2D posterior histograms in the rest of this paper show
 that the parameters are not independent, and that their PDFs are
 certainly not Gaussian.  Techniques based on the assumption of
 Gaussian errors, with correlation matrix measured around the best
 fit, are not accurate and likely underestimate some errors.
 Therefore, uncertainties must be estimated with care, and eventually
 asymmetric errors must be adopted in case of large asymmetries
 observed in the posterior PDF.

 To compress the posterior PDFs and provide a convenient way of
 comparing them, we use the median and the standard deviation
 estimators. It has been shown \citep{simardphd} that the median is
 the most robust estimator for unimodal asymmetric distributions ---
 which is usually the kind of distribution we have for our parameters
 --- whereas the mean estimator is valid only if the distribution is
 close to Gaussian. The more samples we have, the less we are affected
 by outliers.

%
%

\section{Lens potential parameter degeneracies} 

\label{simulation} 
 In this section, we present and interpret the degeneracies observed
 in galaxy cluster strong lensing models. Degeneracies will always
 appear in strong lensing modelling because the lensing only
 constrains the mass inside an Einstein radius.  Unfortunately in
 parametric models, the parameters involved in the computation of the
 mass inside the Einstein radius are rarely orthogonal and strongly
 degenerate. 

 In the literature, we have found several papers presenting parameters
 degeneracies \citep[see e.g.][for illustrations of the NFW
 $r_s$---$\rho_s$
 degeneracy]{zekser2006,rzepecki2007,meneghetti2007b}.  We are finding
 similar results, although we are going beyond most of the previous
 study by exploring many more parameters. 

 In this section, we use the same potential to simulate and recover
 the cluster-scale halo, respectively a PIEMD, a NFW and a S\'ersic
 potential. Fitting the data by the true model never happens in
 practice. However, the presented degeneracies always appear and
 simple models are required for a proper understanding. 

 In section \ref{sec:evid}, we will use different models for the
 simulation and the recovery in order to compare the limits of each
 model given the data.

%
%

\subsection{Description of the simulation} 

\subsubsection{The mass models}

 We simulate a cluster of galaxies comprising a cluster-scale halo,
 and 78 galaxy-scale subhalos that perturb the lensing signal. The
 cluster-scale halo is modelled successively by a PIEMD, a NFW and a
 S\'ersic potential whose input parameters are reported in
 Table~\ref{tab3pot}. The galaxy-scale subhalos are modelled by PIEMD
 potentials with vanishing core radius.  The cluster is placed at
 redshift $z=0.2$. Hereafter, we will refer to each model  as the
 PIEMD, the NFW and the S\'ersic models.  

 The galaxy-scale subhalo distribution follows the galaxy distribution
 in the cluster Abell~2390 in a region of 200 kpc around the cluster
 centre. This is two times larger than the radius of the outermost
 images in our simulation. Thus, we account for the shearing effect
 produced by outer galaxies.  The selected galaxies are part of the
 cluster red-sequence and therefore are assumed to be cluster members.

 The galaxy-scale subhalos $r_{cut}$ and $\sigma_0$ are scaled with
 the scaling relations \eref{eqscaling}. A constant M/L ratio is
 assumed. We consider the scaling parameters $r_{cut}^\star=18$~kpc
 and $\sigma_0^\star=200$~km/s as the input values for our
 simulations.  These values correspond to measured values obtained
 through galaxy-galaxy lensing in Abell~2390 \citep{natarajan2006}.
 The apparent K-band magnitude of an $L^*$ galaxy at the cluster
 redshift is $M^\star=17.05$ (in AB magnitude) \citep{depropris1999}.
 The galaxy magnitudes come from observations of Abell~2390 in the
 K-band \citep{jullo2007}, and are used to calculate the true mass
 parameters in the simulations.

 We also include a cD galaxy in the model to produce more systems of
 multiple images in the cluster centre. The cD galaxy is described by
 an individual subhalo modelled by a PIEMD potential with vanishing
 core, and shape parameters matching the light distribution. Its mass
 profile is characterized by $\sigma_0 = 290.$ km/s, $r_{core}=0$ and
 $r_{cut} = 38.$ kpc.  The cluster Einstein radius for a $z=10$
 background source is 30.'' The enclosed mass at this radius is
 $M_{eins}=6.7\ 10^{13}\ M_\odot$, of which the galaxies' contribution
 is about 9\%. 

\begin{table}
\begin{indented}
\item[]\caption{\label{tab3pot} Input parameters for the 3
simulated cluster-scale components. }
\item[]\begin{tabular}{lrclr@{ = }l}
\br
PIEMD & $\epsilon_\Sigma$ & = &  0.3   & $r_{core}$  & 40. kpc \\
      &      $PA$  & = & 127.   & $r_{cut}$   & 900. kpc \\
      &            &   &        & $\sigma_0$  & 950. km/s \\
\mr
NFW  &  $\epsilon_\varphi$ & = &  0.2   & $r_{200}$   & 1800. kpc \\
     &  $PA$       & = &  127.  & $c$         & 6 \\
\mr
S\'ersic & $\epsilon_\varphi$ & = & 0.2  & $R_e$ & 1500. kpc \\
         &    $PA$    & = & 127. & $\Sigma_e$ & $5\ 10^7$ M$_\odot$/kpc$^2$ \\
         &            &  &       & $n$  & 2.8 \\
\mr
$L^\star$ galaxy & $r_{cut}^\star$ & = & 18. kpc & $\sigma_0^\star$ & 200. km/s \\
\br
\end{tabular}
\end{indented}
\end{table}

%
%
\subsubsection{Strong lensing constraints}

 We lens three background sources, $A$, $B$ and $C$, at redshifts
 $z_A=0.6$, $z_B=1.0$ and $z_C=4.0$, through each simulated cluster. We
 adjust the $B$ and $C$ source positions in order to produce the three
 following configurations of multiple images.

\begin{figure} 
\begin{indented} 
\item[]\begin{tabular}{@{}l} 
\includegraphics[width=\linewidth]{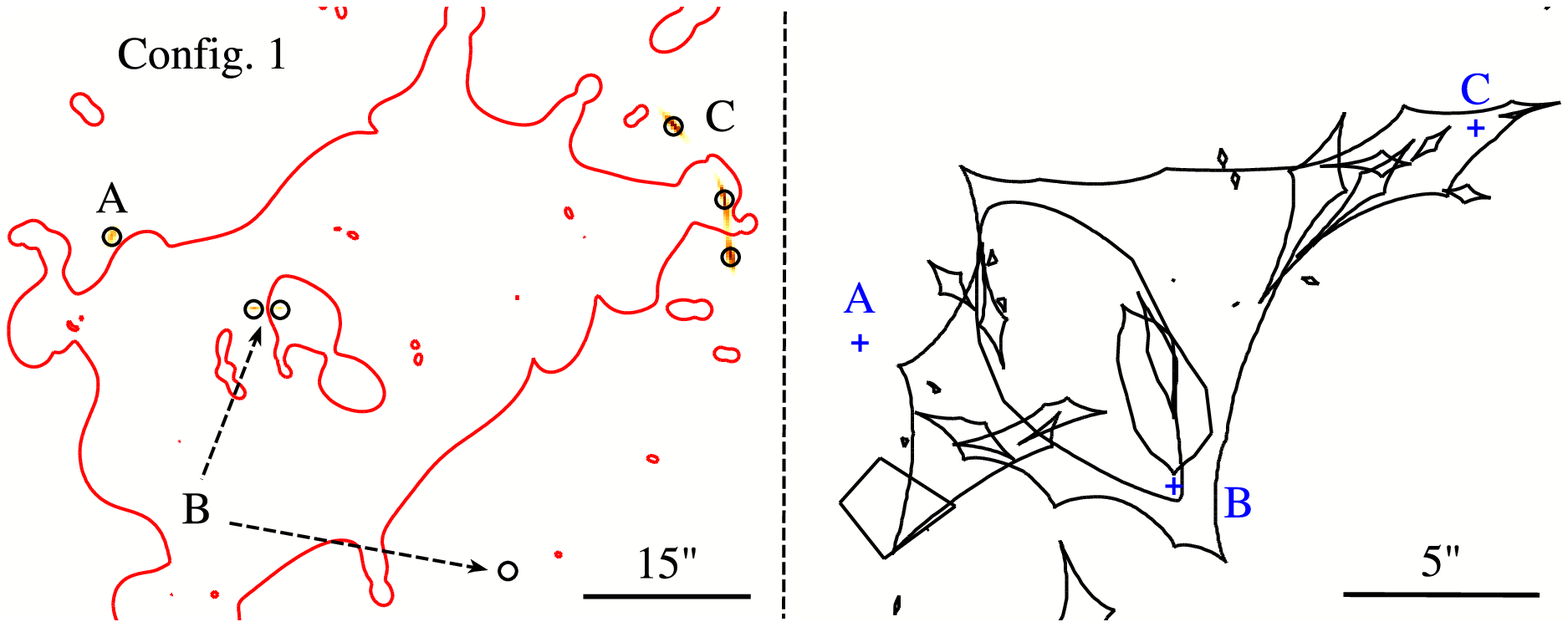} \\ 
\includegraphics[width=\linewidth]{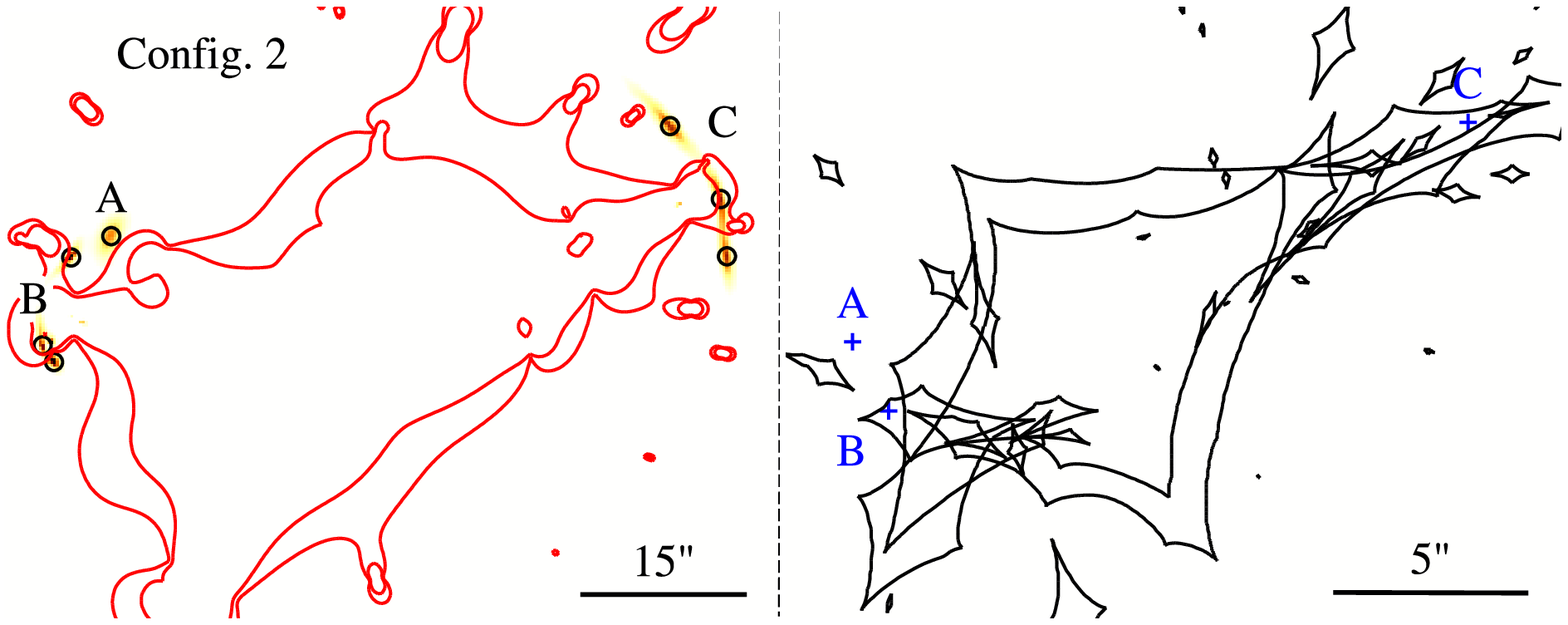} \\ 
\includegraphics[width=\linewidth]{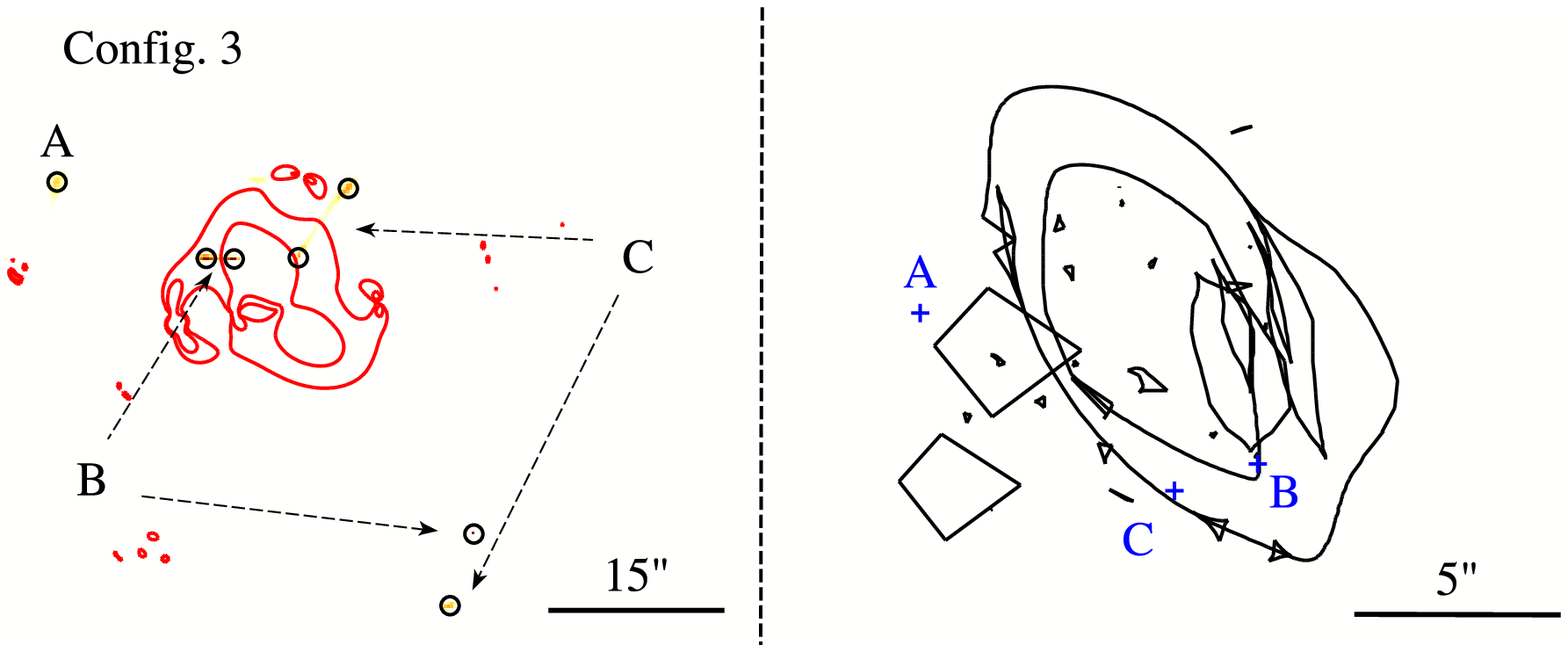}  
\end{tabular} 
\end{indented} 
\caption{ \label{slconfigs} 
 {\bf Left panel:} Image plane for the PIEMD simulated cluster,
 showing the image positions of the systems $A$, $B$ and $C$ at
 redshifts $z_A=0.6$, $z_B=1.0$ and $z_C=4.0$ in the configurations 1,
 2 and 3. The black circles mark the image positions.  The critical
 curves of systems $B$ and $C$ are shown in red. 

 {\bf Right panel:} The corresponding source plane. The blue crosses mark
 the source positions; the caustic curves are plotted in black. The
 plotted caustics for systems $B$ and $C$ are radial and tangential,
 tangential and tangential and radial and radial respectively for
 configurations 1, 2 and 3.  North is up and East is left in both panels. }

\end{figure}

 {\sl Configuration 1}: source $A$ is placed on the North-East side of
 the cluster, but outside of the multiple image region. It therefore
 produces a single image. Also on the East side, but inside the radial
 caustic, source $B$ produces a radial arc system with 3 images. On the
 West side, source $C$ lies along the West naked cusp of the caustics
 and so produces a system with 3 tangential images.

 {\sl Configuration 2}: sources $A$ and $C$ are in the same places as
 in Config.~1, but source $B$ is placed along the East naked cusp and
 so produces 3 tangential images. The second configuration therefore
 constrain mainly the enclosed mass in the outer part of the cluster
 ($100<r<200$ kpc).

 {\sl Configuration 3}: sources $A$ and $B$ are at the same place as
 in Config.~1,  but source $C$ is placed close to the radial caustics
 and therefore produces a second radial system of 3 images on the West
 side of the cluster. The third configuration then preferentially
 constrains the inner part of the mass profile ($r<100$ kpc).

 The source and image positions in the three configurations are
 presented in Figure~\ref{slconfigs}, along with the critical and
 caustic curves for sources at redshift $z_B=1.0$ and $z_C=4.0$.
 Gaussian noise of FWHM 0.1''\ was added to the image positions to
 mimic the observational uncertainties. All the predicted images are
 used for the parameter recovery unless their lensing magnification is
 lower than 1. In practice, such images are never observed (too faint
 or blended in the cD flux).

 Config.~1 constrains the cluster central and outer regions, Config.~2
 only constrains the outer region and in Config. 3, the 4 radial
 images strongly constrain the cluster central region on both the East
 and the West sides. 

%
%
\subsection{PIEMD posterior PDF analysis}
\label{sec:piemd}

 First, we fit the PIEMD model with a PIEMD potential for the
 cluster-scale halo. For each of the three configurations of
 multiple images, we recover the cluster-scale halo parameters
 ($\epsilon$, PA, $r_{core}$, $r_{cut}$ and $\sigma_0$), as well as
 the galaxy-scale subhalos scaling parameters $\sigma_0^\star$ and
 $r_{cut}^\star$.  For each parameter, we assume a uniform prior with
 50\% errors around its input value. In this case the computed
 posterior  PDF is merely proportional to the likelihood PDF.  The cD
 galaxy subhalo parameters are fixed to their input value in order to
 avoid annoying additional degeneracies with the cluster-scale halo
 parameters.  We therefore constrain 7 free parameters with 8
 constraints. The $\chi^2$ is computed in the image plane, although we
 observed no difference between the source plane $\chi^2$ and the
 image plane $\chi^2$.

\begin{figure*} 
\begin{indented} 
\item[]\begin{tabular}{ccc} 
\includegraphics[width=0.28\linewidth]{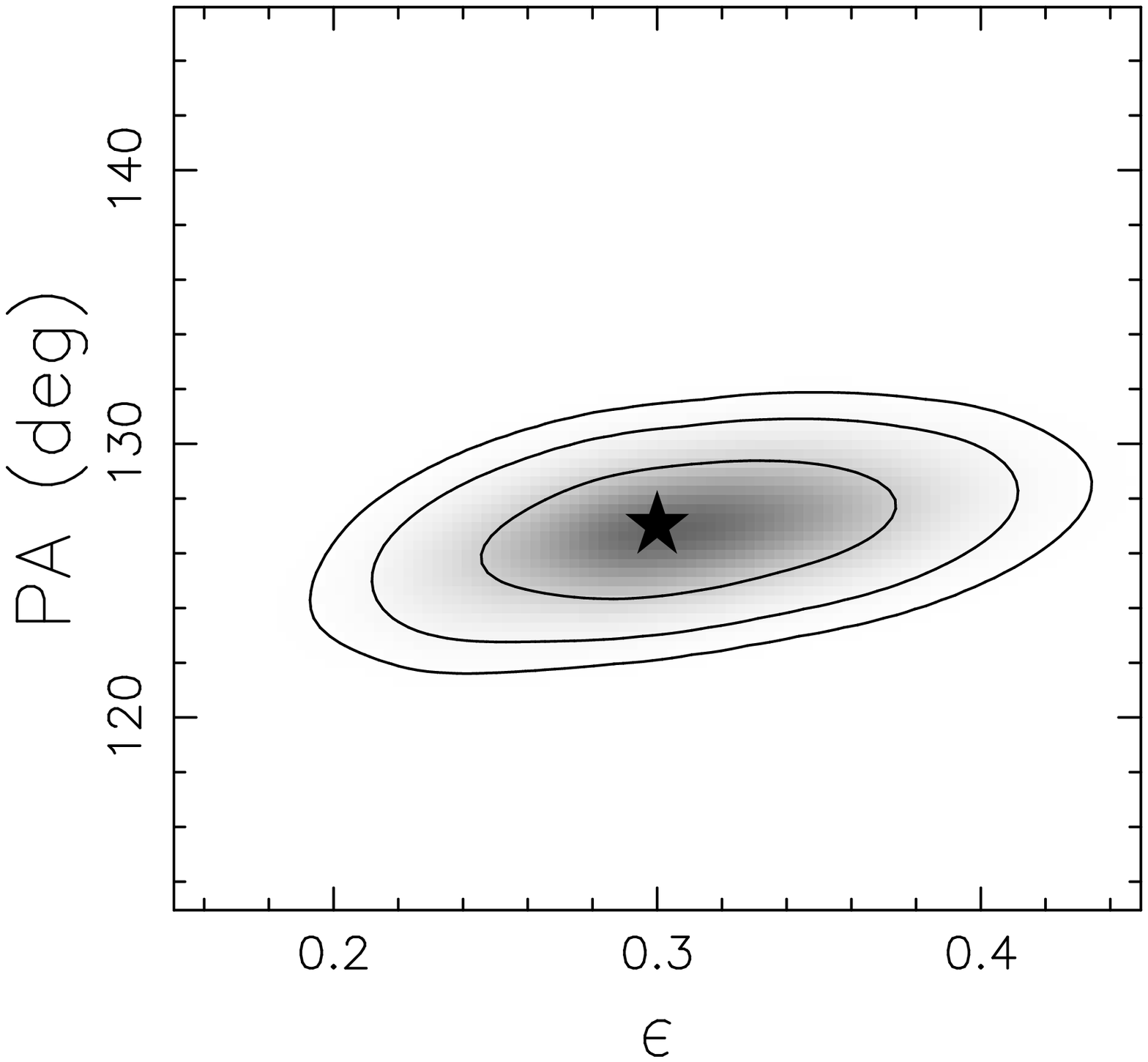} & 
\includegraphics[width=0.28\linewidth]{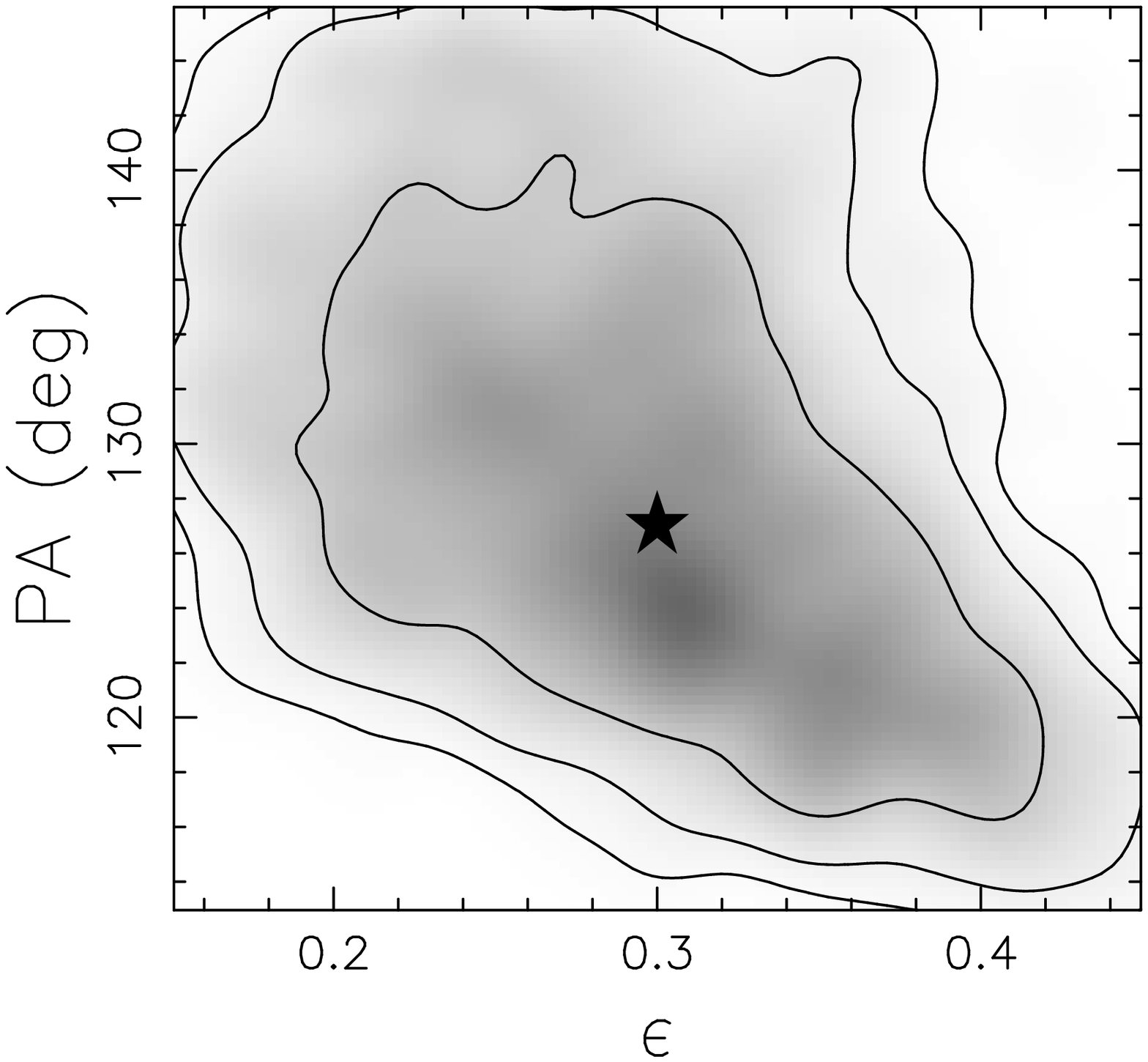} & 
\includegraphics[width=0.28\linewidth]{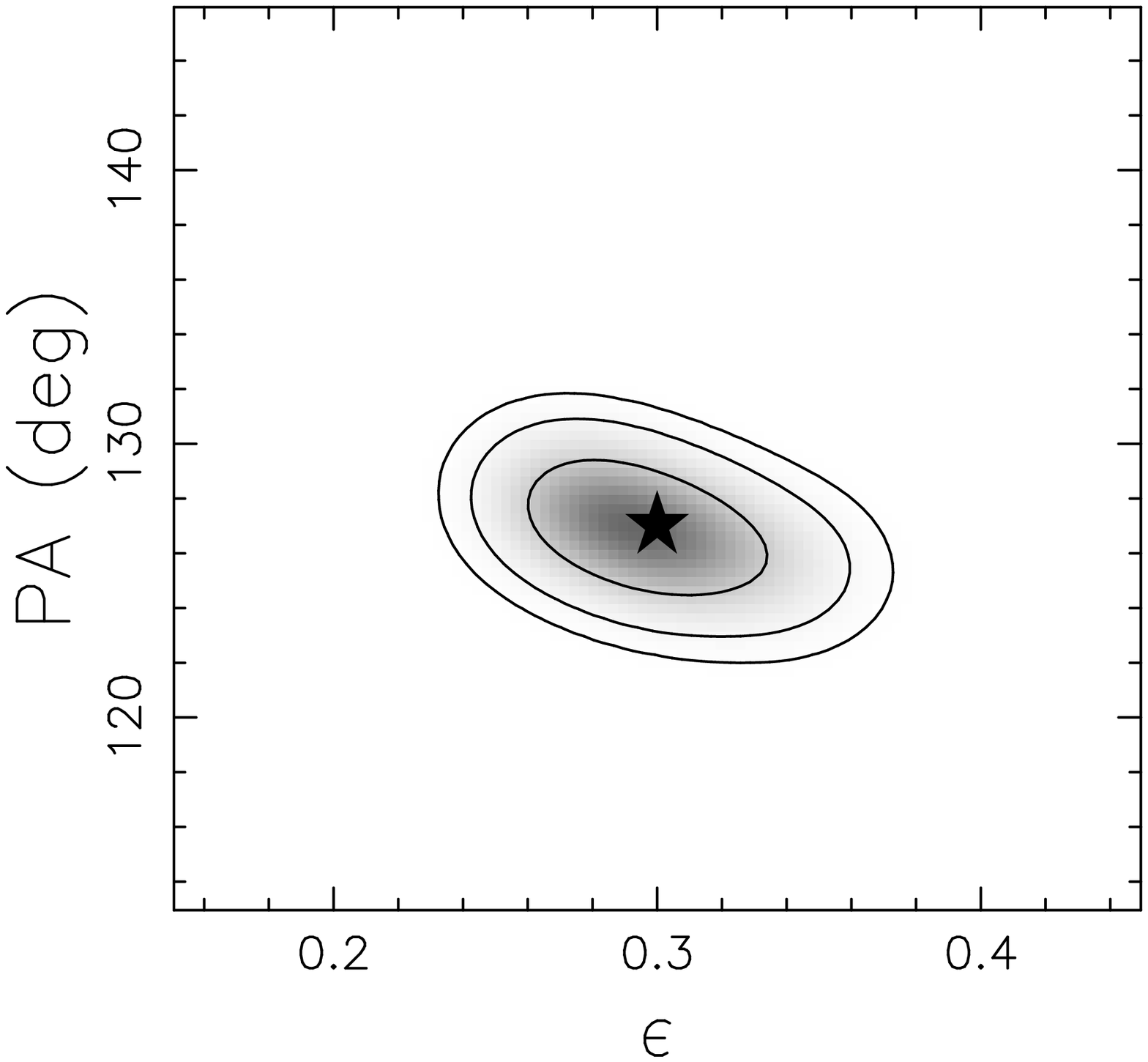} \\ 
\includegraphics[width=0.28\linewidth]{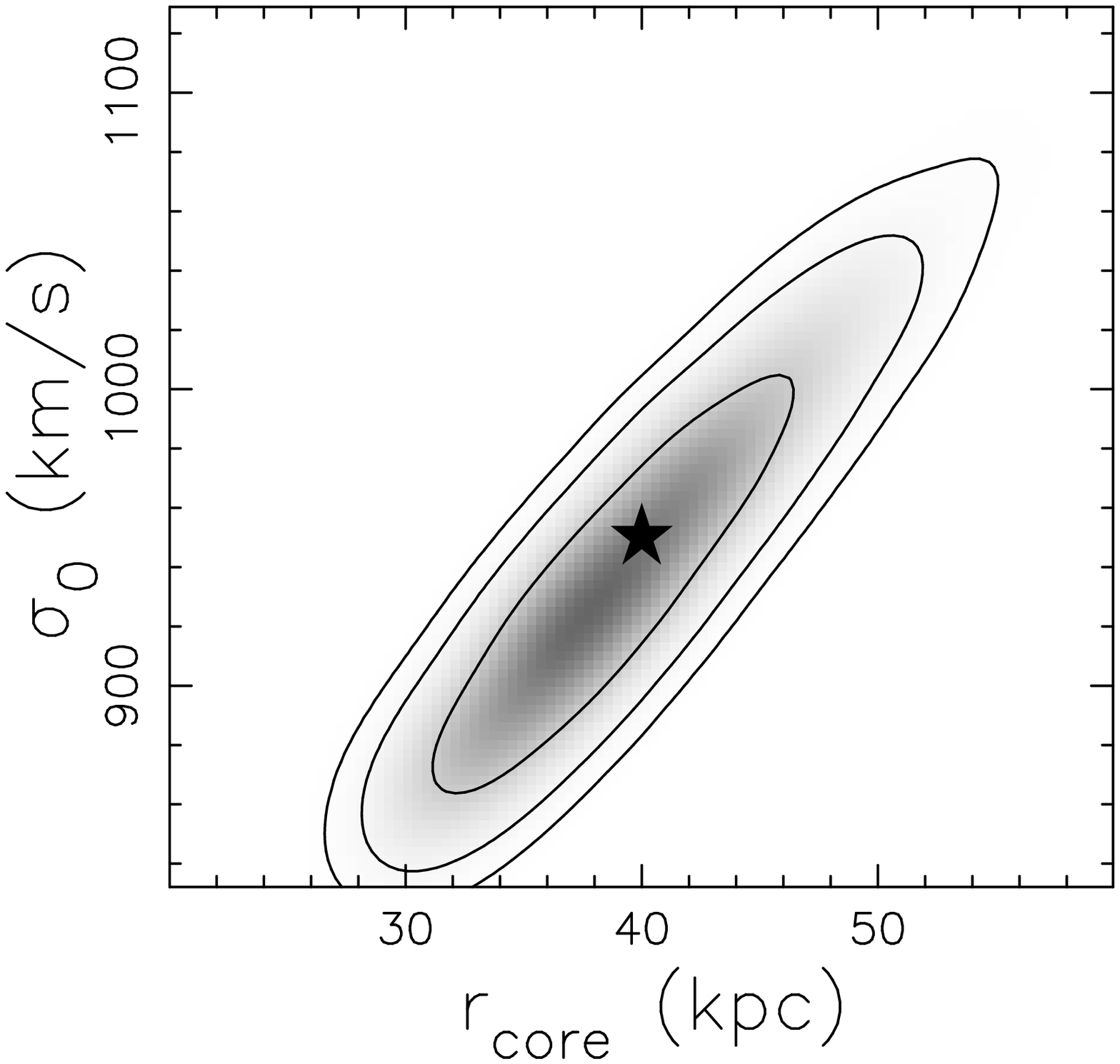} & 
\includegraphics[width=0.28\linewidth]{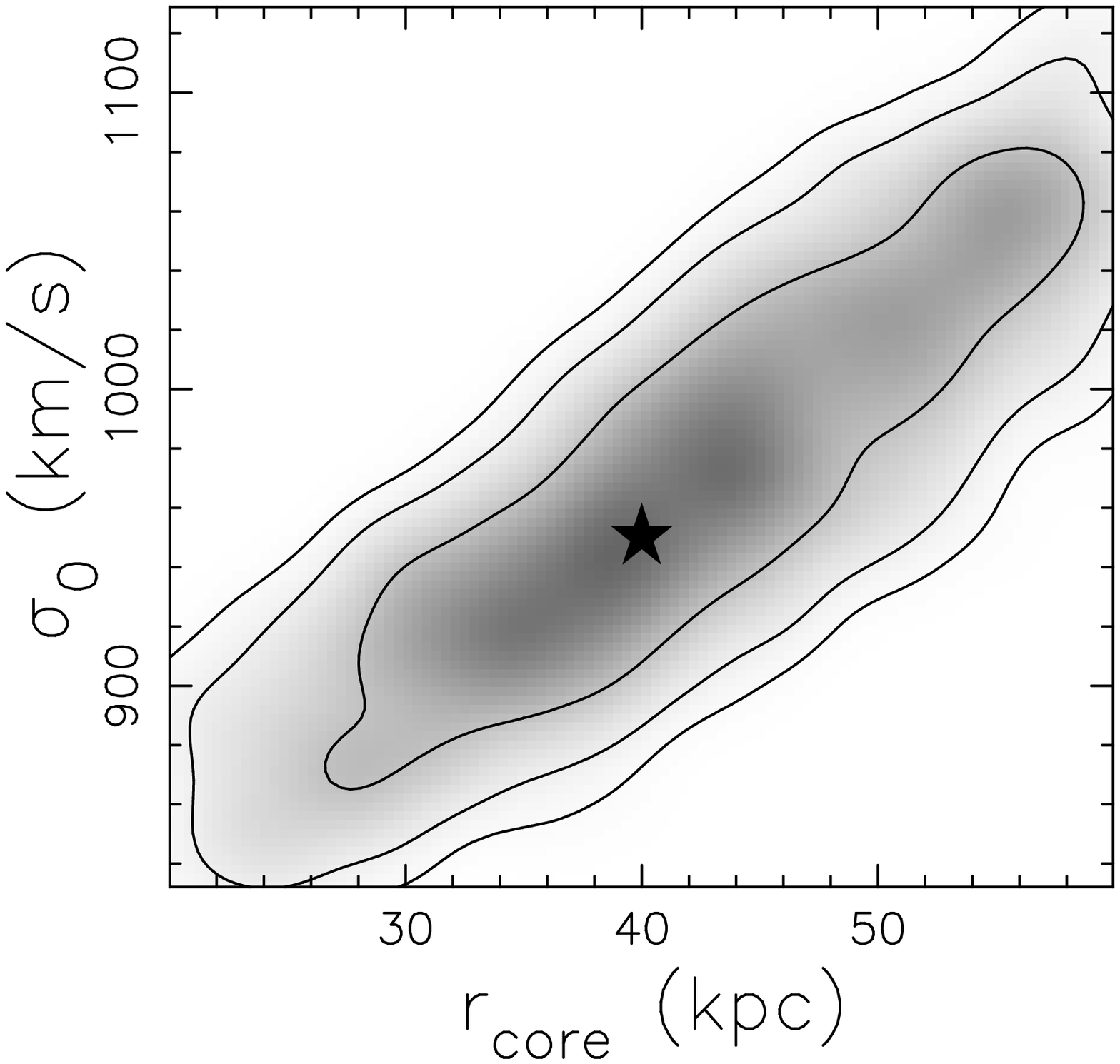} & 
\includegraphics[width=0.28\linewidth]{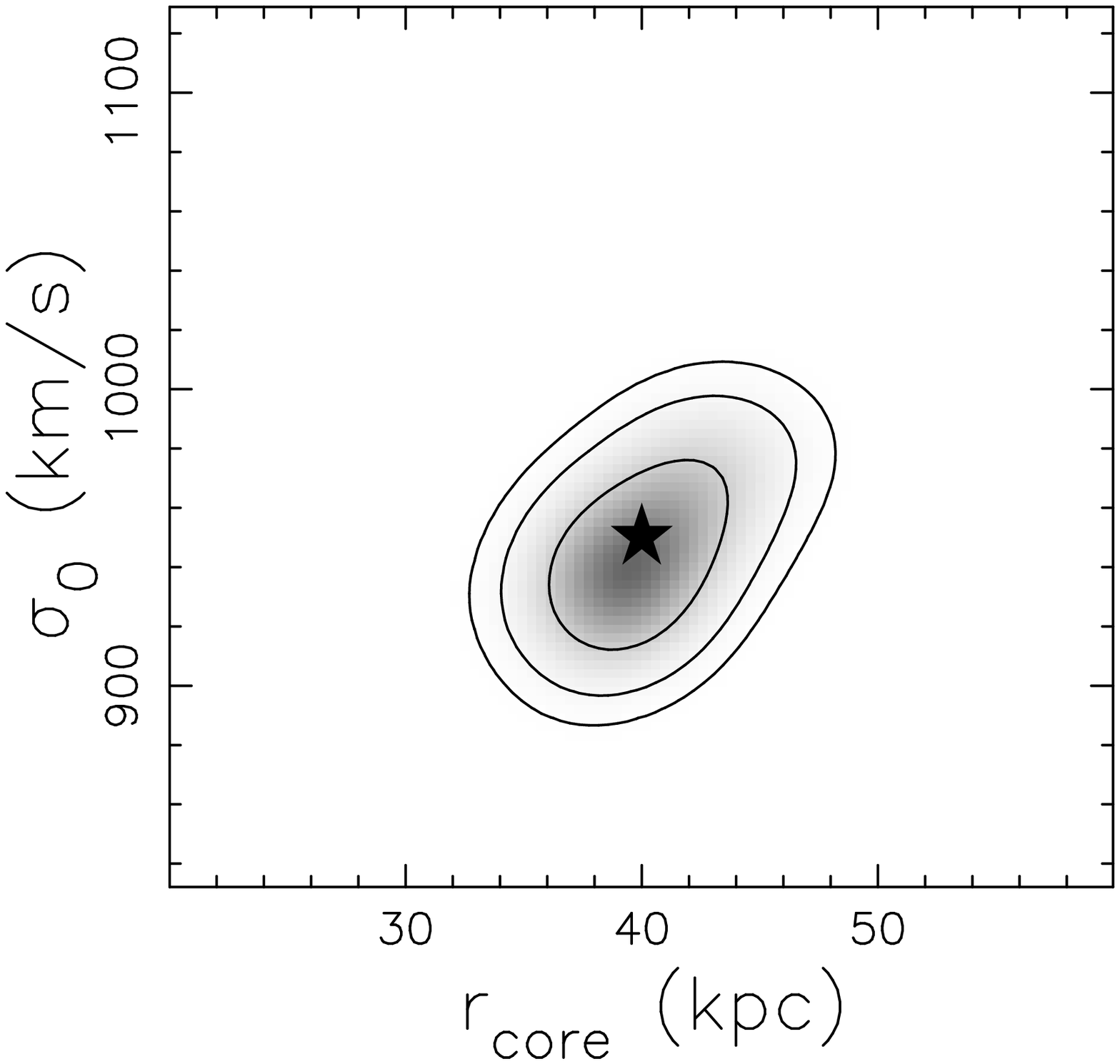} \\ 
\includegraphics[width=0.28\linewidth]{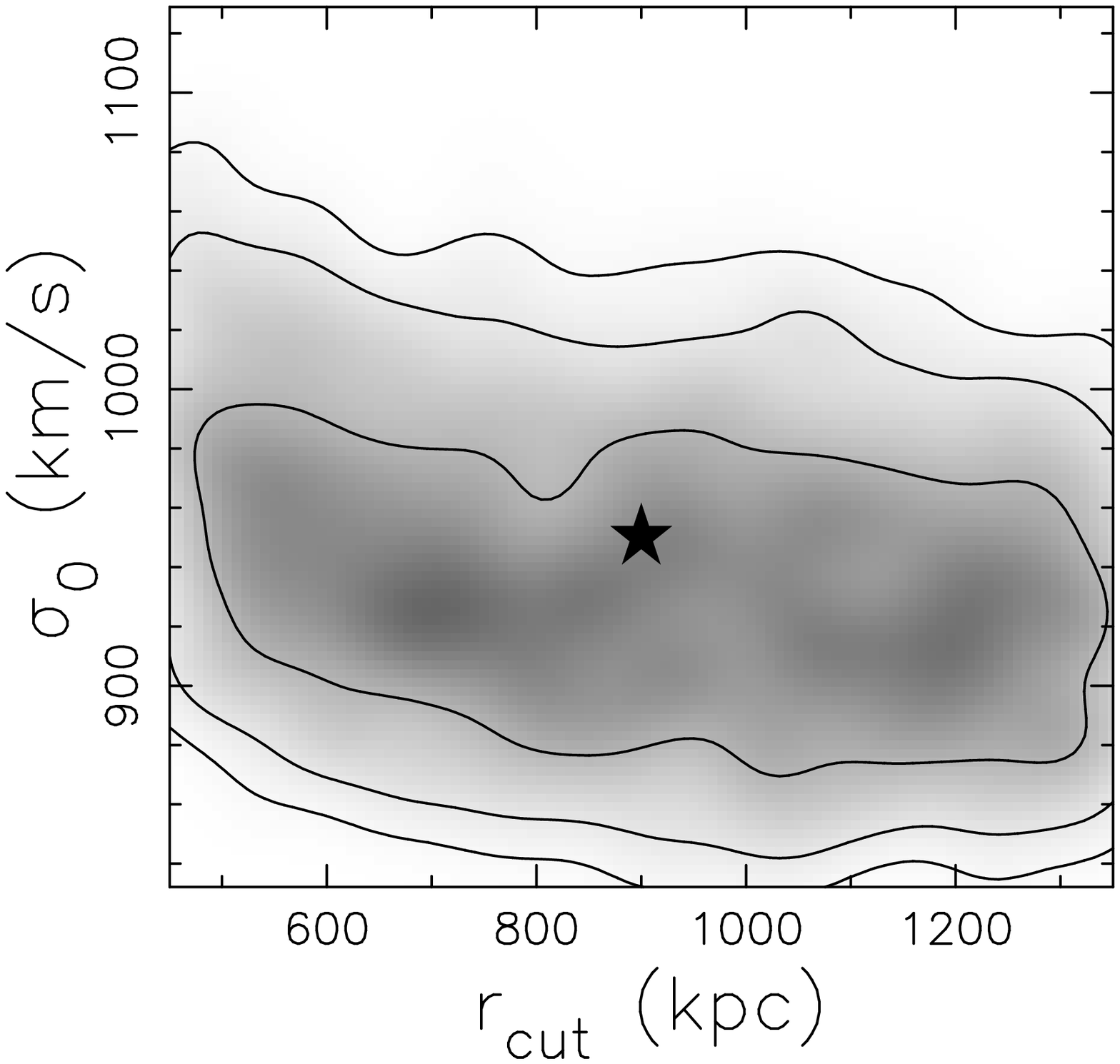} & 
\includegraphics[width=0.28\linewidth]{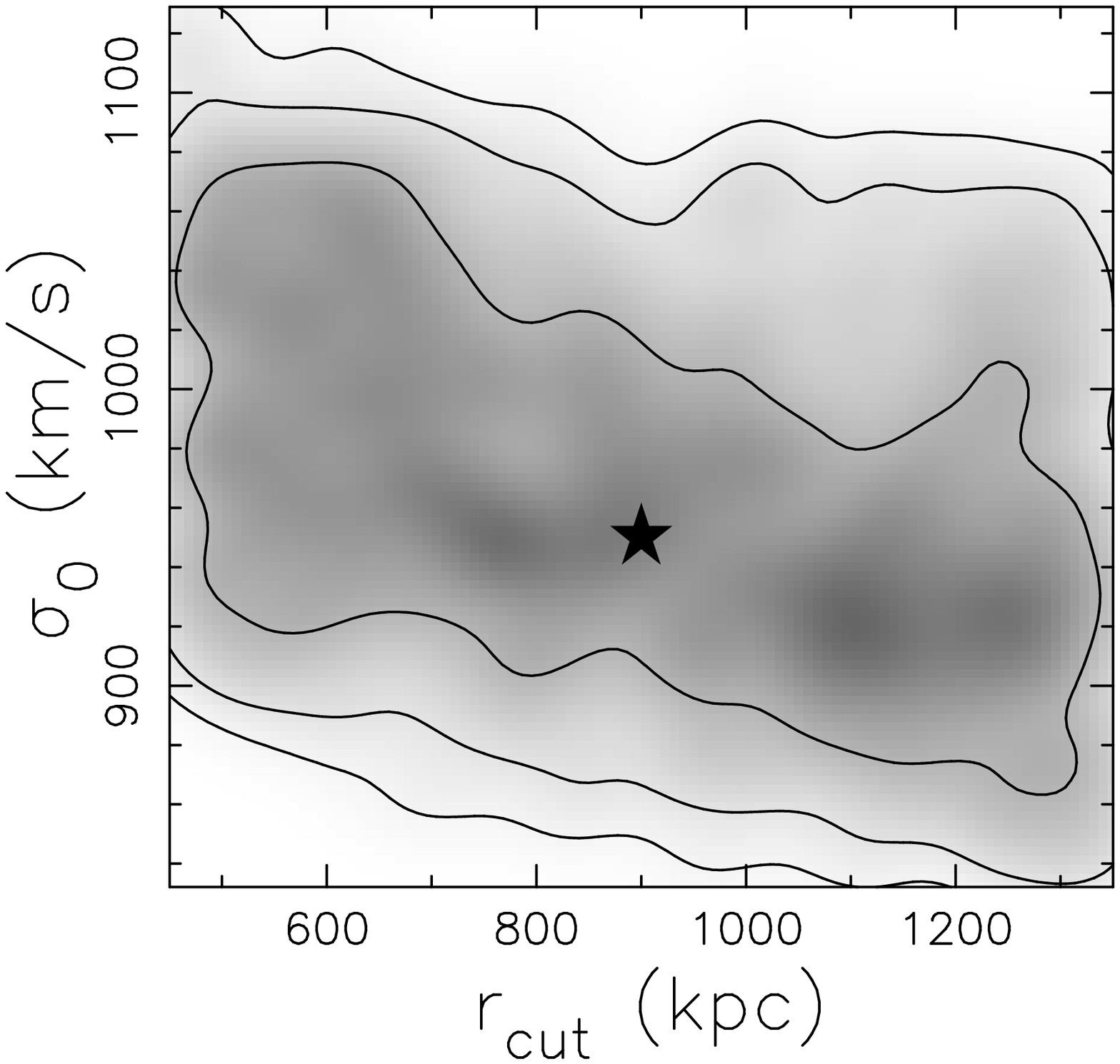} & 
\includegraphics[width=0.28\linewidth]{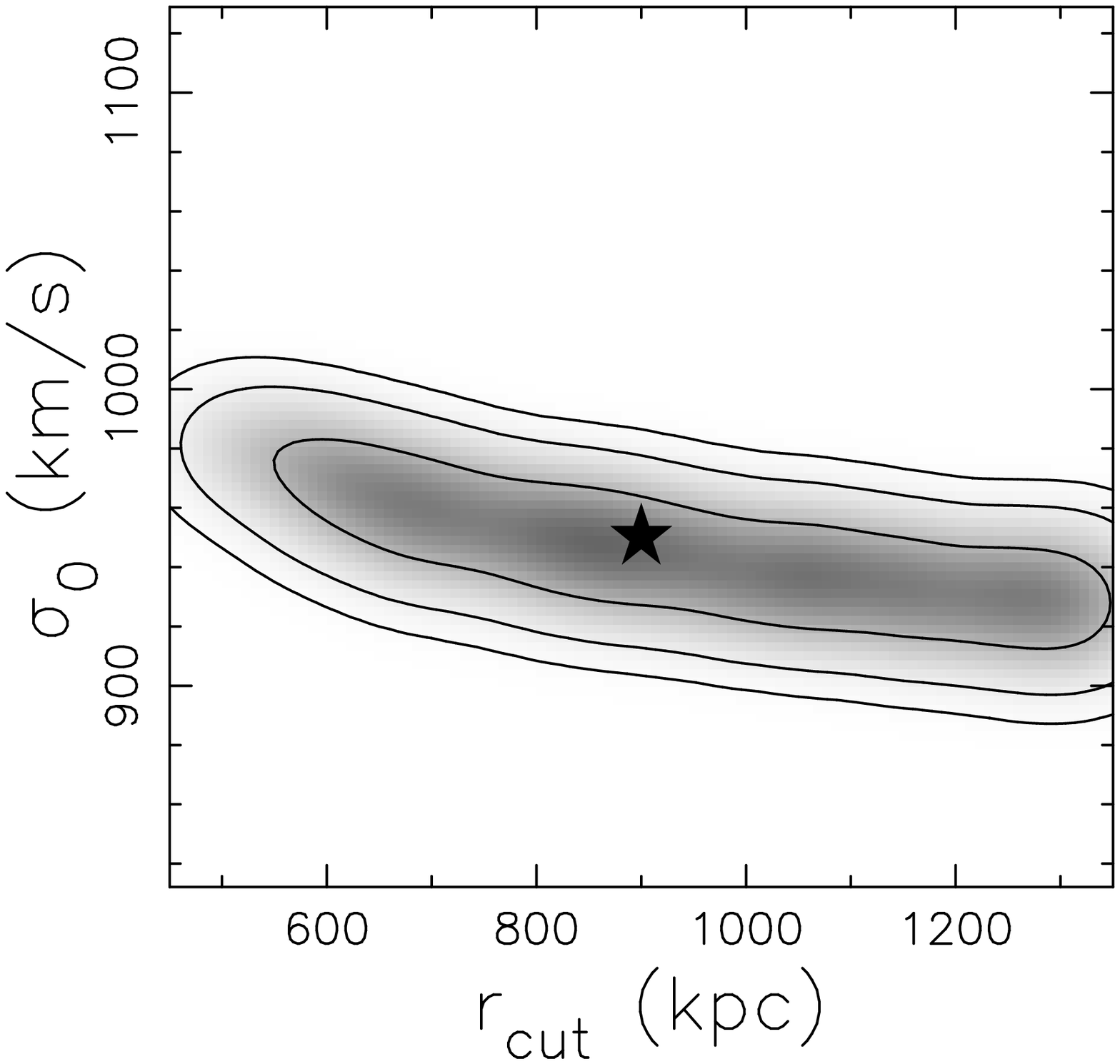} \\ 
\includegraphics[width=0.28\linewidth]{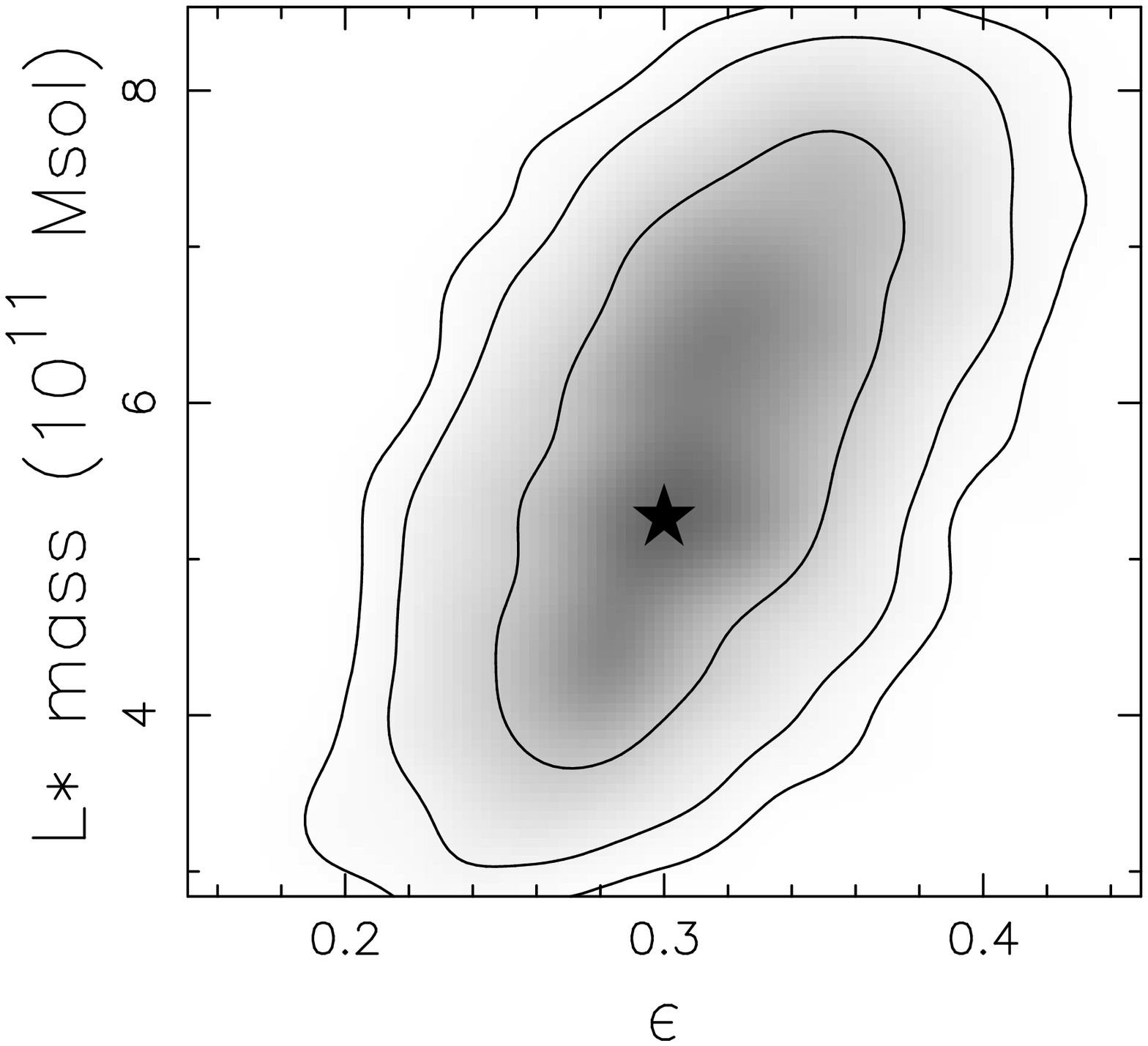} & 
\includegraphics[width=0.28\linewidth]{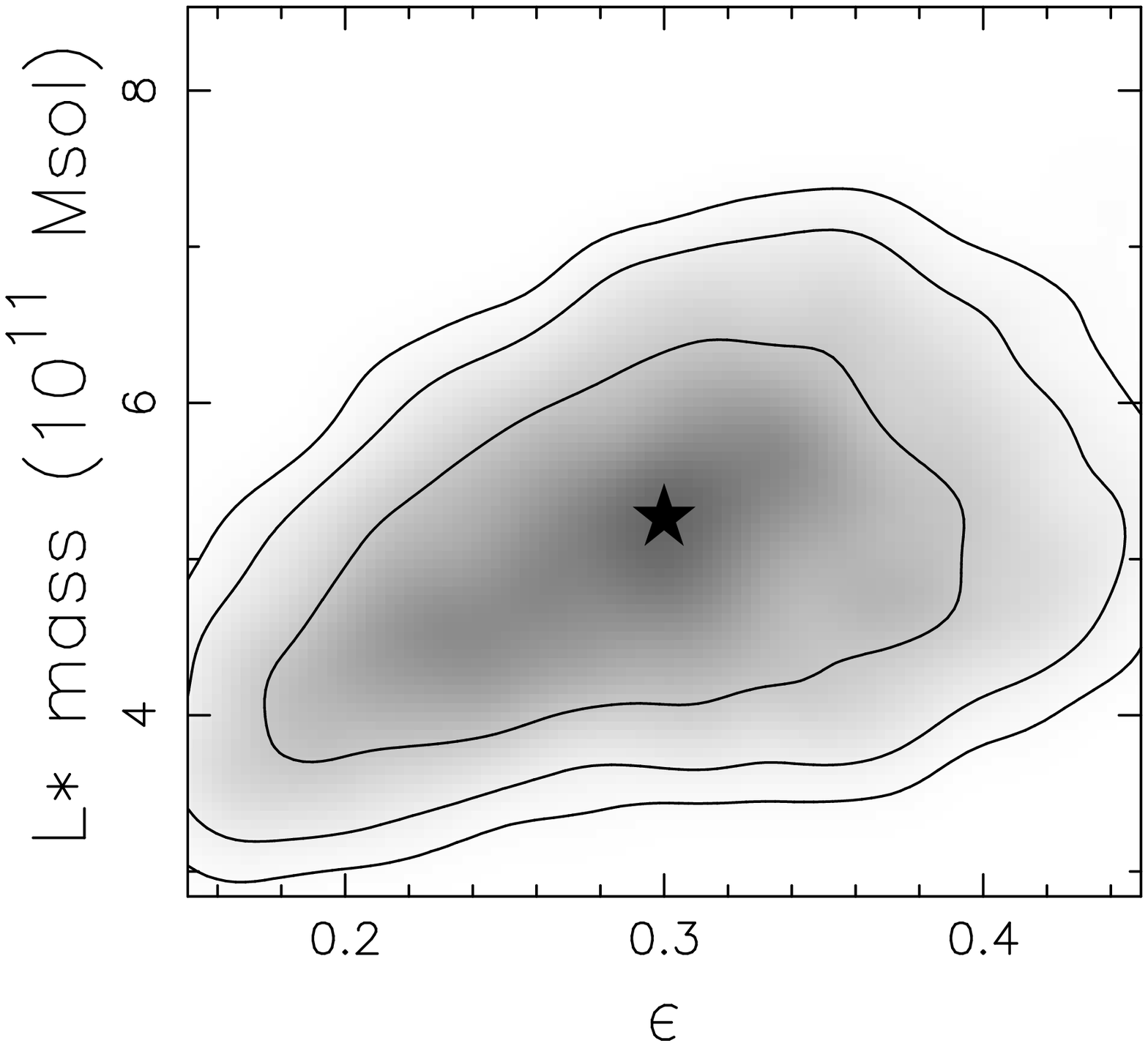} & 
\includegraphics[width=0.28\linewidth]{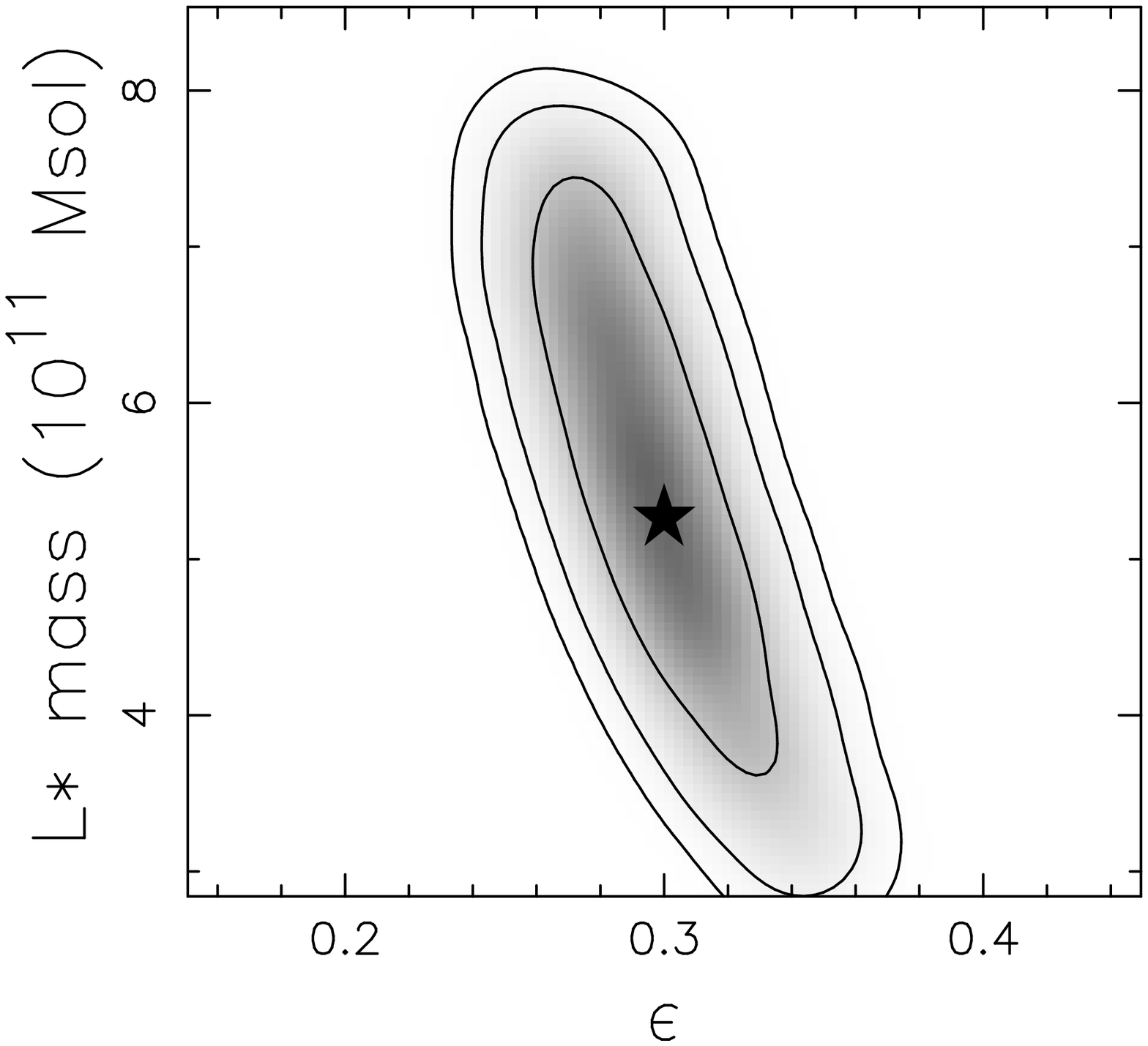} \\ 
\includegraphics[width=0.28\linewidth]{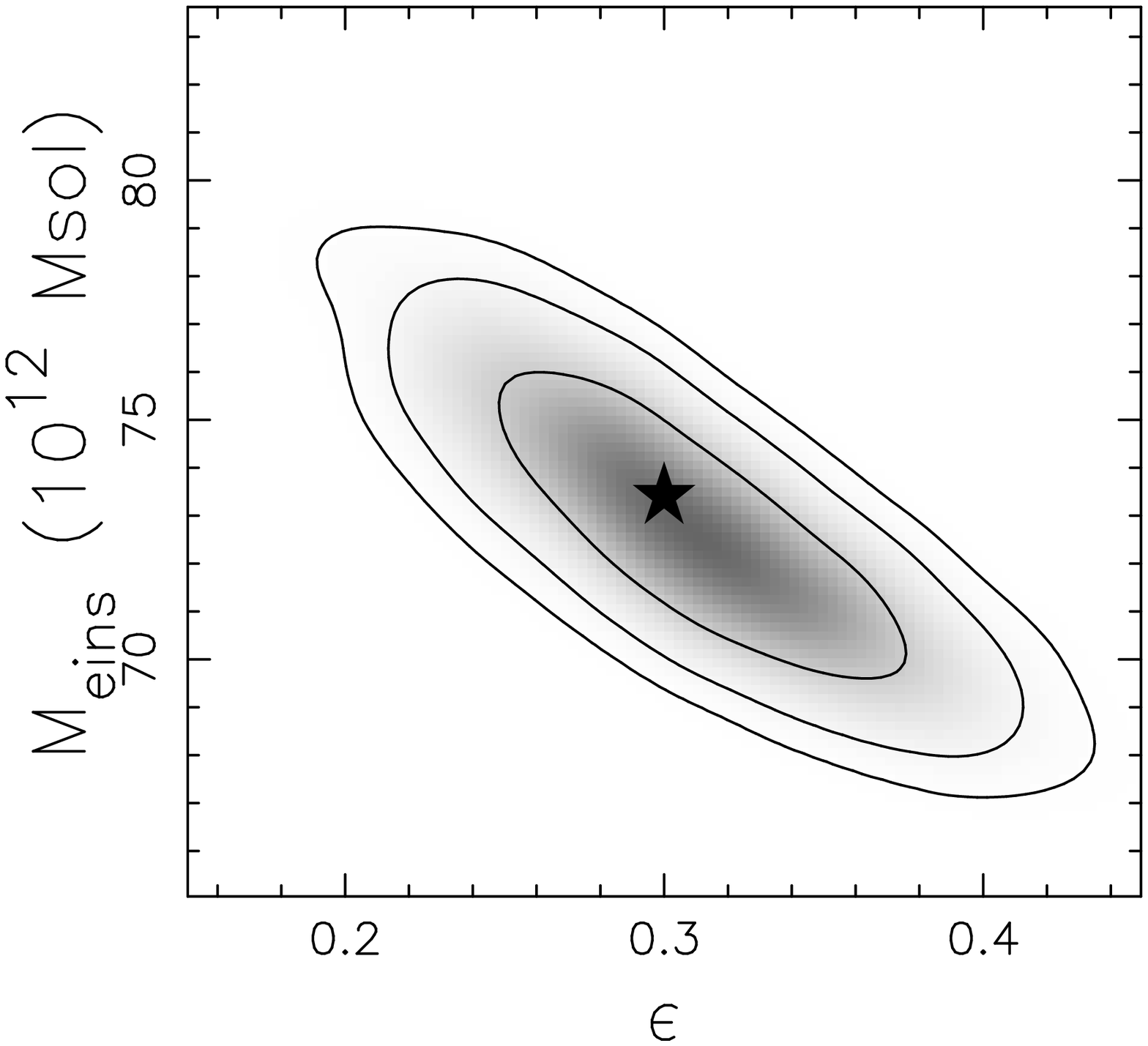} & 
\includegraphics[width=0.28\linewidth]{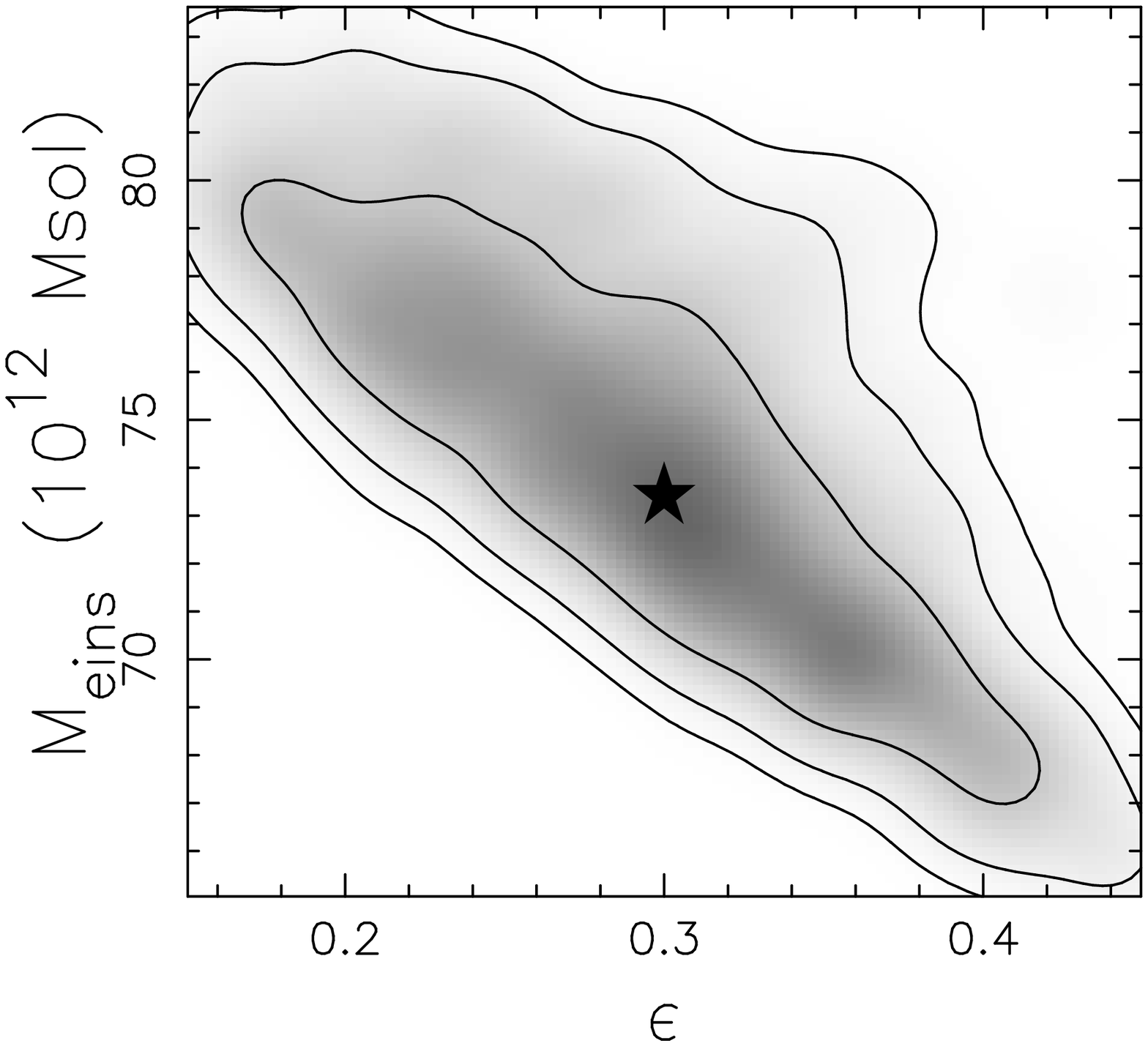} & 
\includegraphics[width=0.28\linewidth]{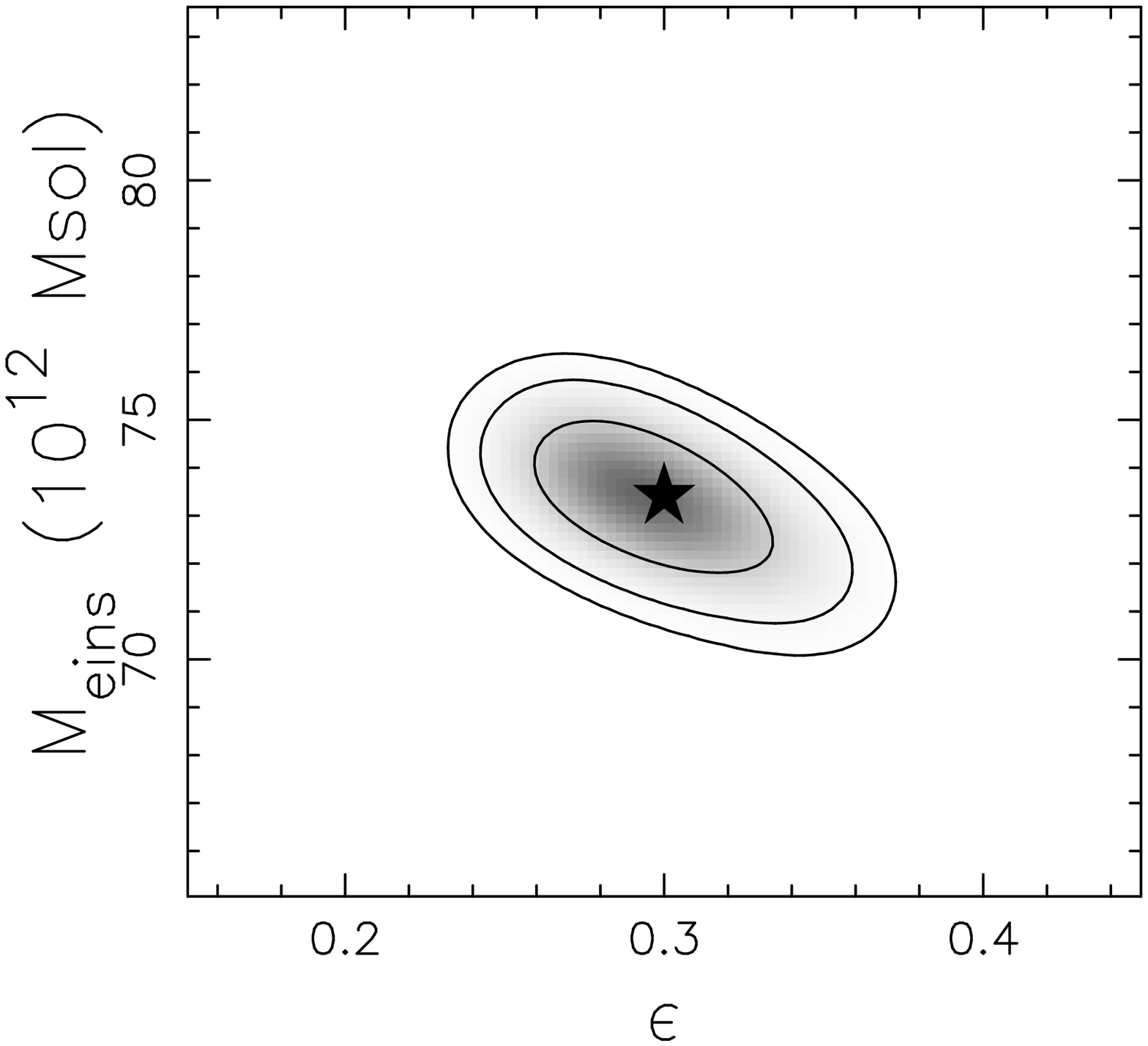} \\ 

\includegraphics[width=0.28\linewidth]{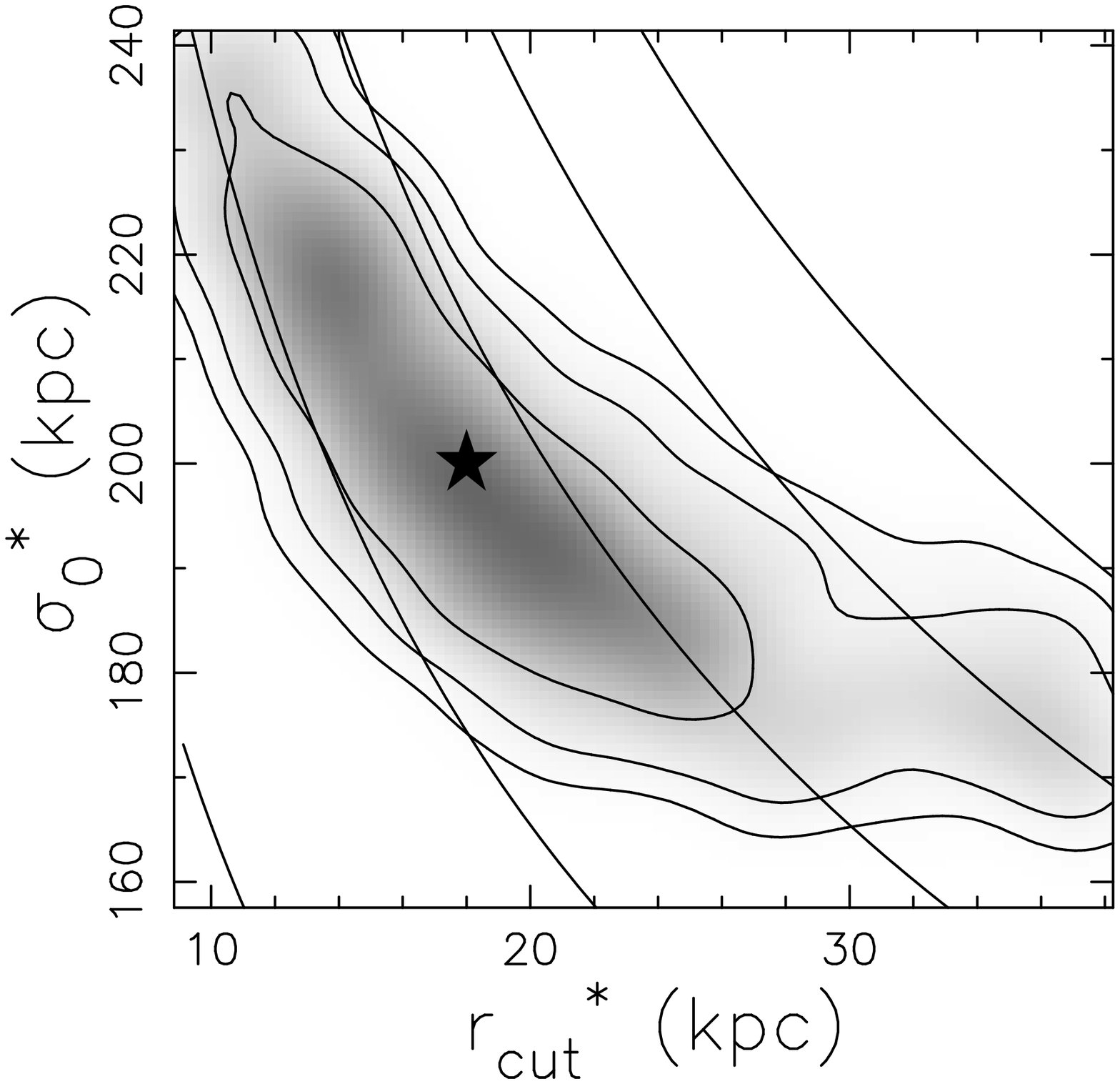} & 
\includegraphics[width=0.28\linewidth]{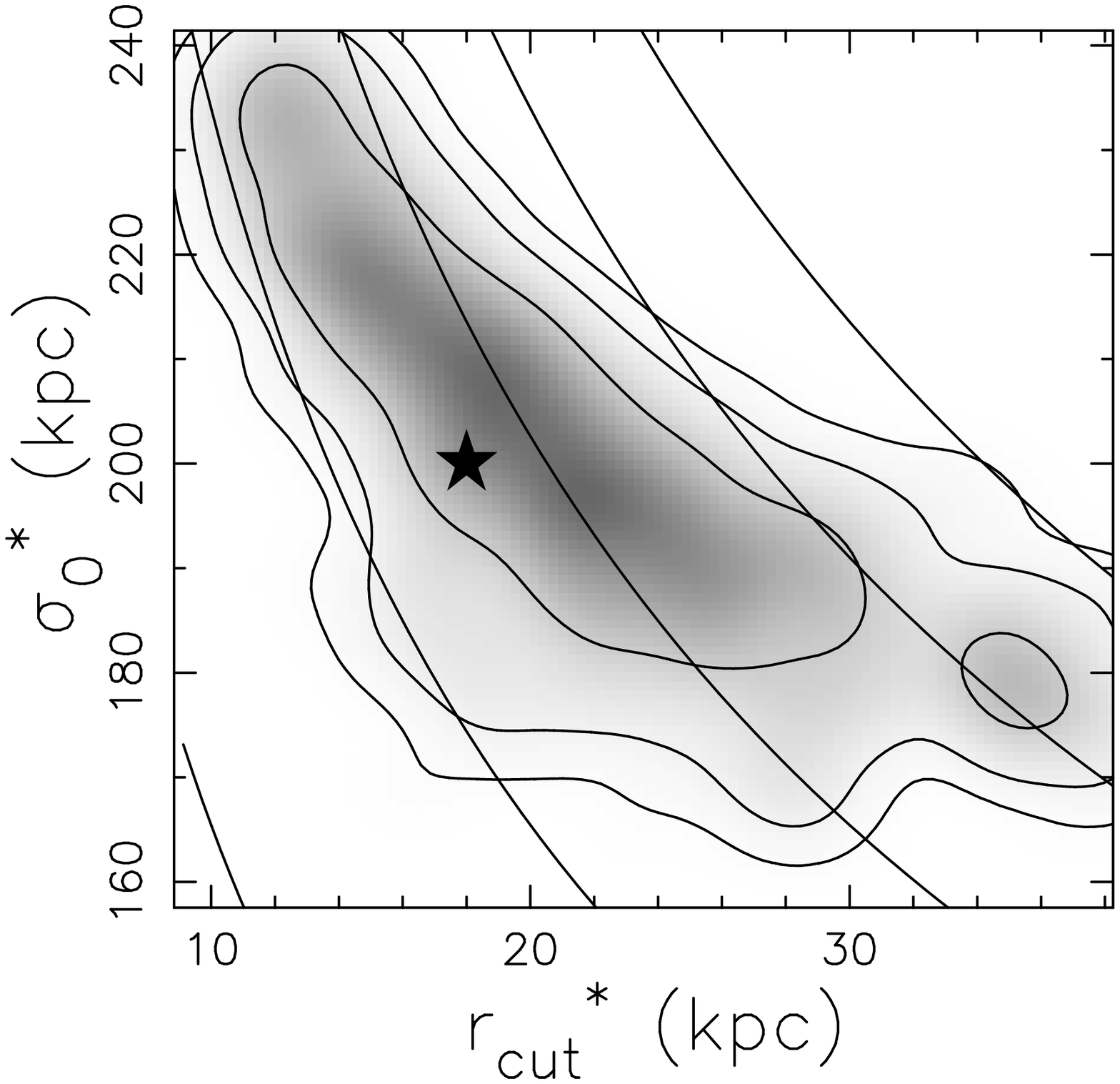} & 
\includegraphics[width=0.28\linewidth]{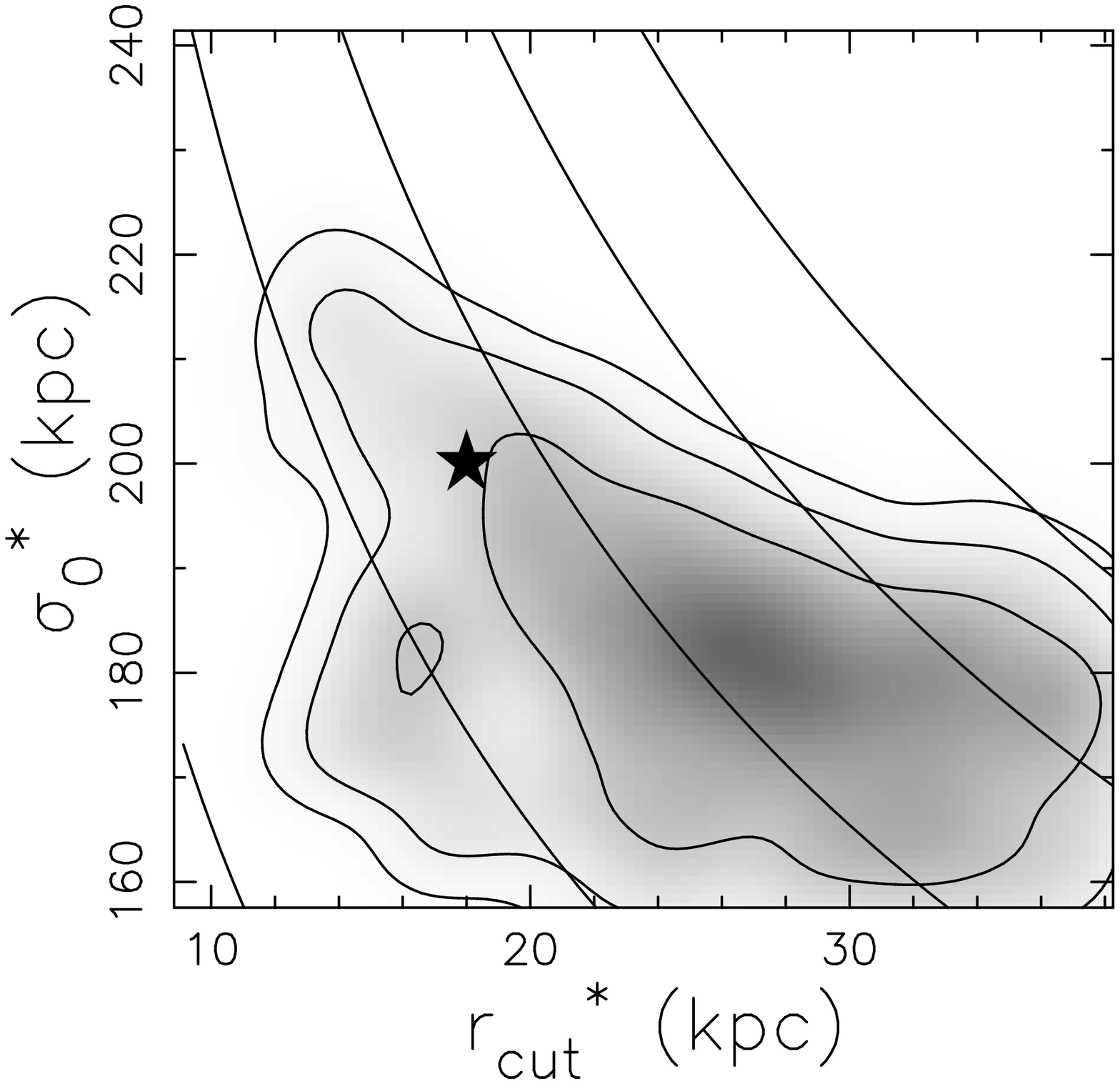} \\ 
\includegraphics[width=0.28\linewidth]{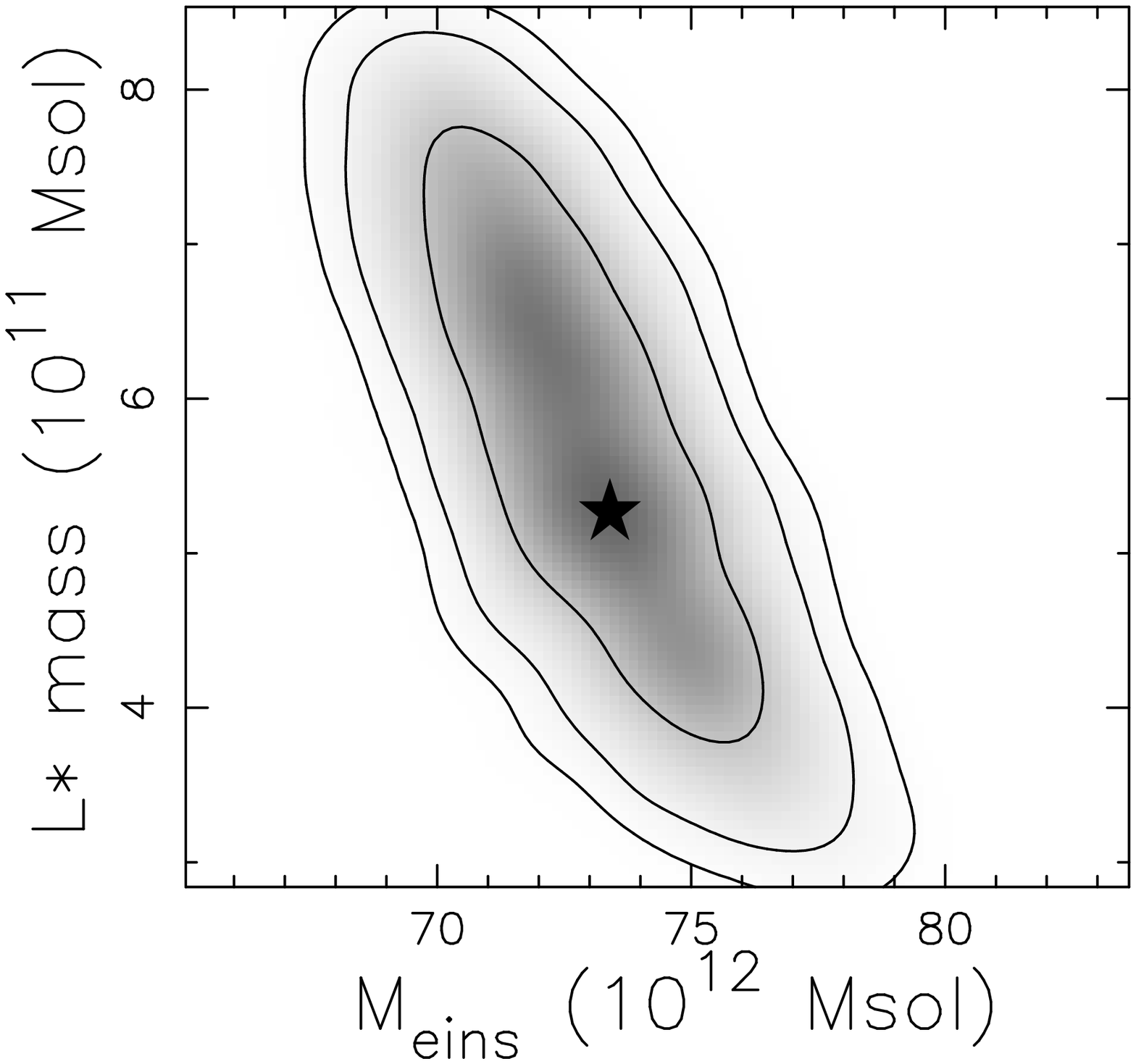} & 
\includegraphics[width=0.28\linewidth]{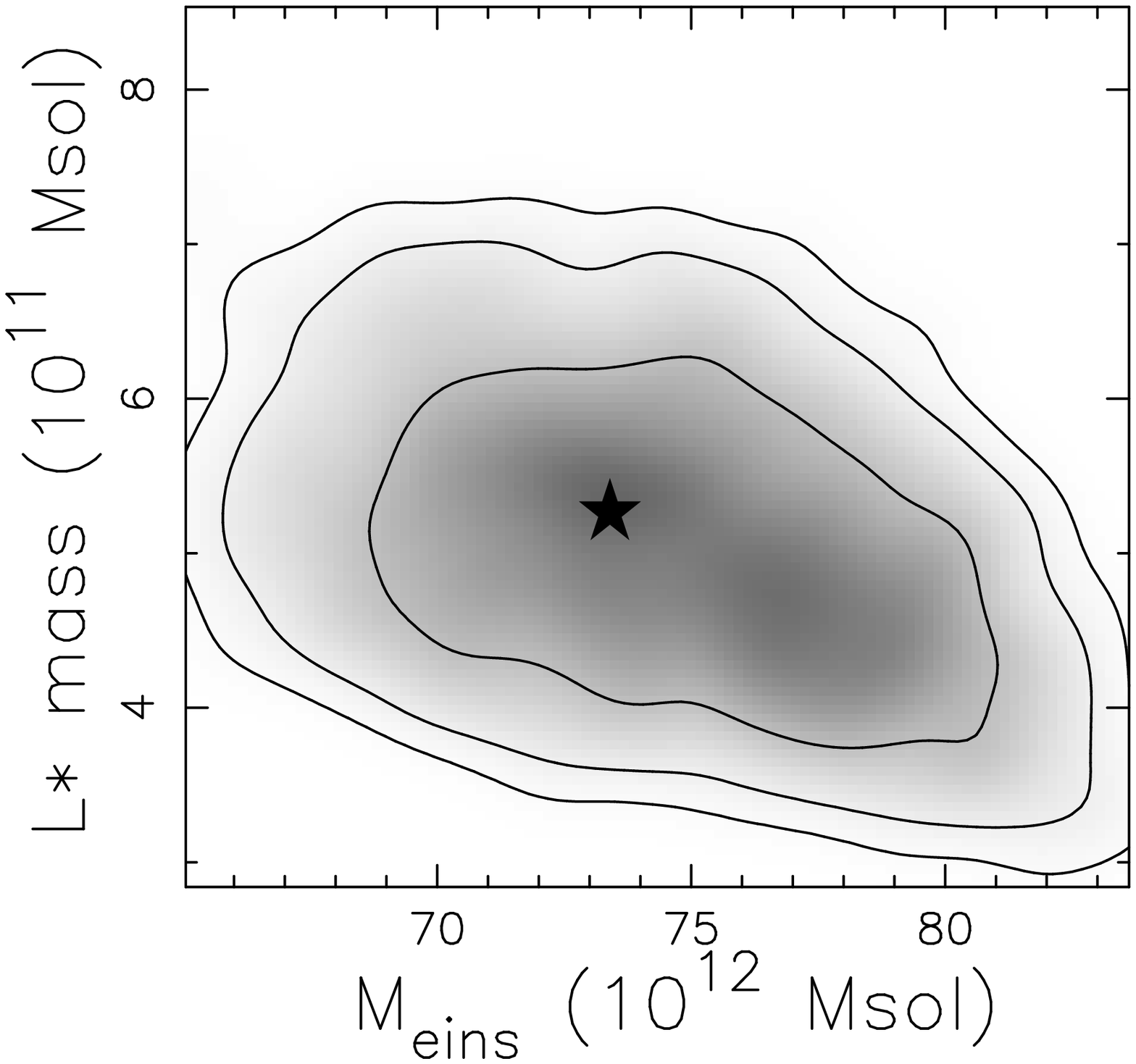} & 
\includegraphics[width=0.28\linewidth]{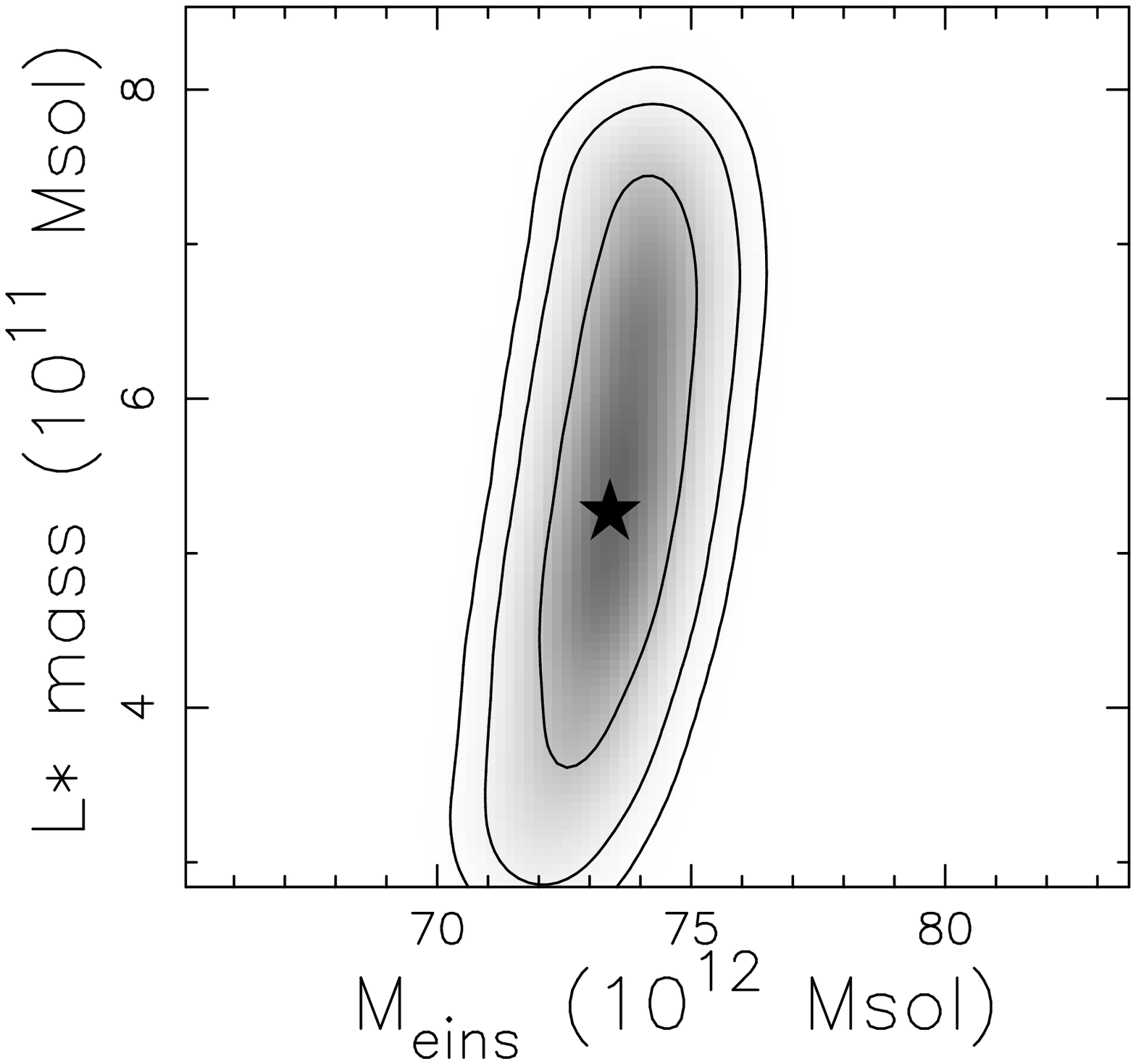} \\ 
\end{tabular} 
\end{indented} 
\caption{\label{figpiemd} 
 2D marginalized posterior PDFs for the parameters of the
 cluster-scale halo modelled with a PIEMD potential obtained, from
 left to right, with multiple image configurations 1, 2 and 3
 respectively. The 3 contours stand for the 68\%, 95\% and 99\% CL.
 The input values to the simulation are marked by the stars. The mass
 of an $L^\star$ galaxy is the total mass for a circular profile. The
 plotted contours in the $r_{cut}^\star$--$\sigma_0^\star$ plot are
 isodensity contours. The cluster mass $M_{eins}$ is the inferred
 total enclosed mass (i.e.  galaxy subhalos and cluster-scale halo)
 within the Einstein radius (30'').  }

\end{figure*} 

 The obtained posterior PDF is marginalized
 (by making a histogram in two dimensions and ignoring the samples' other
 parameters), and plotted in
 Figure~\ref{figpiemd}. The estimated (median) parameters are given in
 Table~\ref{tabdegepiemd}. In every configuration, the input
 values are recovered well, but strong degeneracies appear.

 First, we note that the posterior PDF is more compact in Config.~3
 than in Config.~1 and 2, in concordance with the number of radial arcs
 in each configuration. This is in agreement with the results of
 \citet{miralda1995a}, who showed that the combination of radial arcs
 and their counter image provides a stringent constraint on the profile
 shape as well as the enclosed mass.  

 Second, the velocity dispersion tightly correlates with the core
 radius, and, to a lesser extent, with the cut-off radius. This is a
 mathematical degeneracy that appears when the mass enclosed by the
 Einstein radius is maintained constant (or in this case, 
 constrained tightly by the data).  Indeed, for a PIEMD
 potential, the enclosed mass is given by \citep{limousin2005}:
\begin{equation} 
\label{eq:piemdma}
M_{aper}(<R) = \frac{\pi r_{cut} \sigma_0^2}{G} \left( 1 - 
	\frac{ \sqrt{r_{cut}^2 + R^2} - \sqrt{r_{core}^2+R^2} }
	{ r_{cut} - r_{core} } \right)\;.
\end{equation} 
 Thus, for a mass enclosed into a large circle of radius $R \sim r_{cut}$, 
 we derive
 $\sigma_0^2 \propto 1/r_{cut}$. At a smaller radius, assuming $r_{core} \ll
 r_{cut}$, the 3D density
 approximates $\rho = \rho_0 / ( 1 + R^2/r_{core}^2)$ and the
 corresponding enclosed mass becomes 
\begin{equation}
M_{aper}(<R) = \frac{\pi \sigma_0^2}{G} \left( \sqrt{r_{core}^2 + R^2}
               - r_{core} \right)\;,
\end{equation}
 For a constant aperture mass, we then obtain $\sigma_0^2 \propto
 (\sqrt{r_{core}^2 + R^2} - r_{core} )^{-1}$, which is also equivalent
 to $\sigma_0^2 \propto \frac{1}{R^2} ( r_{core} + \sqrt{r_{core}^2 +
 R^2} )$,  an increasing function of $r_{core}$ resembling
 the observed degeneracy.

 Third, in Config.~3, the cluster-scale cut-off radius is slightly
 better constrained than in Config.~1 or 2. Since strong lensing
 cannot probe directly the surface density at the cut-off radius, this
 result is just a product of the aperture mass definition
 \ref{eq:piemdma} and the stringent constraints obtained for
 $r_{core}$ and $\sigma_0$. 

 Fourth, we observe changes in the slopes of the ellipticity--PA, the
 ellipticity--$L^\star$ mass and the $M_{eins}$--$L^\star$ mass
 degeneracies between Config.~1., 2 and 3. 

 This more effect is due to a subtle interaction 
 between the cluster-scale halo and the galaxy-scale
 subhalos' mass distributions during the inference. 
 In particular, in Config.~2, we suggest that when the ellipticity increases,
 alignment of the cluster with the giant arcs $B$ and $C$ is favoured.
 However, in Config.~1 and 3, this
 behaviour is not so clear, probably because of the presence of radial
 arcs in the central region. 

 Finally, in every configuration, the scaling relation parameters
 $r_{cut}^\star$ and $\sigma_0^\star$ are strongly degenerate, with
 the degeneracy closely following the constant mass contours
 over-plotted with solid lines.  In Table~\ref{tabdegepiemd}, we note
 that strong lensing cannot predict the $L^\star$ cut-off radius to
 better than 24\% accuracy, nor $\sigma_0^\star$ with better than 6\%
 accuracy.

 Although strong degeneracies have been highlighted so far for a
 cluster-scale halo modelled by a PIEMD potential, the aperture mass
 error at the Einstein radius is always smaller than 5\% and even
 reaches 0.8\% in Config.~3. (see Table~\ref{tabdegepiemd}). 

 In section \ref{sec:evid}, we show that the same precision can also
 be achieved when the input and the fitted models are different.

\begin{table*}[!h]
\caption{\label{tabdegepiemd}
 Parameter recovery for a cluster-scale halo modelled by a
 PIEMD potential, given 3 different strong lensing
 configurations.  The errors are given at 68\% CL. The $L^\star$
 galaxy masses are given for a circular mass component with identical
 dynamical parameters. }

\begin{indented}
\item[]\begin{tabular*}{\linewidth}[c]{@{}l@{}rr@{ }rr@{ }rr@{ }r} 
\br 
& Input & \multicolumn{2}{c}{Config.1} & \multicolumn{2}{c}{Config.2} & \multicolumn{2}{c}{Config.3} \\
\mr
$\epsilon$ & 0.3 &  0.31 & $\pm$0.04 &  0.30 & $\pm$0.06 &  0.29 &
$\pm$0.02 \\
$PA$ (deg) & 127. &  127.2 & $\pm$0.9 &  128.5 & $\pm$7.5 &  127.2 & $\pm$0.8 \\
$r_{core}$ (kpc) & 40. &  38.8 & $\pm$4.7 &  41.7 & $\pm$9.3 &  39.8 & $\pm$1.9 \\
$\sigma_0$ (km/s) & 950. &  937.3 & $\pm$43.9 &  966.4 & $\pm$59.8 &  946.5 & $\pm$16.6 \\
$r_{cut}$ (kpc) & 900. &  907.8 & $\pm$253.7 &  894.6 & $\pm$264.7 &  936.5 & $\pm$235.7 \\
$r_{cut}^\star$ (kpc) & 18. &  18.5 & $\pm$6.5 &  19.6 & $\pm$6.5 &
25.9 & $\pm$6.3 \\
$\sigma_0^\star$ (km/s) & 200. &  196.9 & $\pm$16.8 &  
199.0 & $\pm$16.5 &  180.4 & $\pm$12.2 \\
$M_{L^\star}$ ($10^{11}\ M_\odot$) & 5.26 &  5.3 & $\pm$1.2 &  5.1 & $\pm$0.8 &  5.7 & $\pm$1.1 \\
$M_{eins}$ ($10^{12}\ M_\odot$) & 73.4 &  73.0 & $\pm$2.0 &  74.6 & $\pm$3.8 &  73.5 & $\pm$0.6 \\
\br
\end{tabular*} 
\end{indented}
\end{table*}

%
%
\subsection{NFW posteriors distribution analysis}

 Now, we fit the NFW model with a NFW potential for the cluster-scale
 halo. Given the three configurations of multiple images, we perform
 the recovery of the cluster-scale halo parameters ($\epsilon$, PA,
 $c$ and $r_s$) as well as the  galaxy-scale subhalo scaling
 parameters $\sigma_0^\star$ and $r_{cut}^\star$.  Again, we assume
 uniform priors for the parameters, with a width of 50\% centred on
 the input values; the cD galaxy subhalo parameters are again fixed.
 We constrain 6 free parameters with 8 constraints.

\begin{figure*} 
\begin{indented} 
\item[]\begin{tabular}{ccc} 
\includegraphics[width=0.28\linewidth]{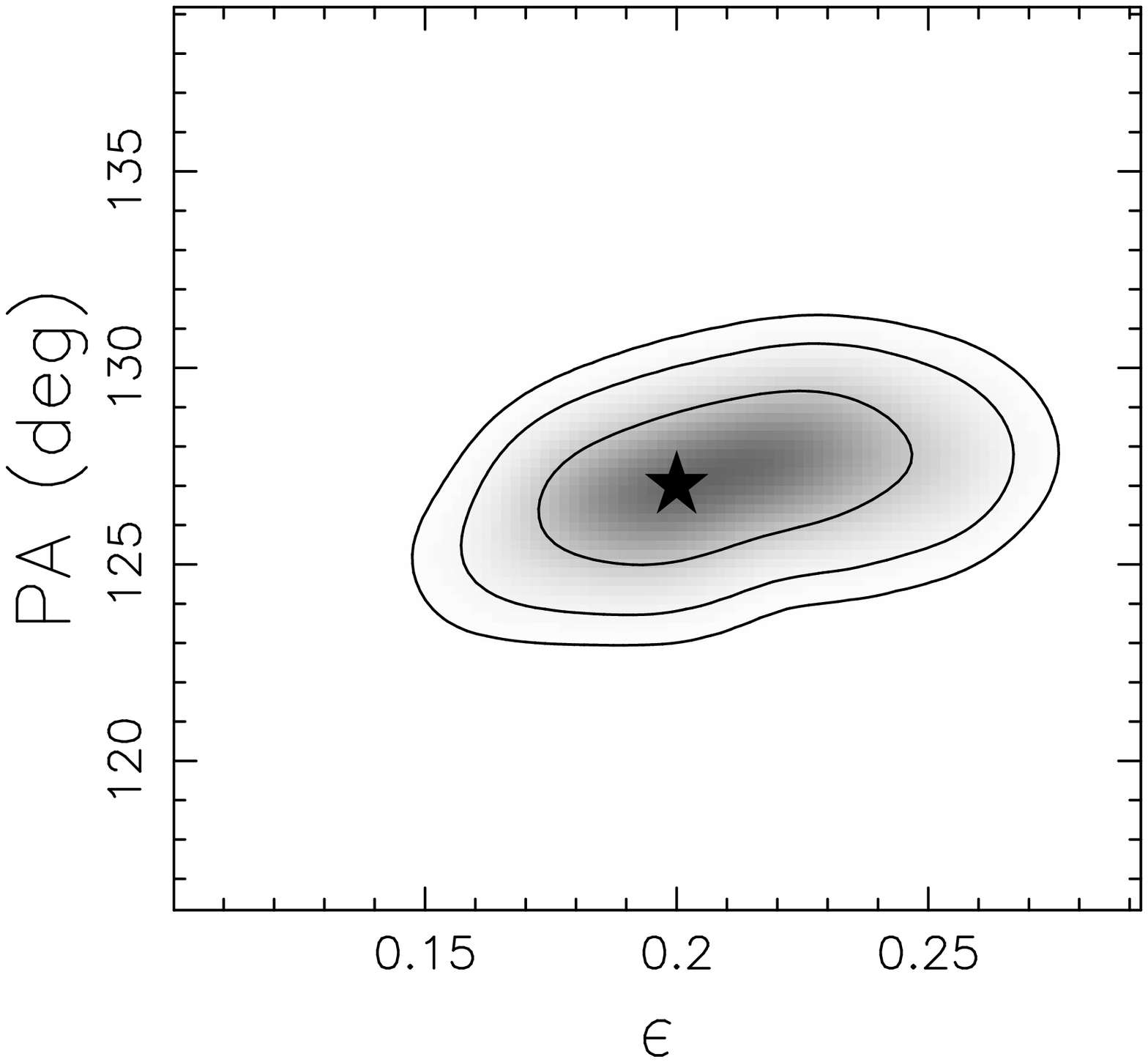} &
\includegraphics[width=0.28\linewidth]{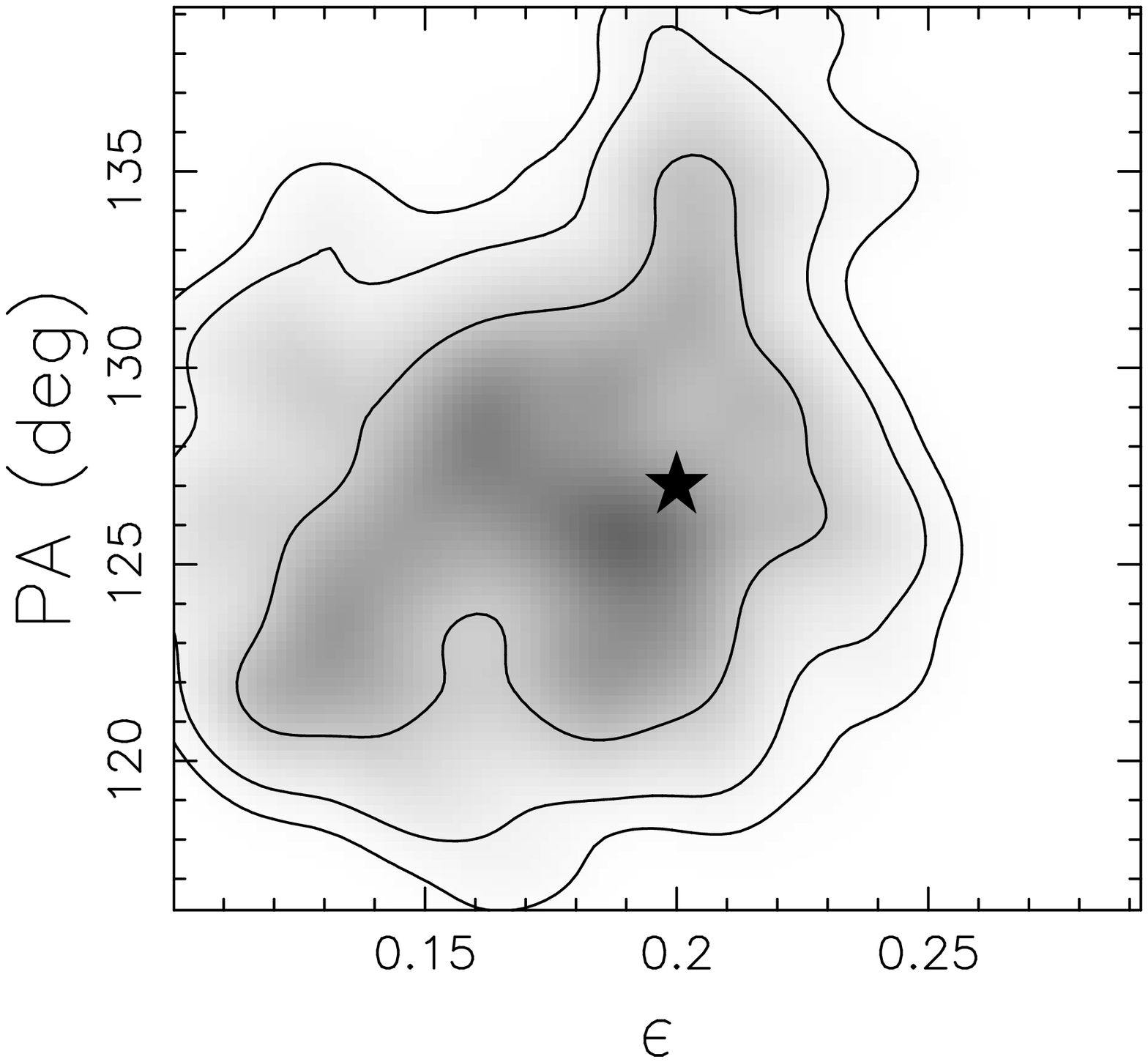} &
\includegraphics[width=0.28\linewidth]{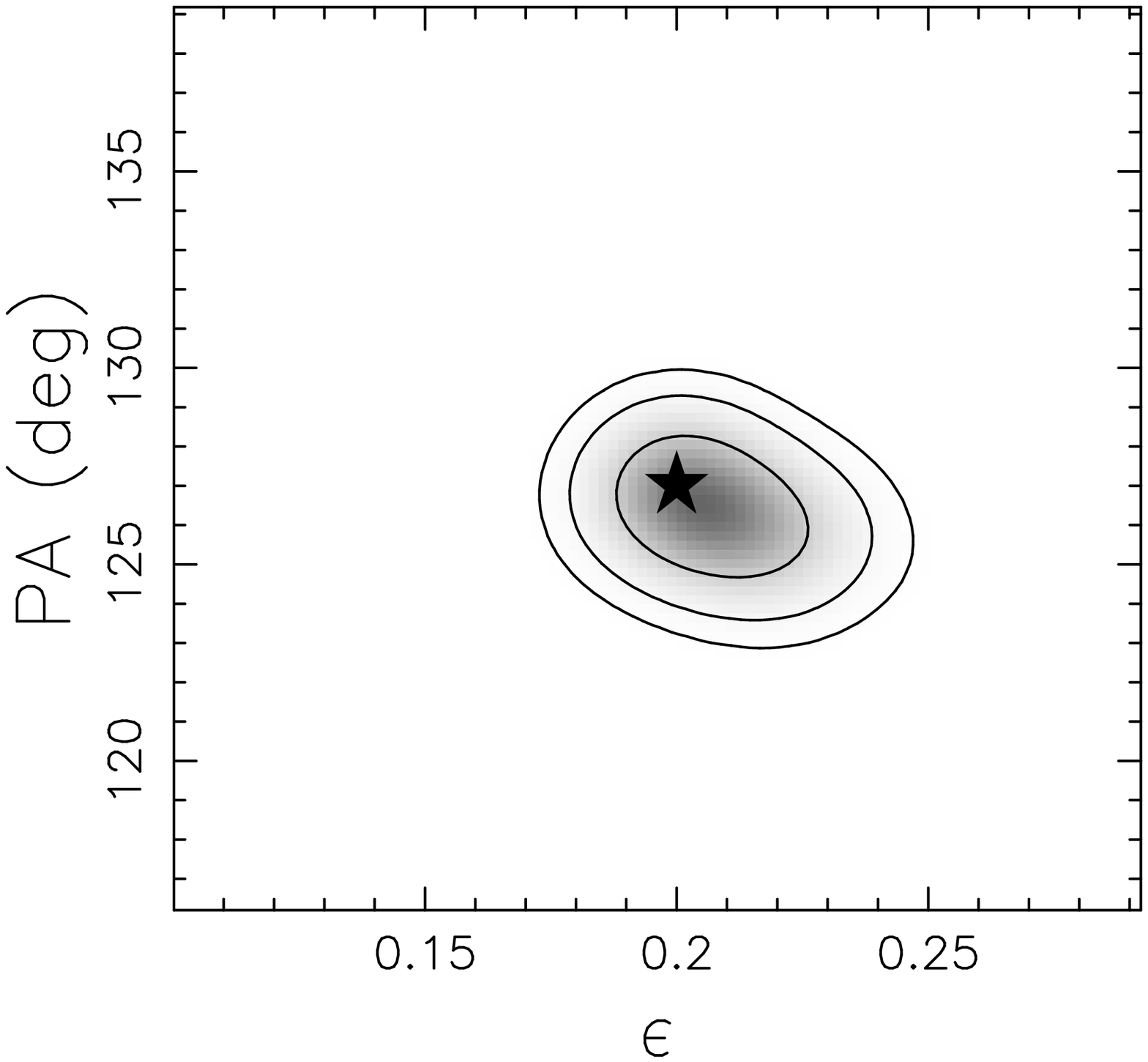} \\
\includegraphics[width=0.28\linewidth]{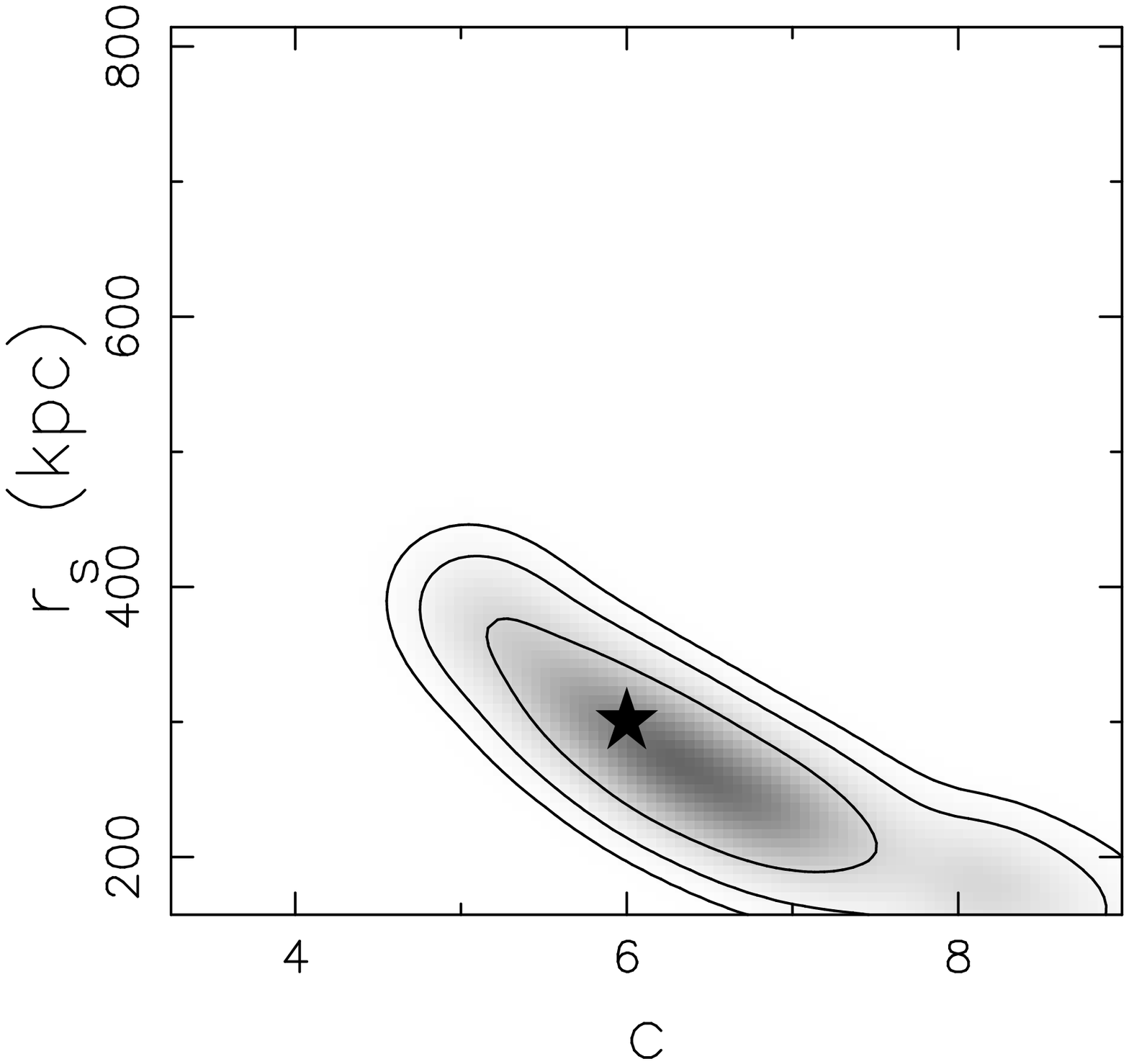} &
\includegraphics[width=0.28\linewidth]{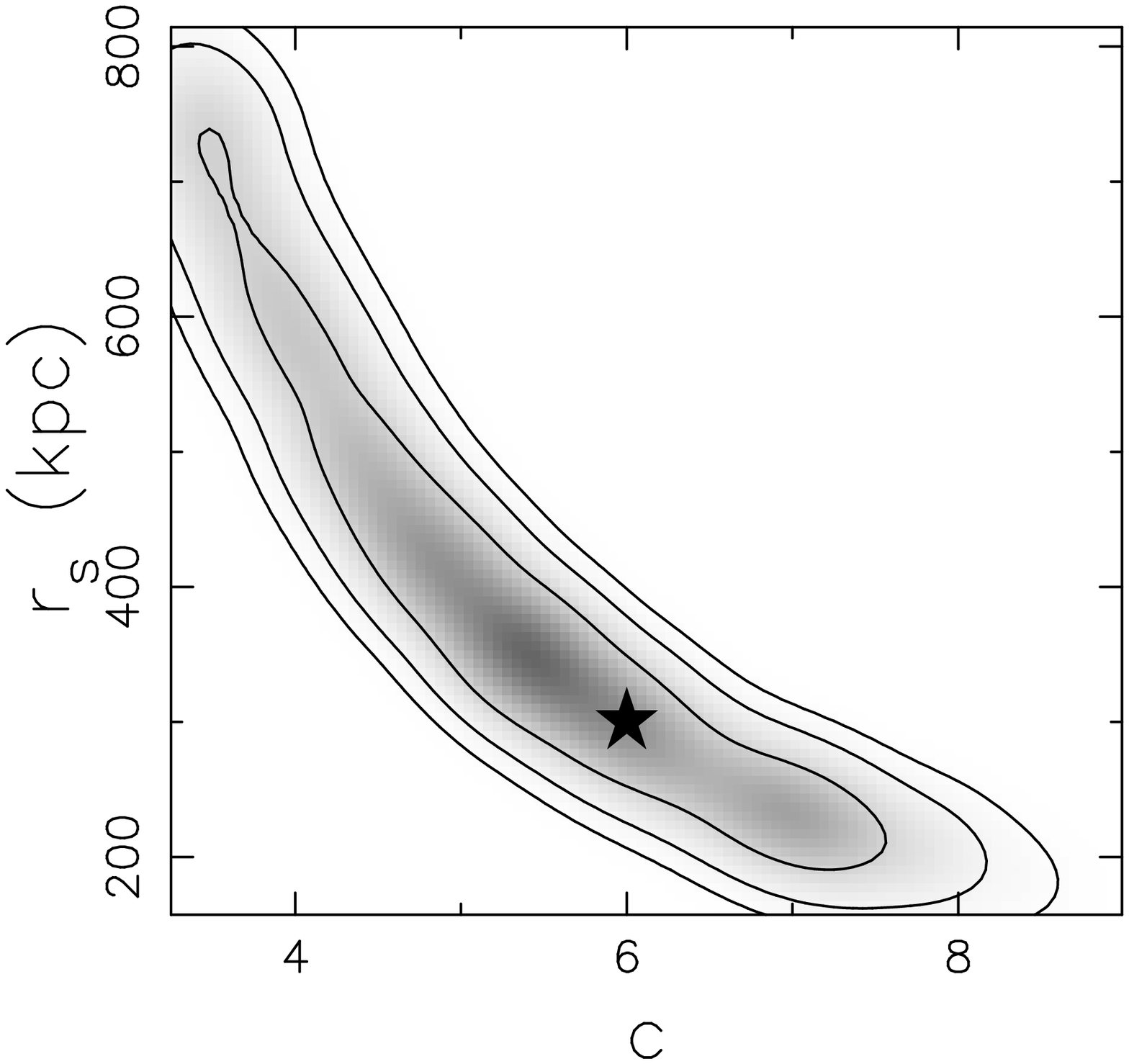} &
\includegraphics[width=0.28\linewidth]{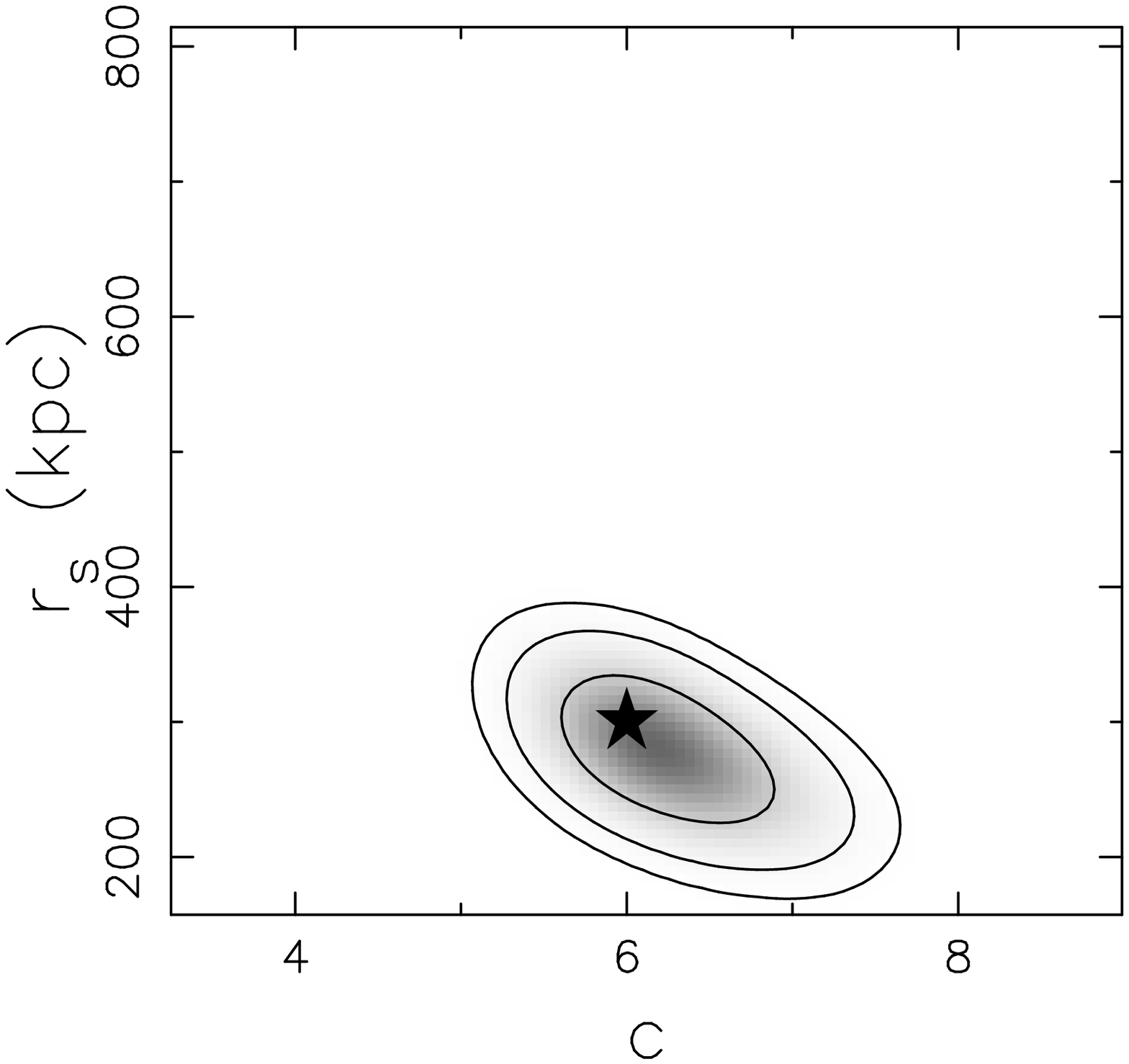} \\
\includegraphics[width=0.28\linewidth]{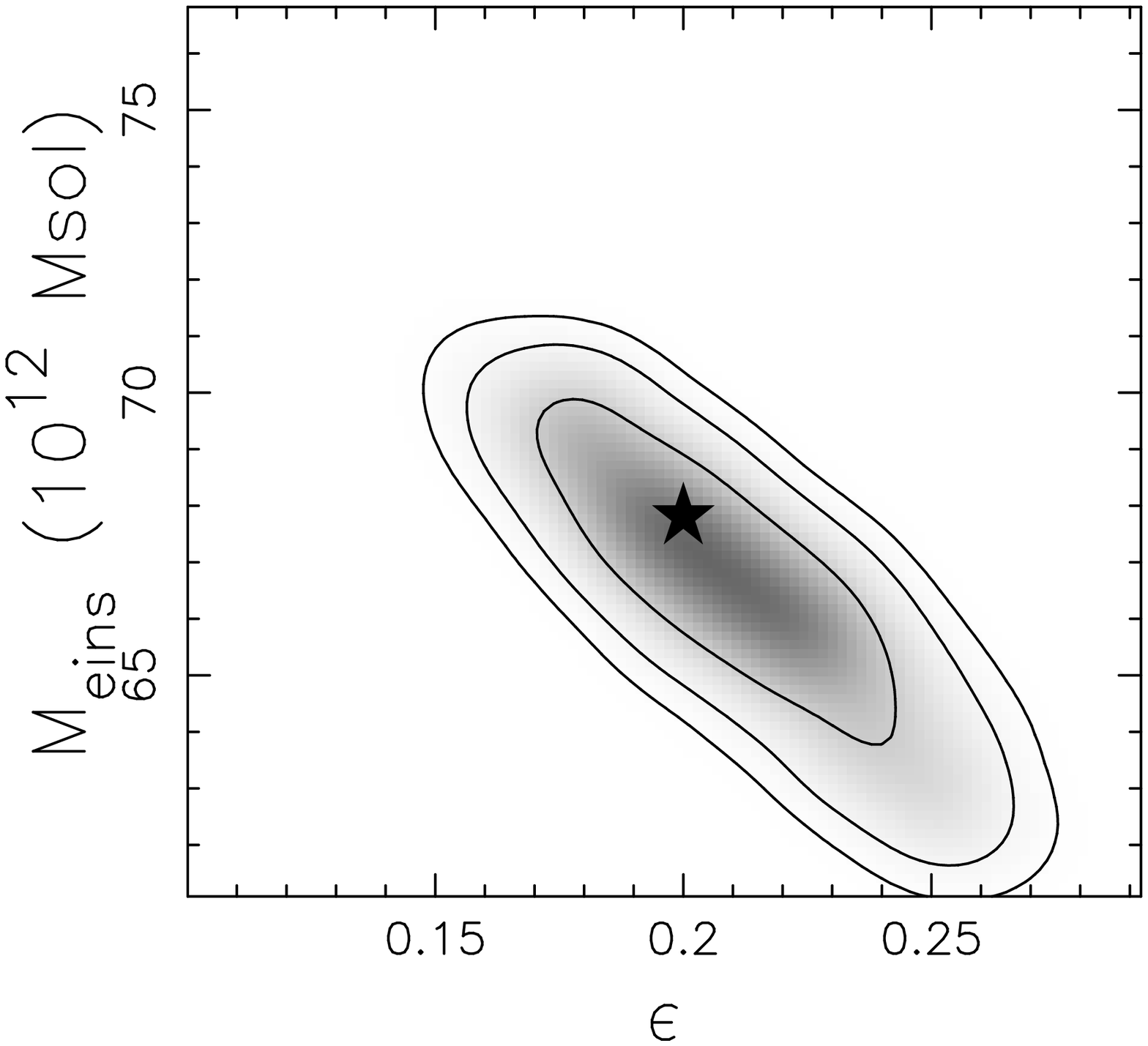} &
\includegraphics[width=0.28\linewidth]{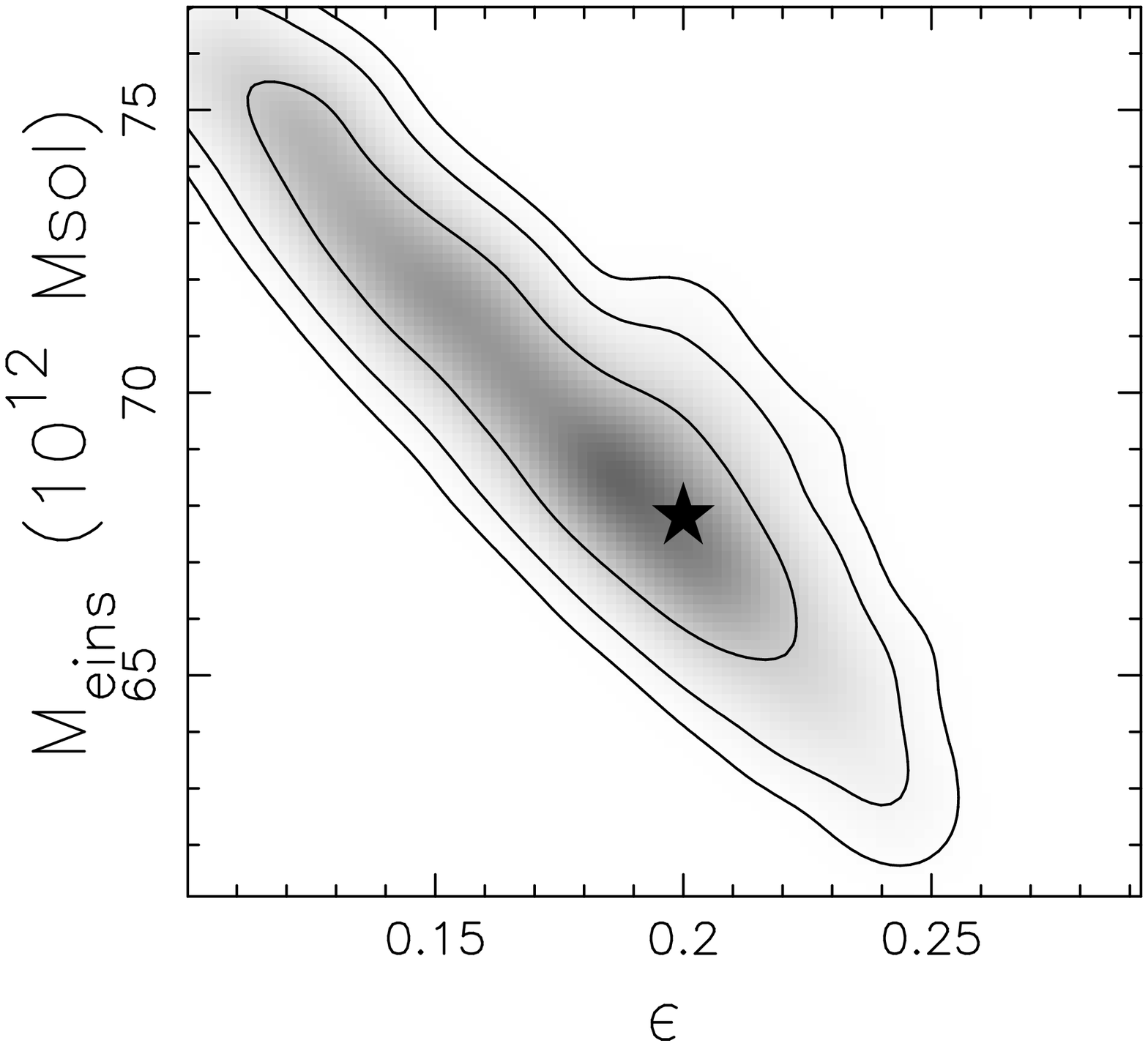} &
\includegraphics[width=0.28\linewidth]{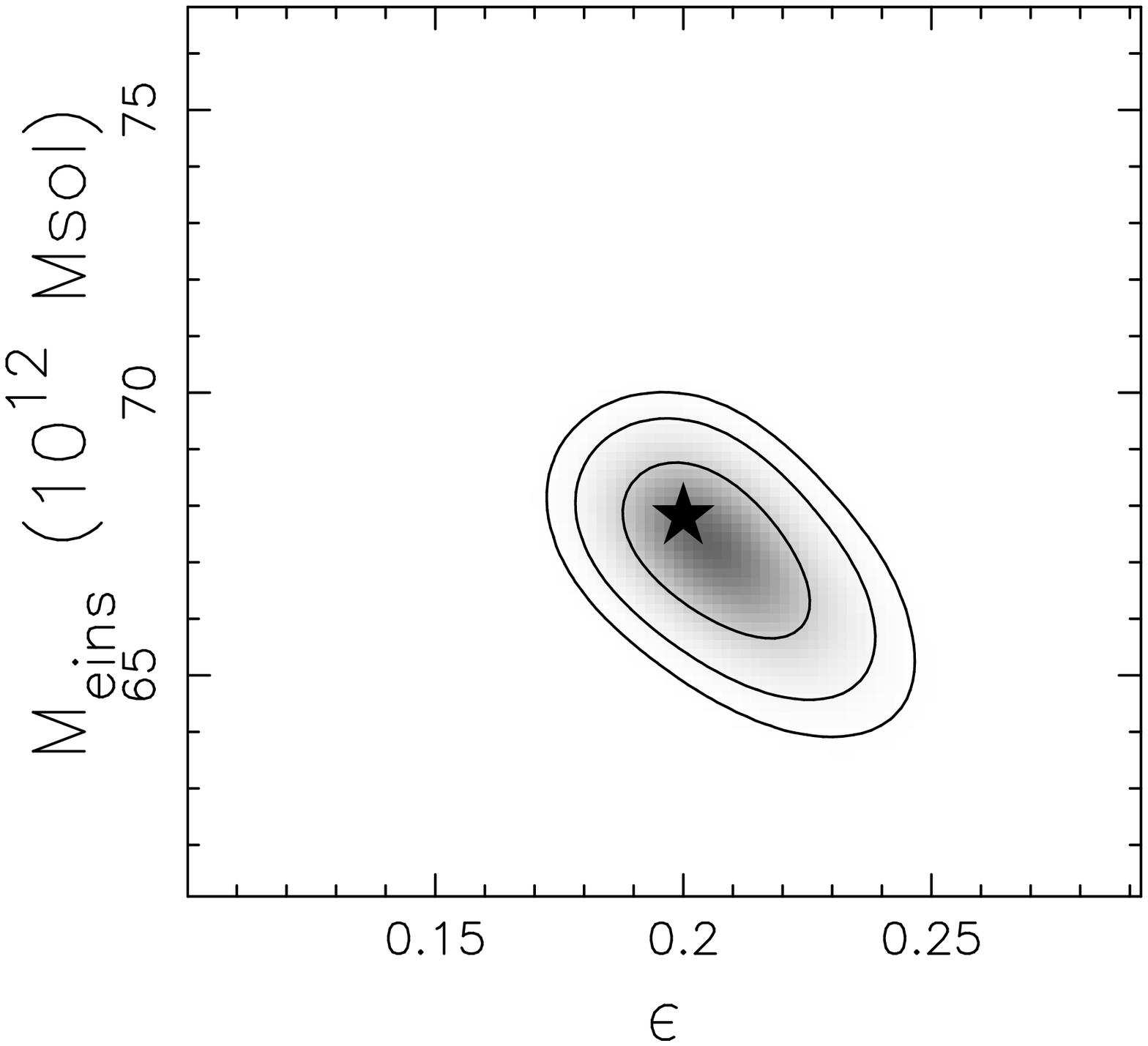} \\
\includegraphics[width=0.28\linewidth]{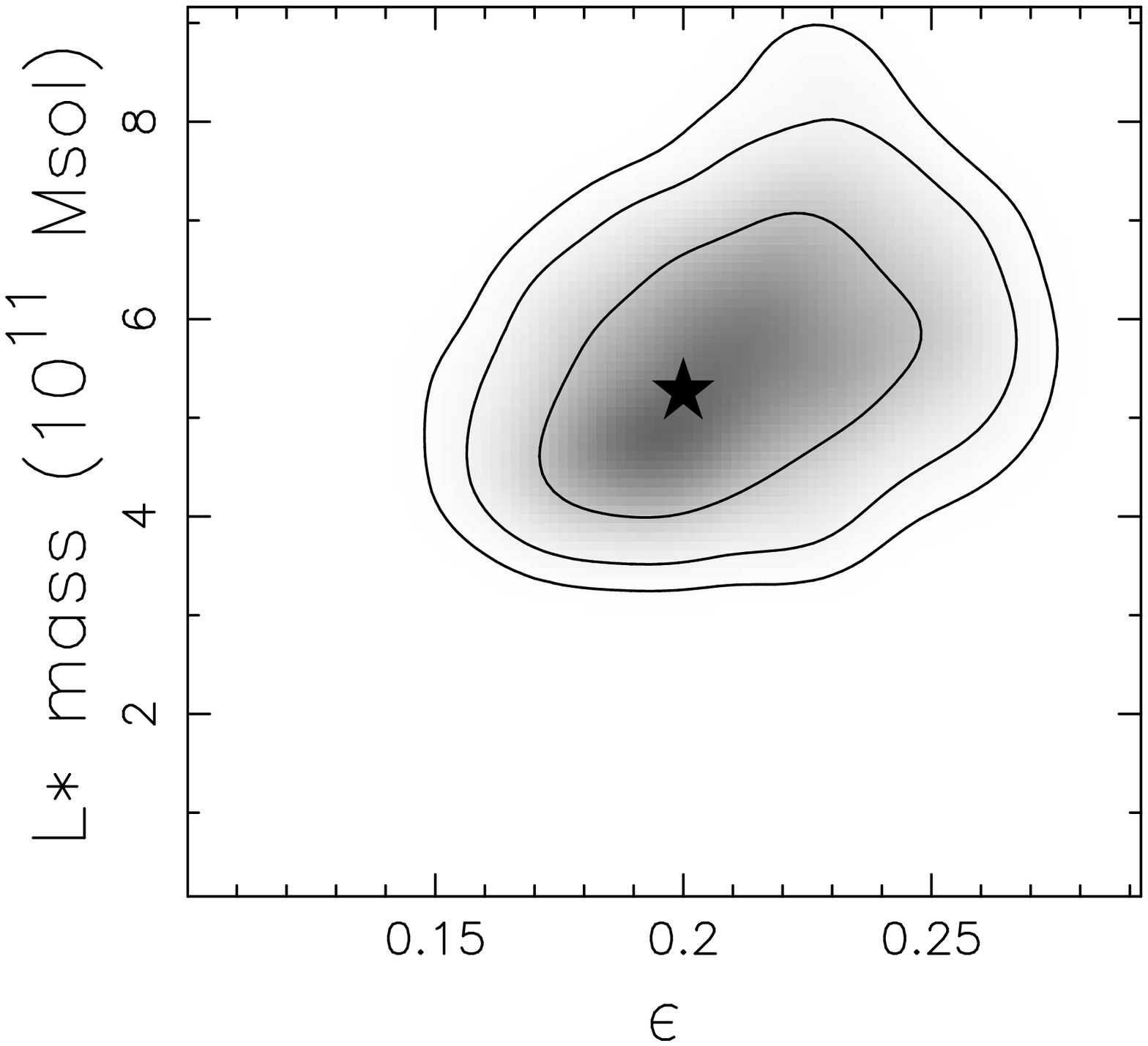} &
\includegraphics[width=0.28\linewidth]{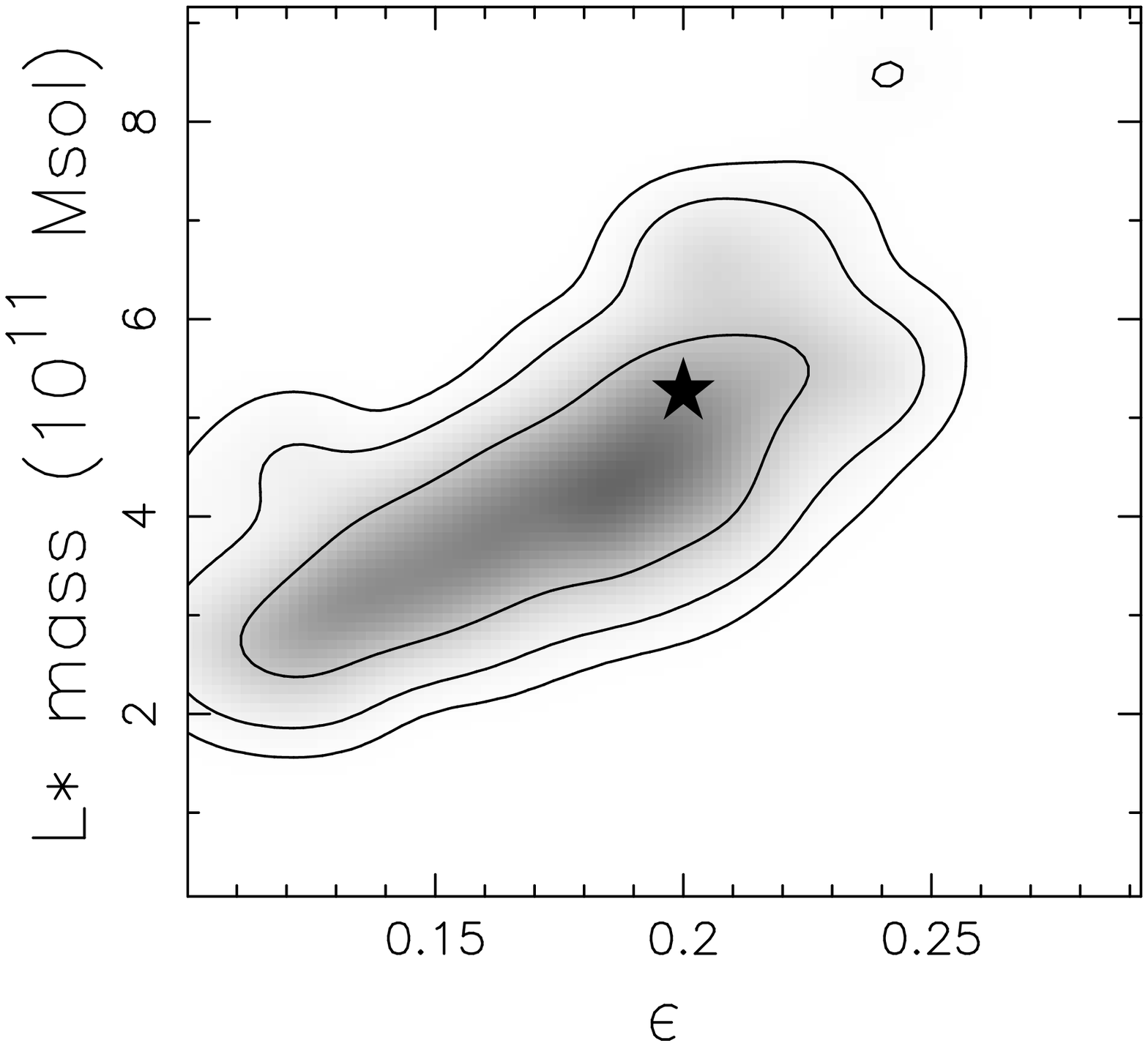} &
\includegraphics[width=0.28\linewidth]{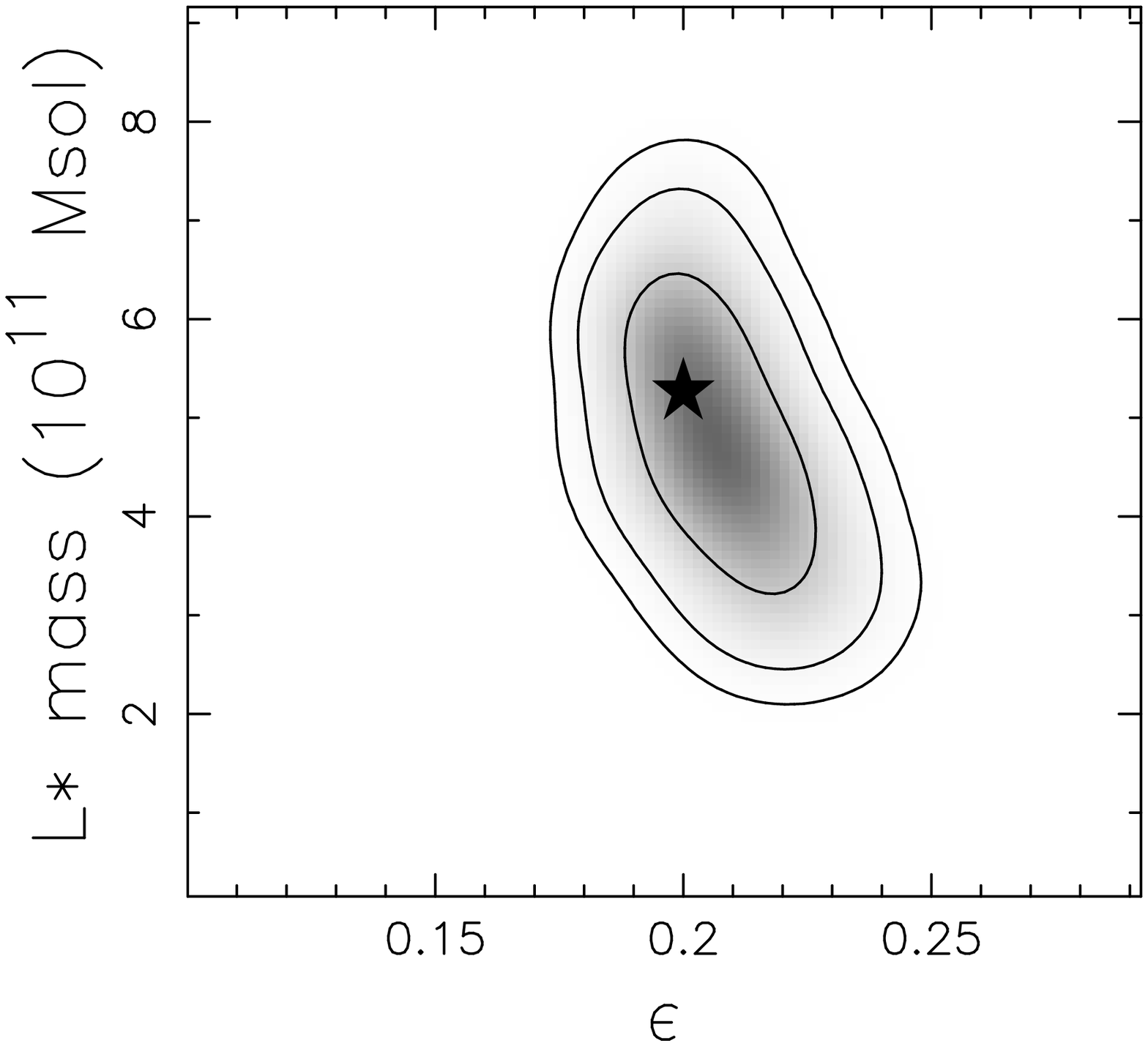} \\
\includegraphics[width=0.28\linewidth]{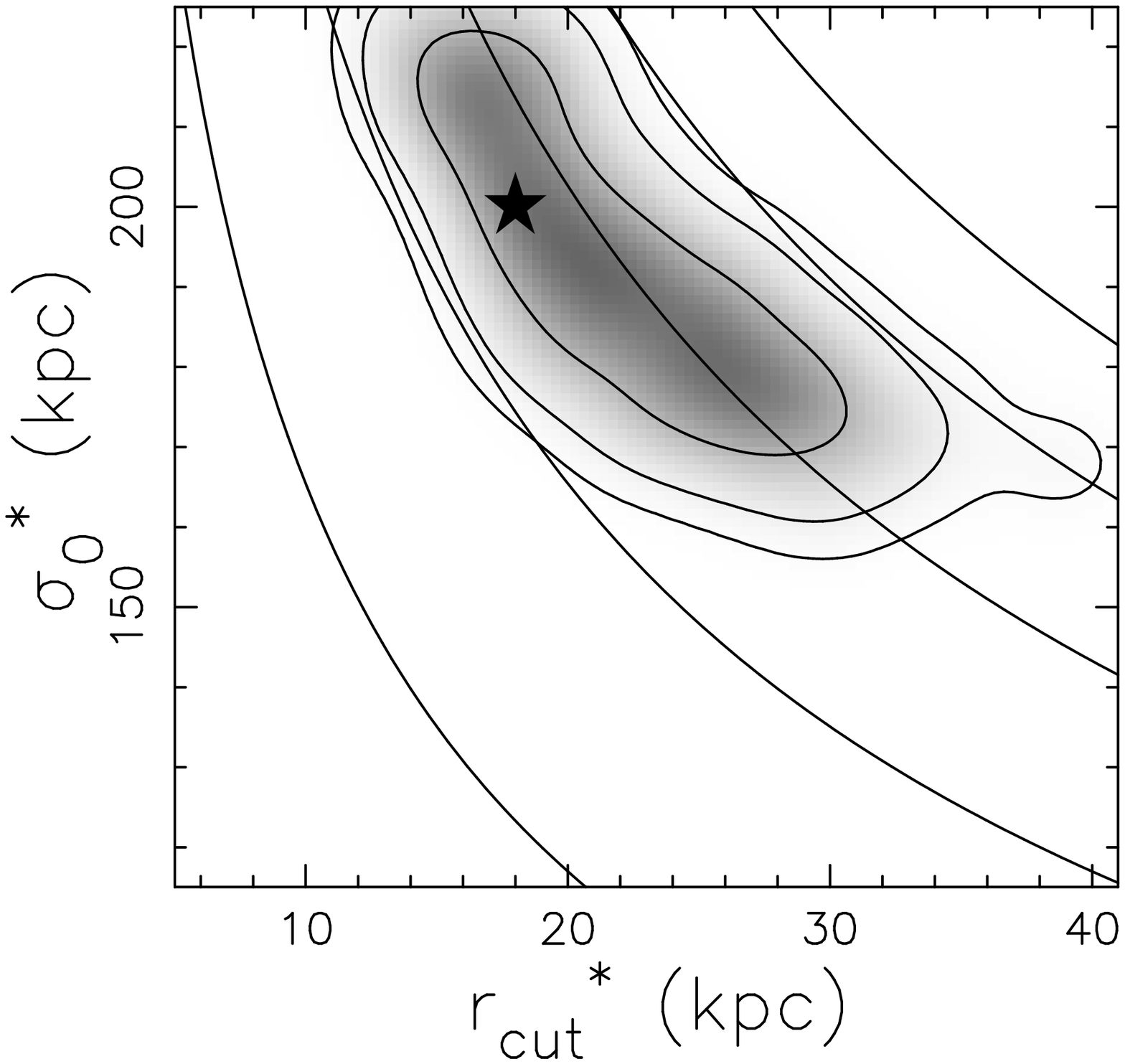} &
\includegraphics[width=0.28\linewidth]{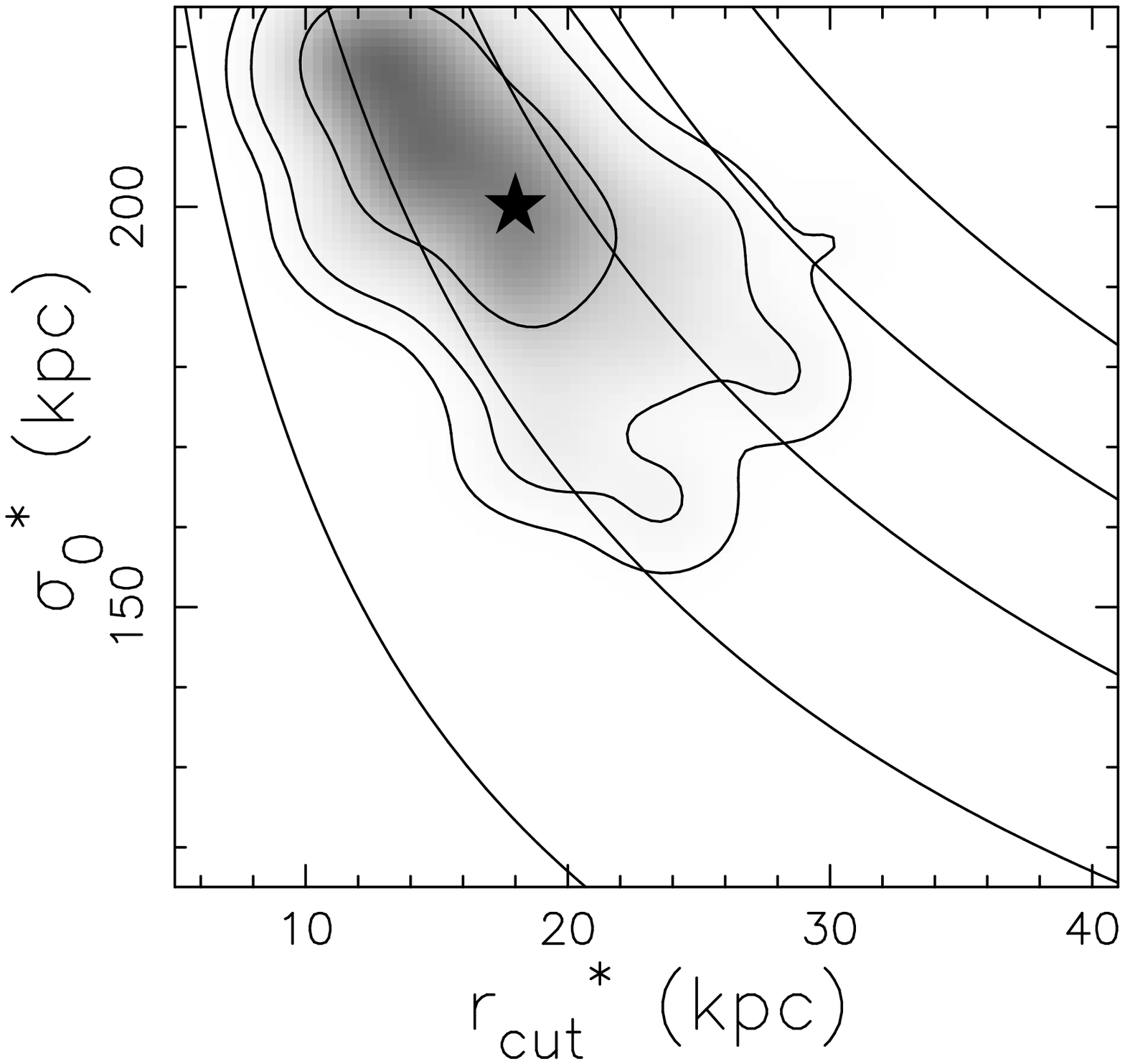} &
\includegraphics[width=0.28\linewidth]{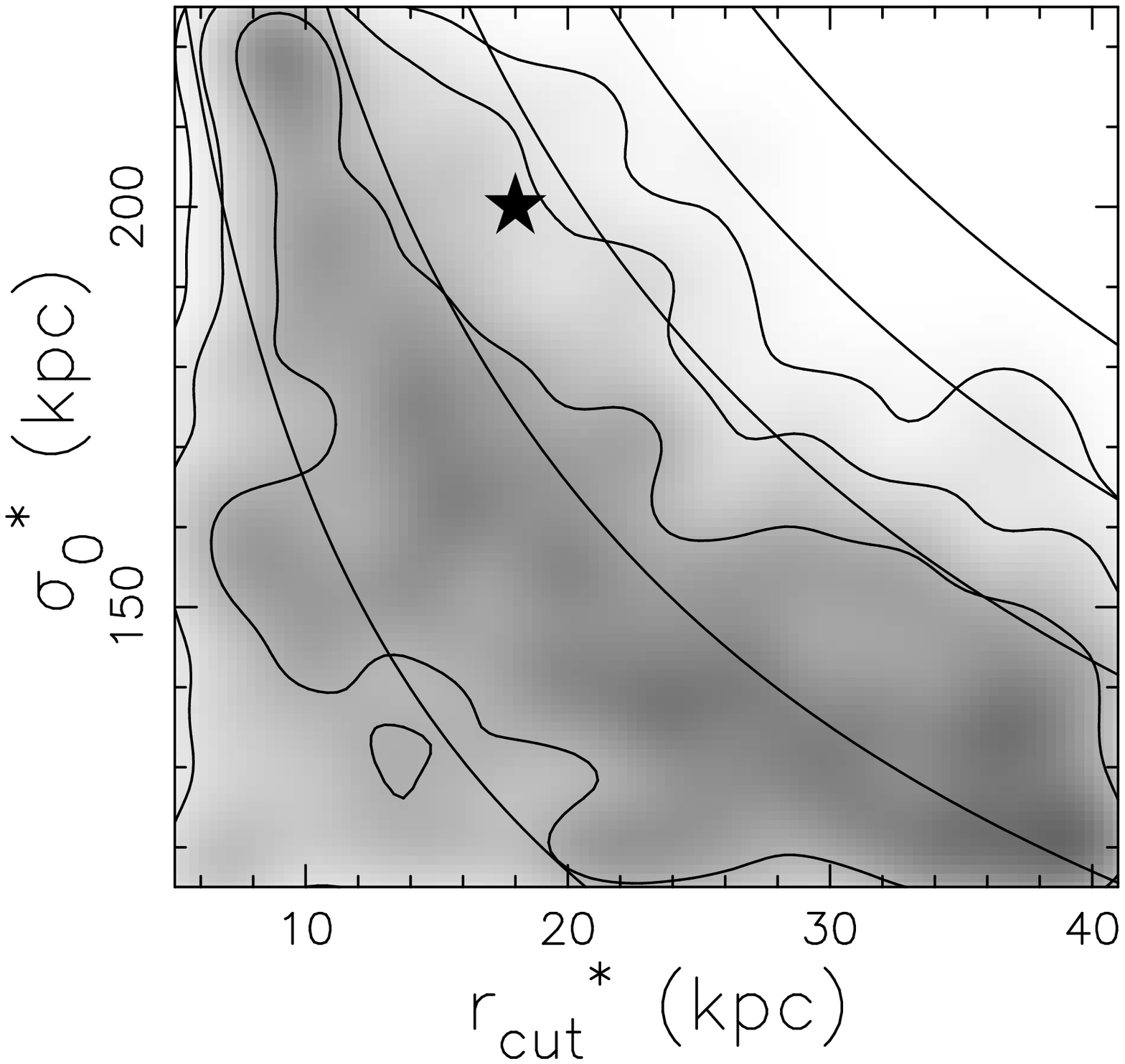} \\
\includegraphics[width=0.28\linewidth]{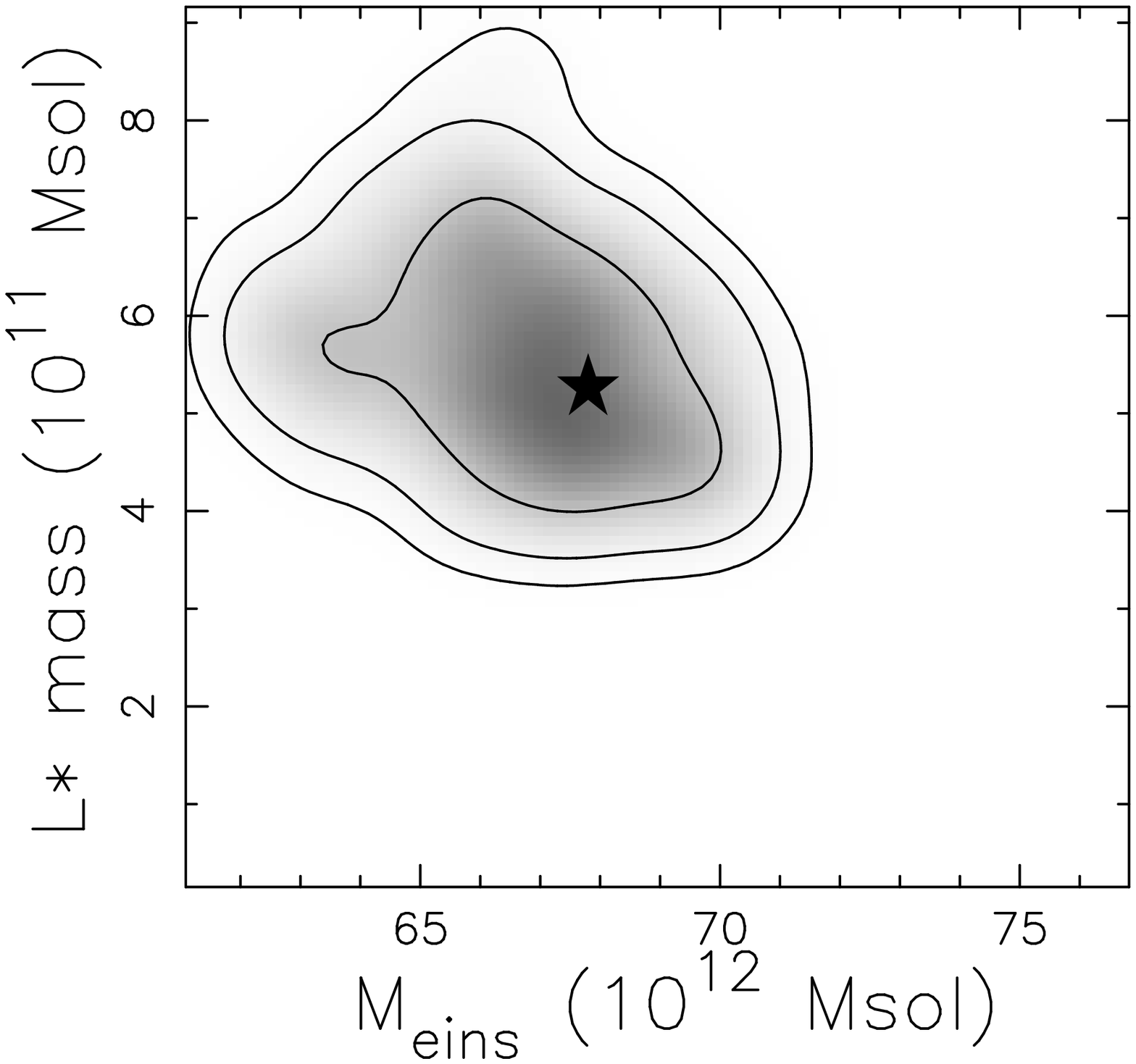} &
\includegraphics[width=0.28\linewidth]{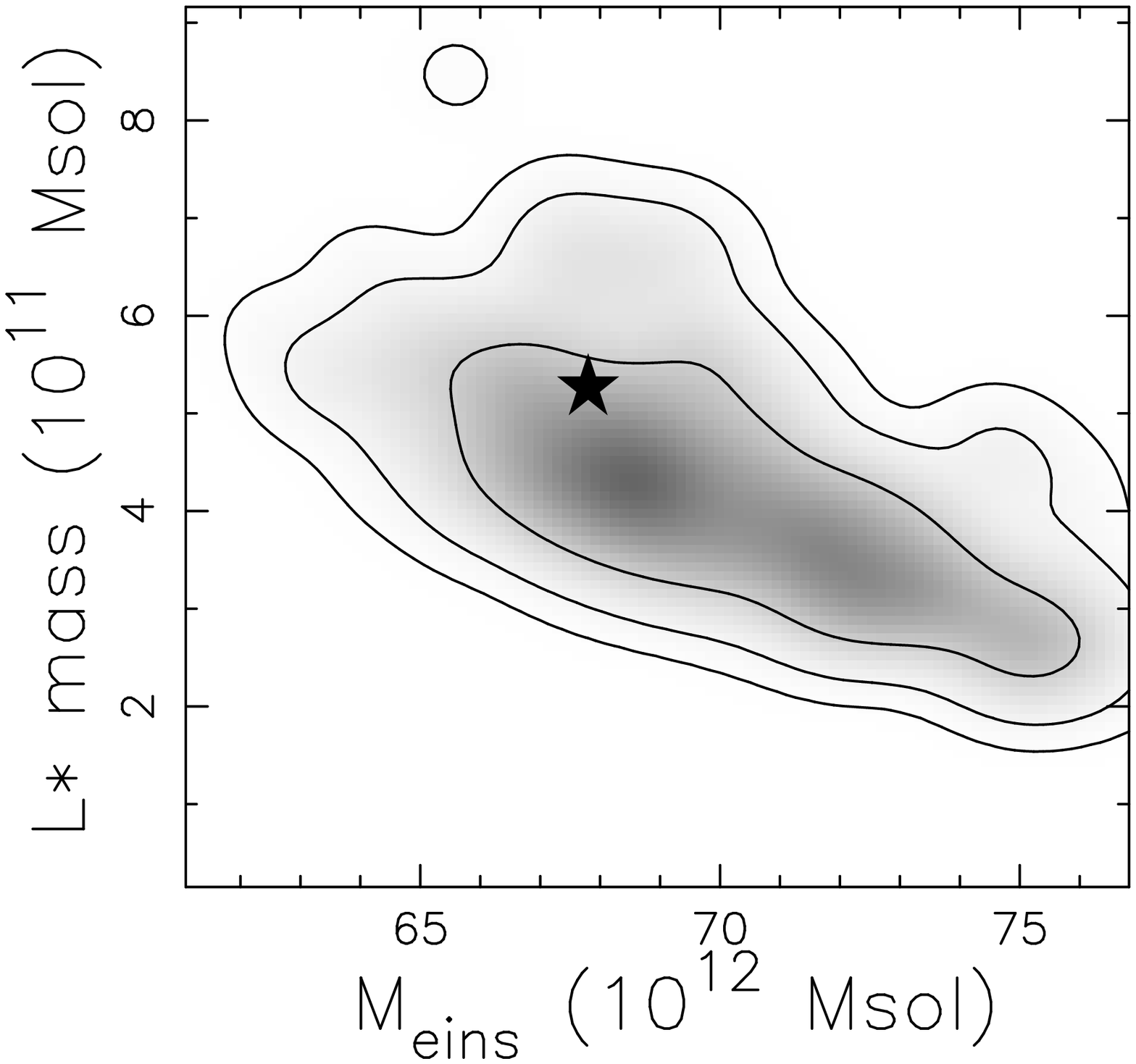} &
\includegraphics[width=0.28\linewidth]{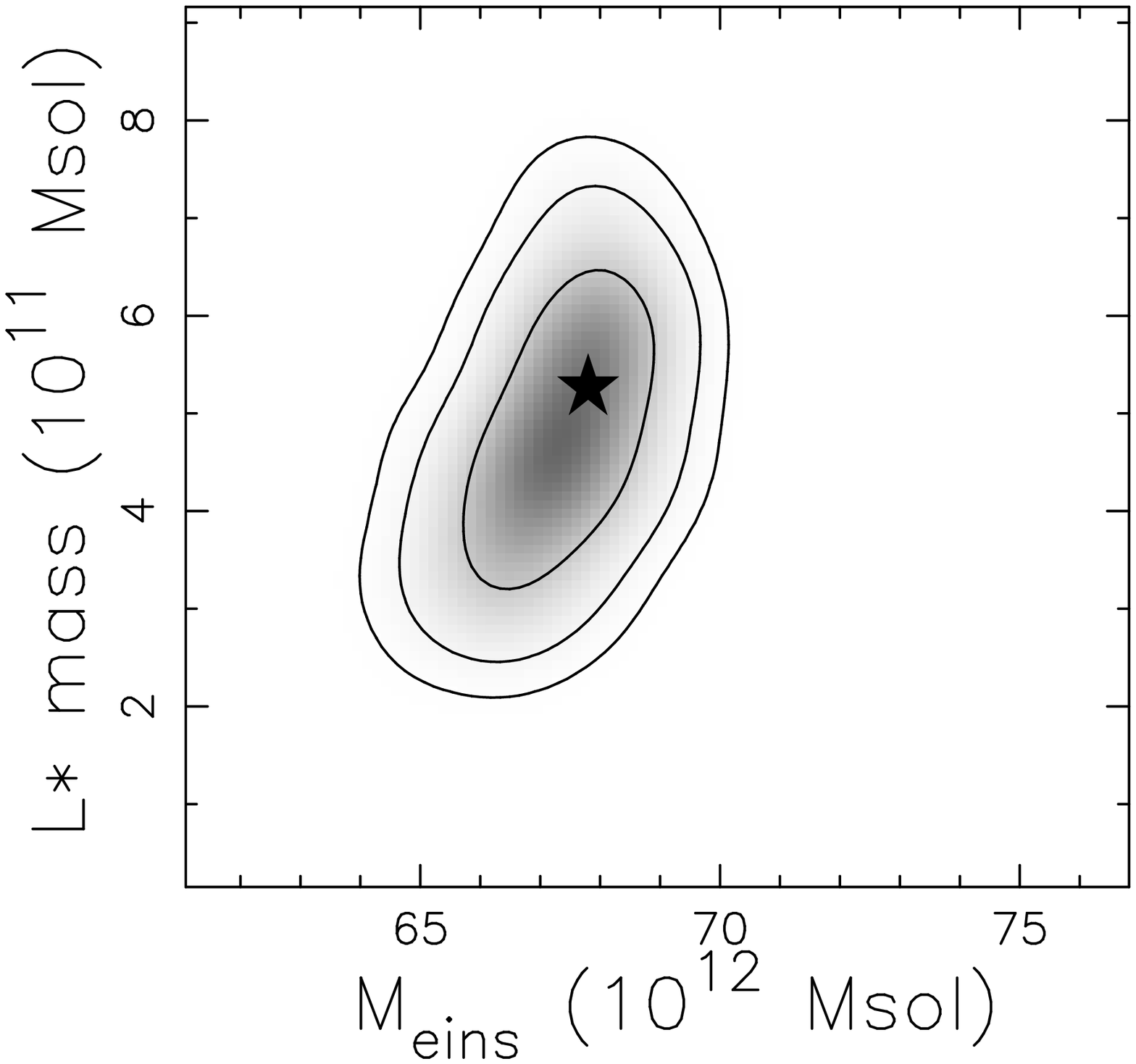} \\
\end{tabular} 
\end{indented} 
\caption{\label{fignfw} 
 2D marginalized posterior PDF of the parameters of the cluster-scale
 halo modelled with an NFW potential, obtained, from left to right,
 with multiple-image configuration~1, 2 and 3 respectively. The 3
 contours stand for the 68\%, 95\% and 99\% CL .  The fiducial values
 are marked by the stars. The mass of a $L^\star$ galaxy is the total
 mass for a circular profile. The plotted contours in the
 $r_{cut}^\star$--$\sigma_0^\star$ plot are the isodensity contours.
 The cluster mass $M_{eins}$ is the total enclosed mass (i.e. galaxy
 subhalos and cluster-scale halo) in the Einstein radius (30'').  }

\end{figure*}

 The obtained posterior PDF is marginalized and plotted in
 Figure~\ref{fignfw}. The (median) estimated parameters are given in
 Table~\ref{tabdegenfw} as well. 

 First, similarly to the PIEMD case, we note that the degeneracies are
 more compact in Config.~3 than in Config.~1 and 2 for which the
 central region of the cluster is less constrained. 

 Second, we note a strong degeneracy between $c$ and the $r_s$. It can
 be fitted by a power law $r_s\propto c^\alpha$ where $\alpha = -1.7$,
 $-1.5$ and $-1.4$ for Config. 1, 2 and 3 respectively. To confirm the
 mathematical origin of this degeneracy, we consider the NFW
 definition of the aperture mass. By solving numerically for~$r_s$
 given~$c$ at constant aperture mass, we manage to reproduce the
 observed degeneracy and measure $\alpha = -1.1$, in relatively good
 agreement with the measured slopes given the uncertainty on the
 aperture mass. 

 Third, the ellipticity, the PA, the $M_{eins}$ and the $L^\star$ mass
 parameters are degenerate in the same manner as in the previous section,
 when the cluster-scale halo was modelled by a PIEMD potential. This
 confirms that these degeneracies are independent of the cluster model,
 and just depend on the lensed image configuration. 

 Finally, in Table~\ref{tabdegenfw}, we note that the $L^\star$
 cut-off radius error is recovered with nearly the same accuracy when
 the cluster-scale halo is modelled by a NFW potential than when
 modelled by a PIEMD potential. This suggest that the scaling relation
 parameters accuracy is model-independent.
 Similarly, the uncertainty on the enclosed mass measured at the 
 Einstein radius is similar to that found when the cluster-scale halo is
 modelled by a PIEMD potential.

\begin{table*}[!h] 
\caption{\label{tabdegenfw} 
 Parameter recovery results for a cluster-scale halo modelled by a
 NFW potential, given 3 different strong lensing
 configurations.  The errors are given at 68\% CL. The $L^\star$
 masses are given for a circular mass component with identical
 dynamical parameters. }

\begin{indented} 
\item[]\begin{tabular*}{\linewidth}[c]{@{}l@{}rr@{ }rr@{ }rr@{ }r} 
\br 
& Input & \multicolumn{2}{c}{Config.1} &
\multicolumn{2}{c}{Config.2} & \multicolumn{2}{c}{Config.3} \\
\mr
$\epsilon$ & 0.2 &  0.21 & $\pm$0.02 &  0.18 & $\pm$0.03 &  
0.21 & $\pm$0.01 \\
$PA$ (deg) & 127. &  127.4 & $\pm$1.0 &  126.6 & $\pm$4.0 &  
126.6 & $\pm$0.6 \\
$c$ & 6. &  6.5 & $\pm$0.9 &  6.4 & $\pm$0.8 &  5.9 & $\pm$0.3 \\
$Scale\ radius$ (kpc) & 300. &  269.3 & $\pm$54.6 &  
367.9 & $\pm$149.9 &  284.7 & $\pm$22.5 \\
$r_{cut}^\star$ (kpc) & 18. &  21.6 & $\pm$4.8 &  16.3 & $\pm$3.9 &
20.6 & $\pm$10.1 \\
$\sigma_0^\star$ (km/s) & 200. &  191.5 & $\pm$15.4 & 
205.6 & $\pm$13.4 &  169.6 & $\pm$27.8 \\
$M_{L^\star}$ ($10^{11}\ M_\odot$) & 5.26 &  5.56 & $\pm$1.7 & 
4.2 & $\pm$1.1 &  4.9 & $\pm$0.9 \\
$M_{eins}$ ($10^{12}\ M_\odot$) & 67.8  & 66.9 &  $\pm$1.8 & 69.5 &  $\pm$2.9 &  67.4 & $\pm$0.8 \\
\br 
\end{tabular*} 
\end{indented} 
\end{table*} 

%
%
\subsection{S\'ersic posterior distribution analysis}

 Finally, we fit the S\'ersic model with a S\'ersic potential for the
 cluster-scale halo.  We perform the recovery of the cluster-scale
 halo parameters ($\epsilon$, PA, $R_e$, $\Sigma_e$ and $n$), as well
 as the galaxy-scale subhalo scaling parameters $\sigma_0^\star$ and
 $r_{cut}^\star$, given the same three configurations of multiple
 images as before.

 Again, we assume uniform priors for the parameters, with widths of 50\%  
 centred on the input values. The cD galaxy subhalo parameters are
 fixed.  We constrain 7 free parameters with 8 constraints.

\begin{figure*} 
\begin{indented} 
\item[]\begin{tabular}{ccc} 
\includegraphics[width=0.28\linewidth]{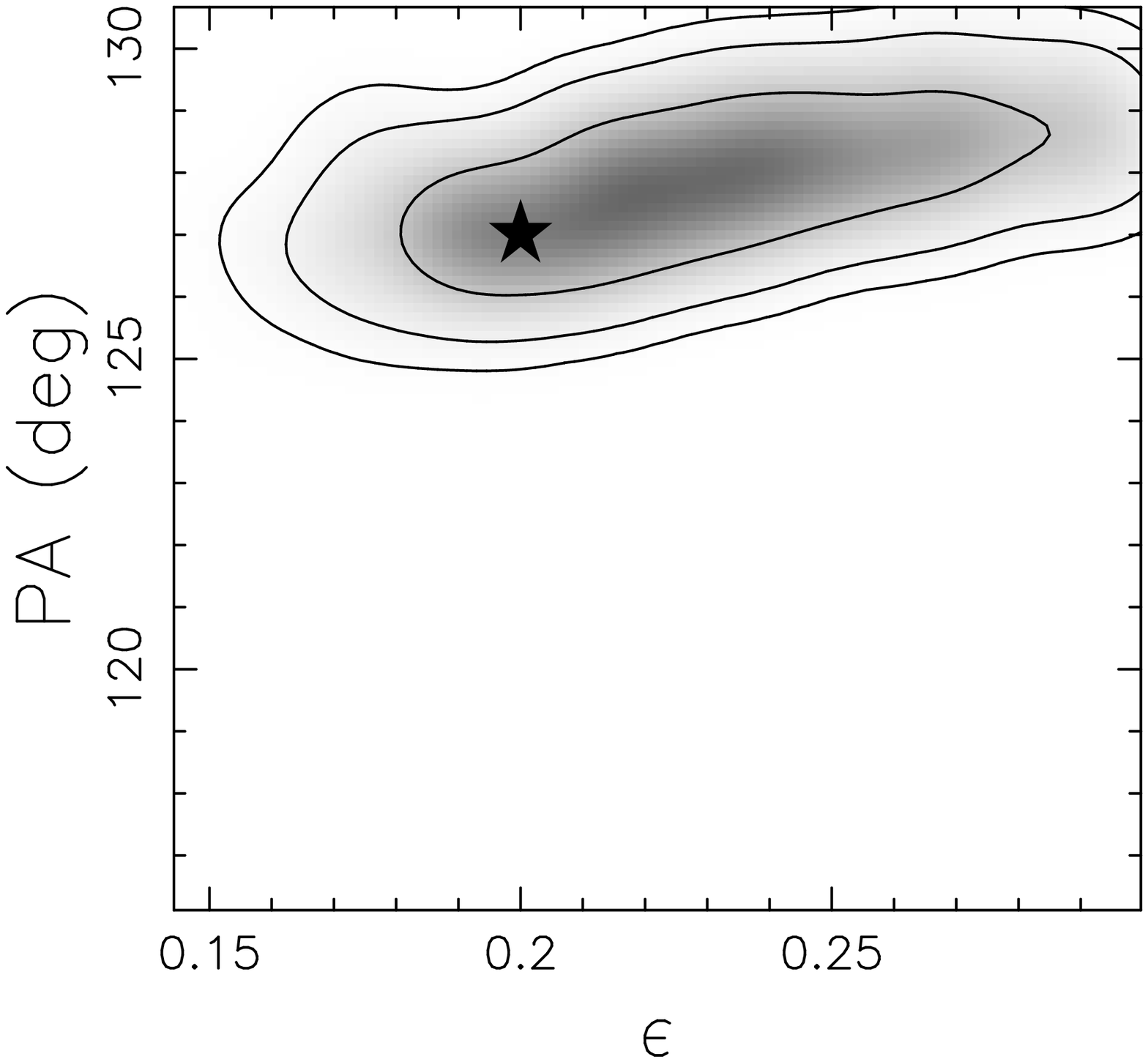} &
\includegraphics[width=0.28\linewidth]{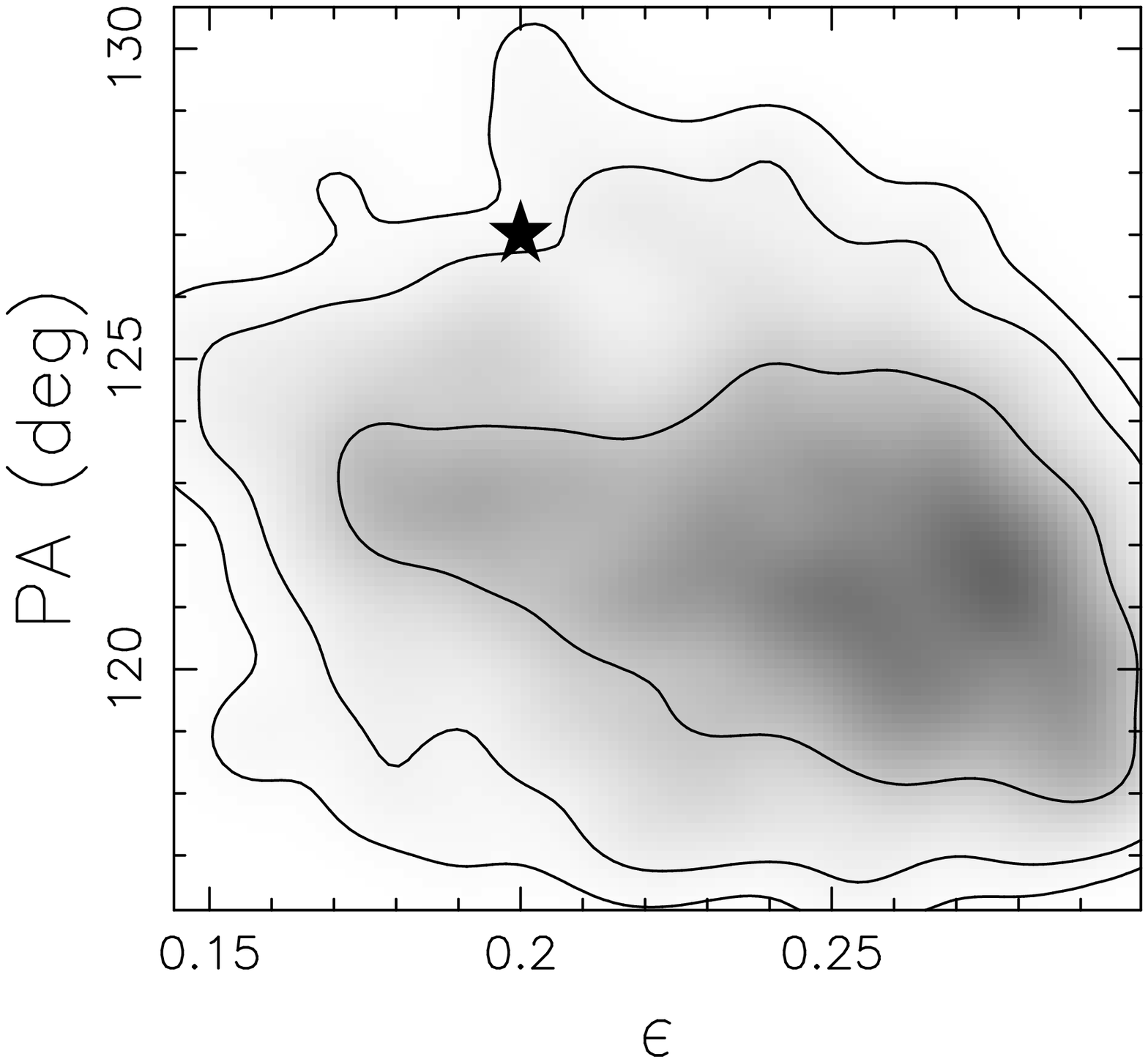} &
\includegraphics[width=0.28\linewidth]{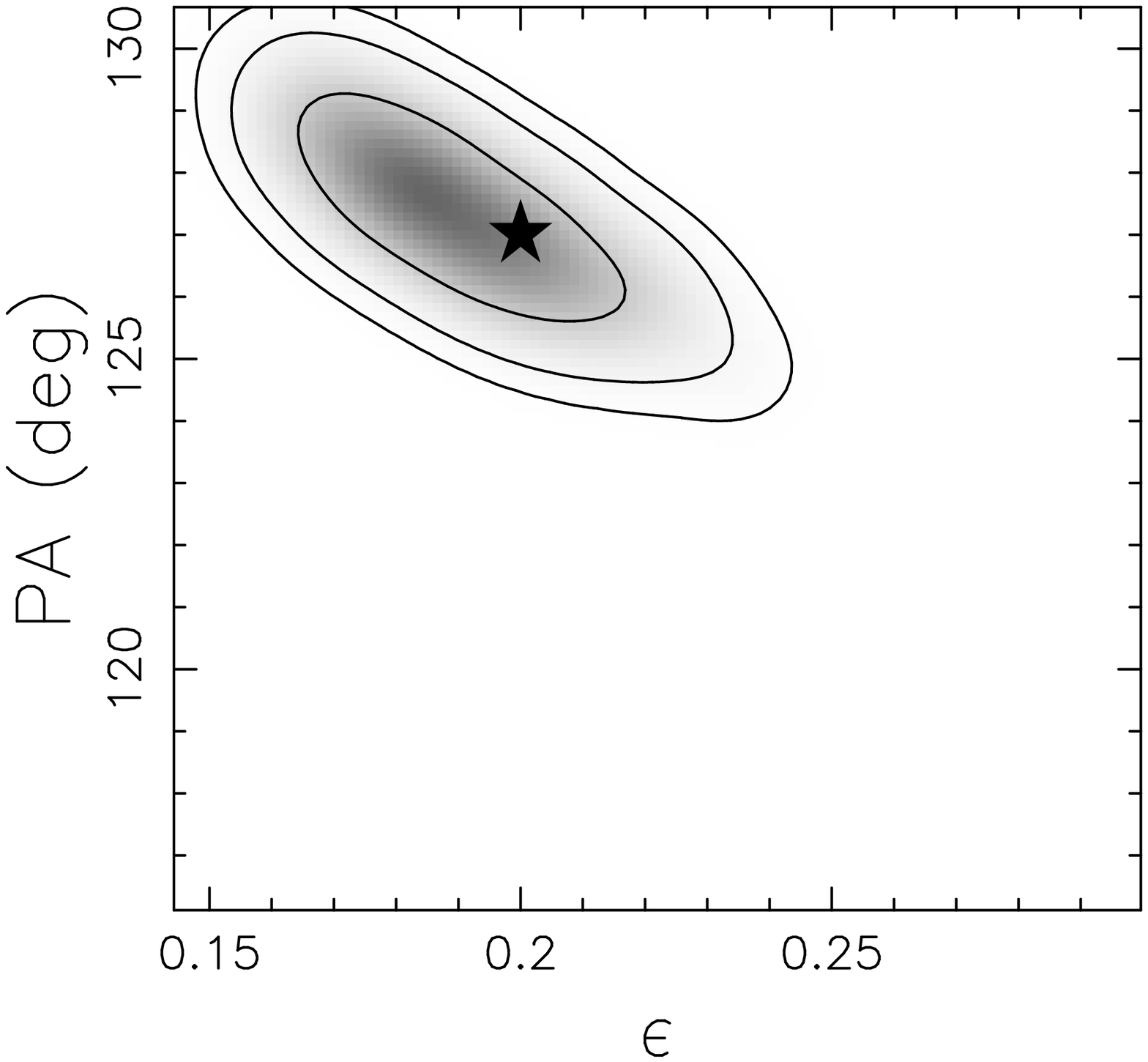} \\
\includegraphics[width=0.28\linewidth]{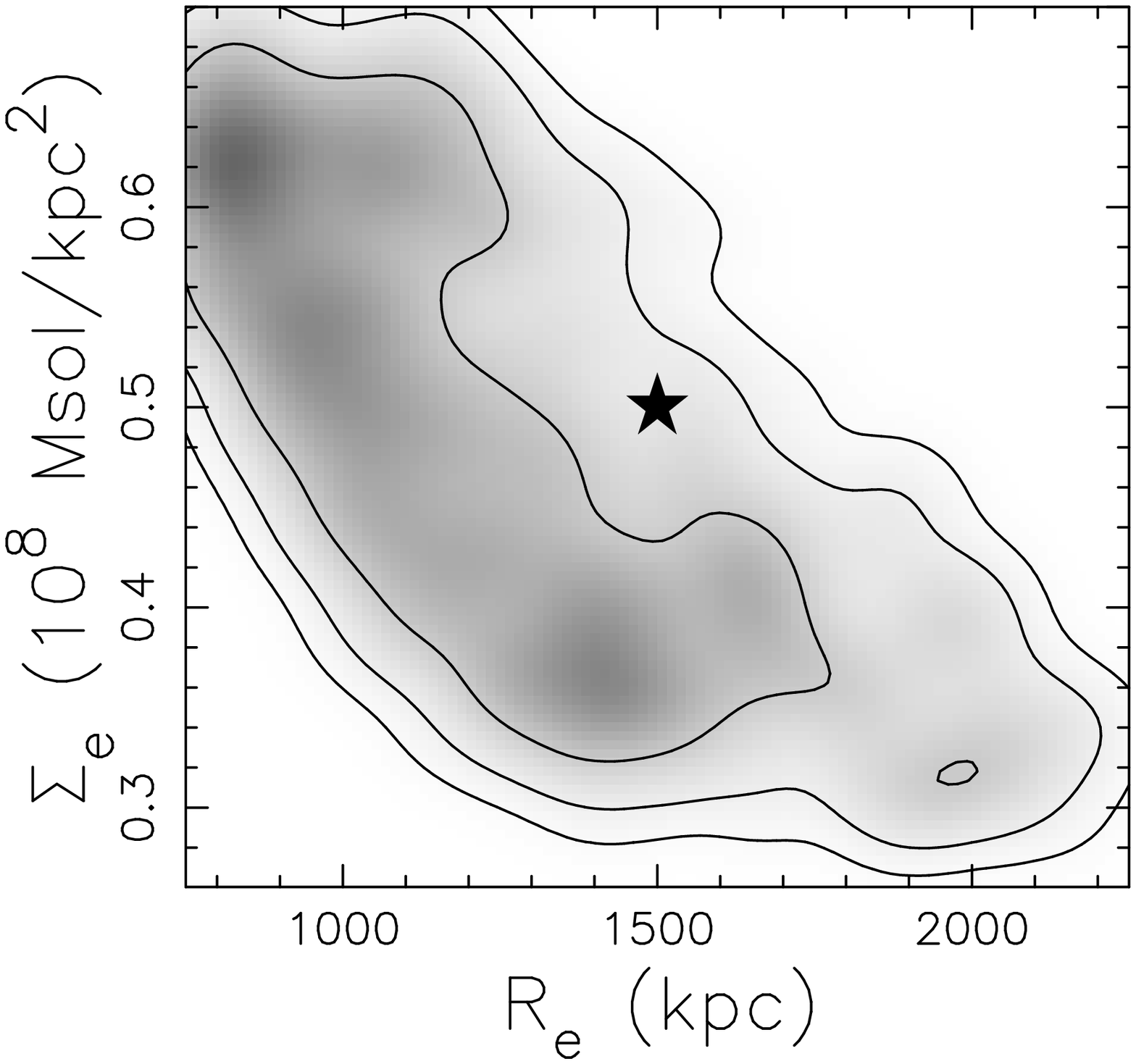} &
\includegraphics[width=0.28\linewidth]{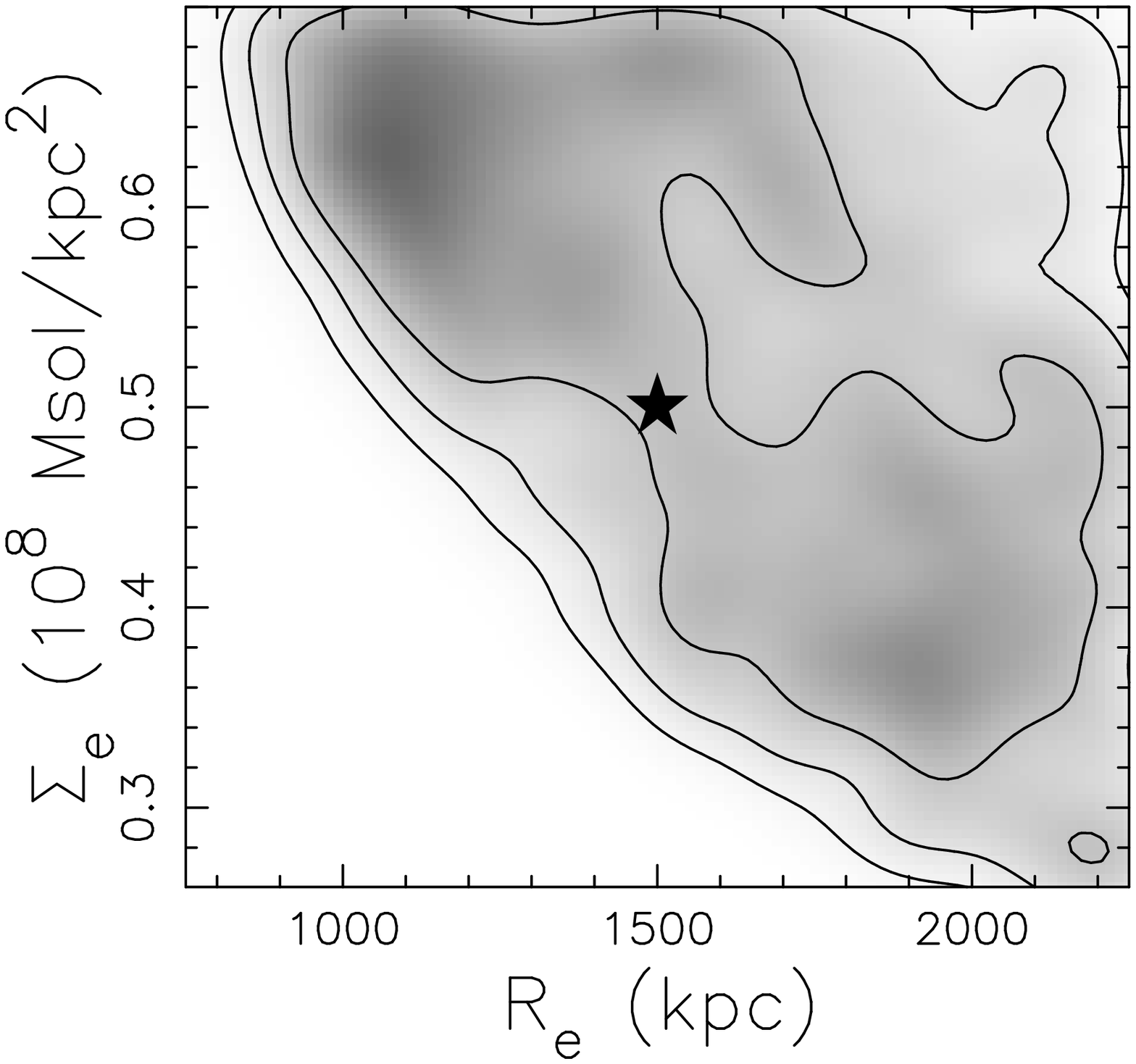} &
\includegraphics[width=0.28\linewidth]{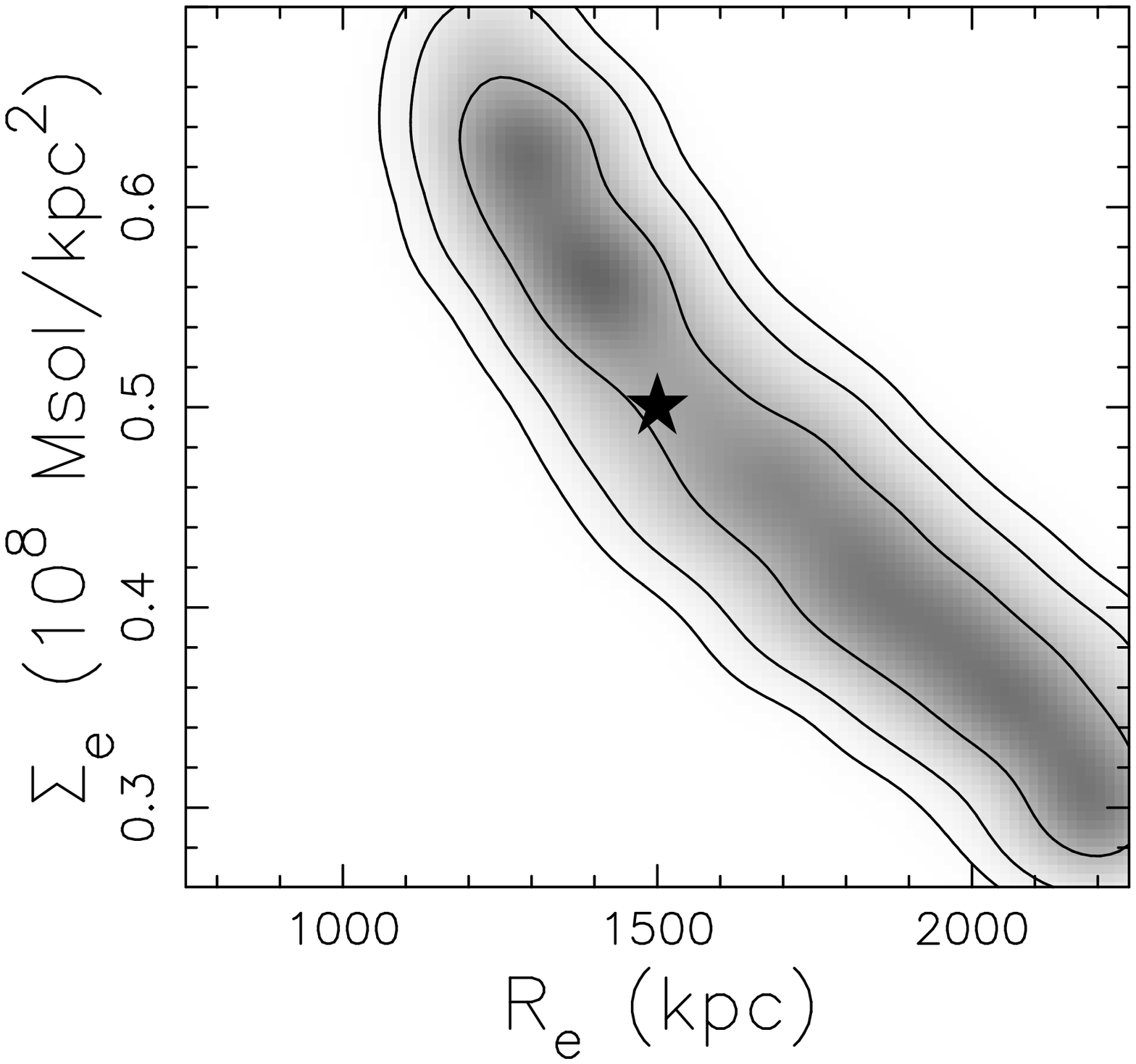} \\
\includegraphics[width=0.28\linewidth]{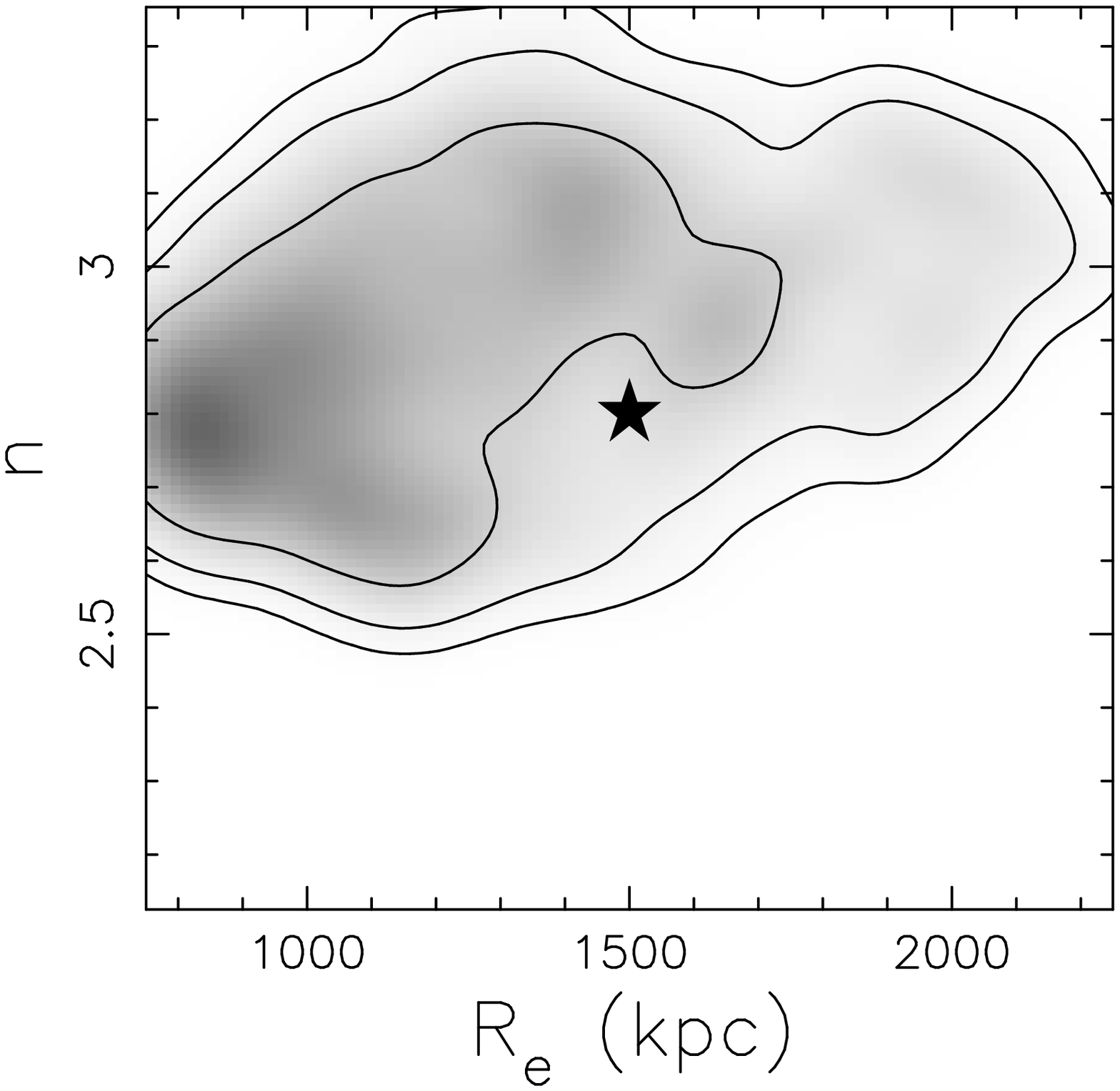} &
\includegraphics[width=0.28\linewidth]{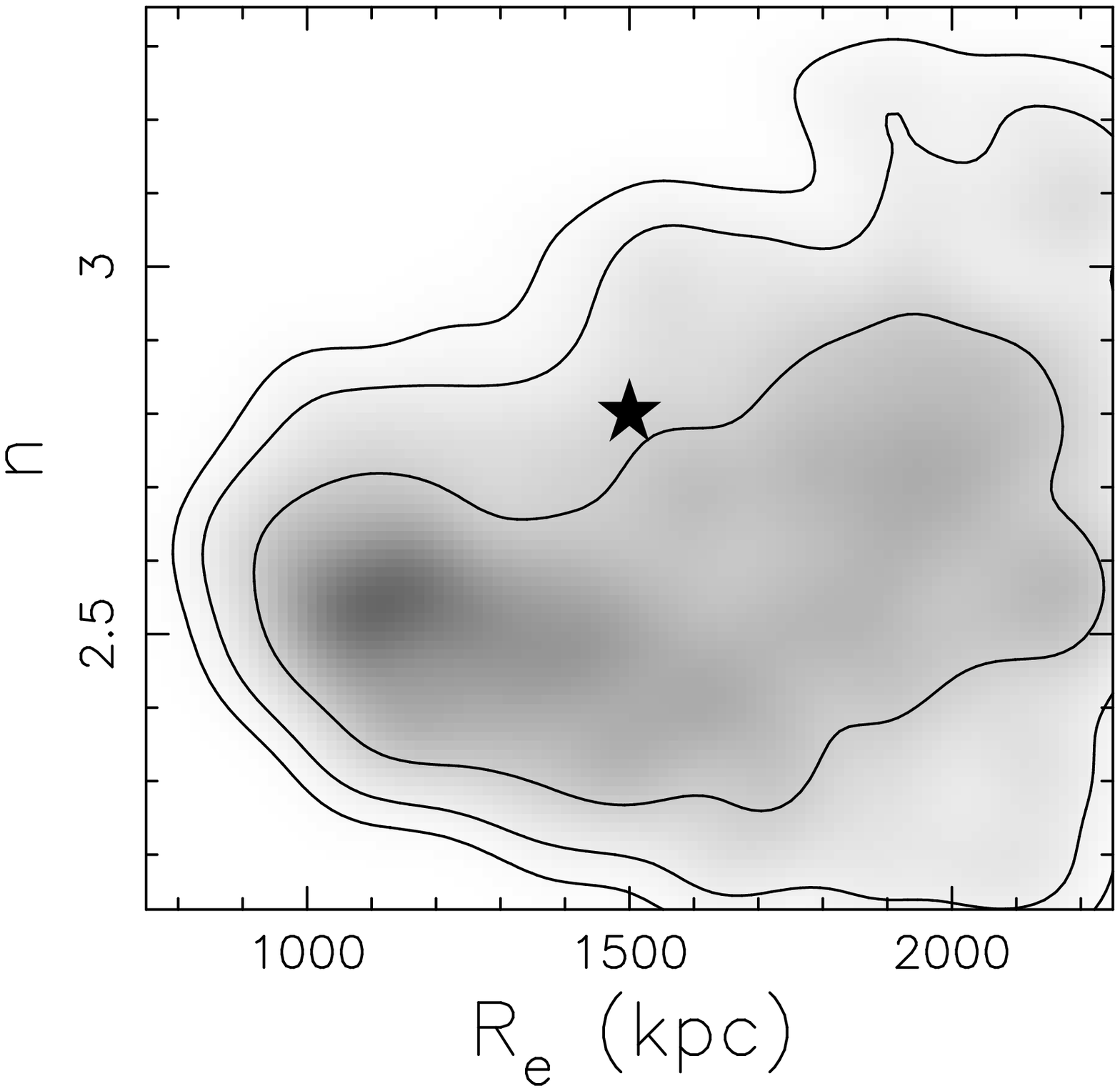} &
\includegraphics[width=0.28\linewidth]{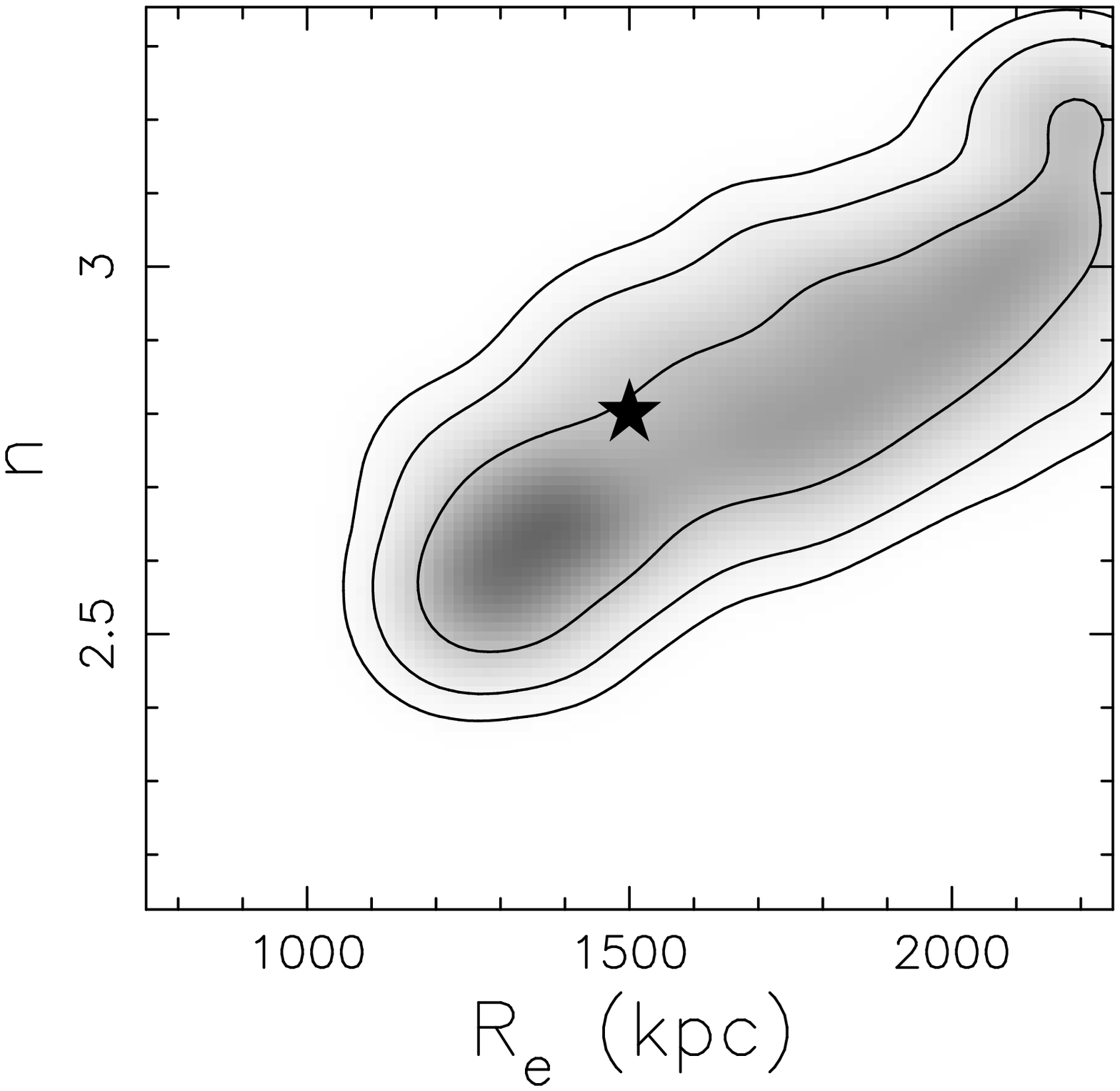} \\
\includegraphics[width=0.28\linewidth]{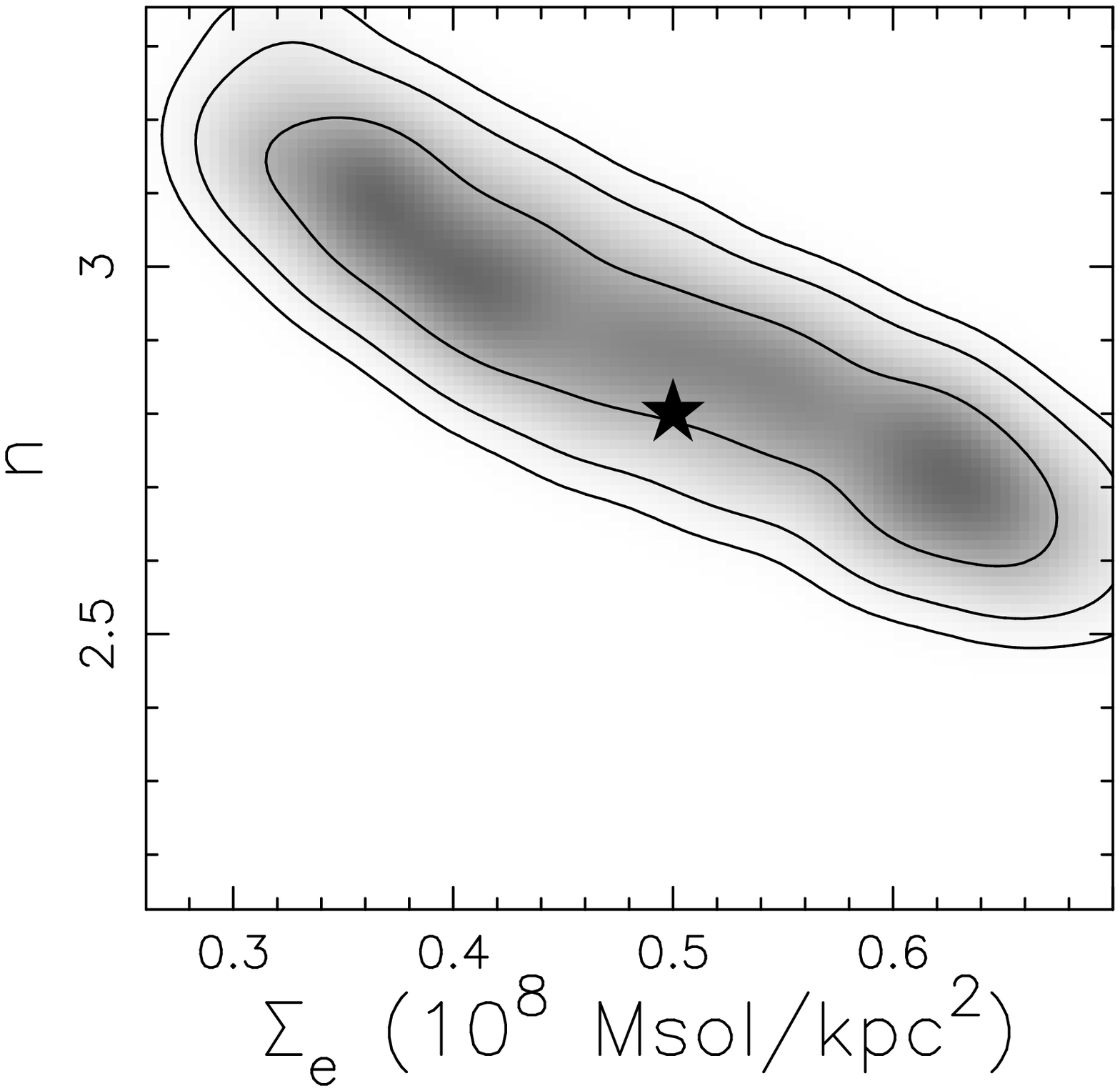} &
\includegraphics[width=0.28\linewidth]{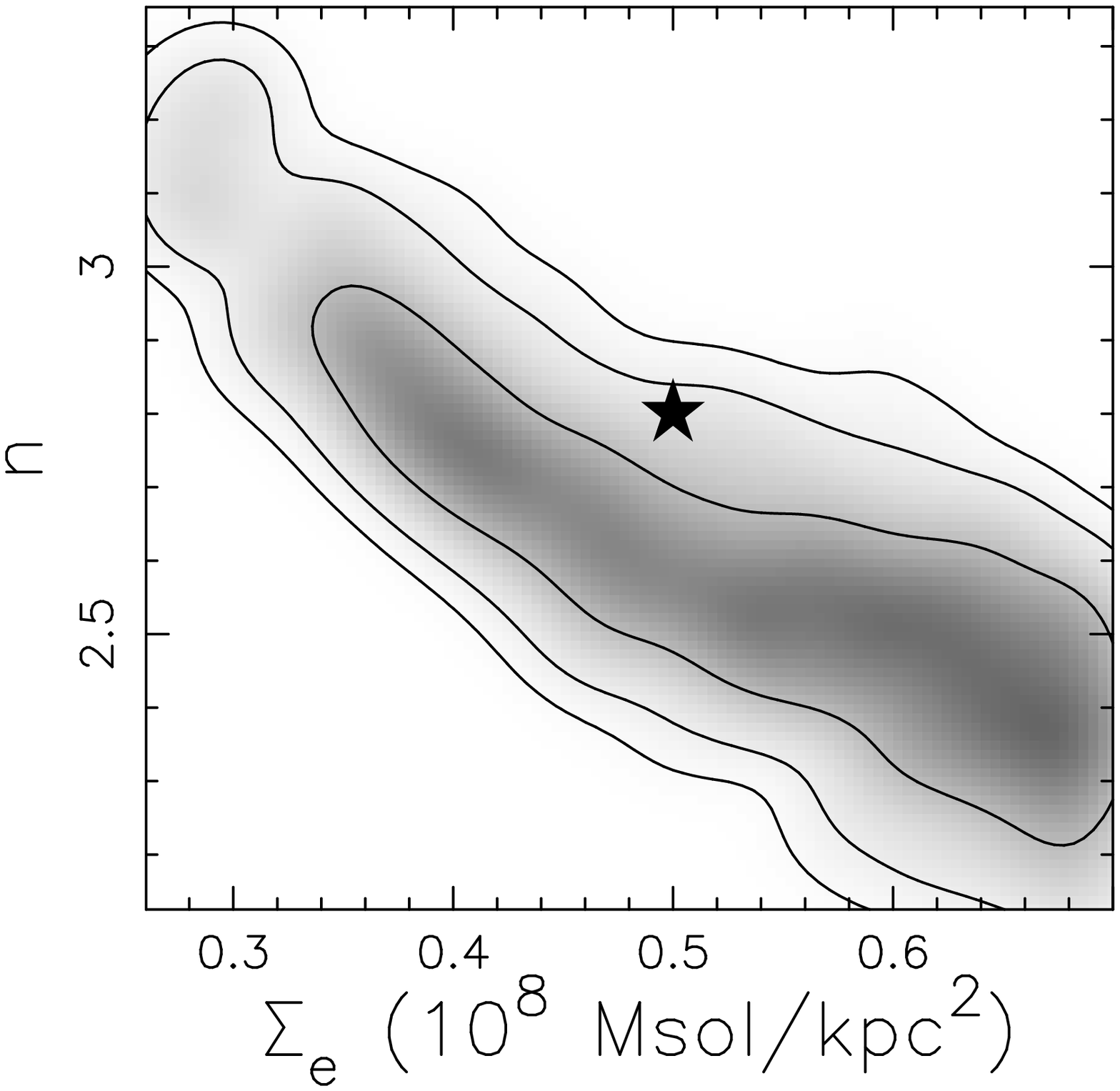} &
\includegraphics[width=0.28\linewidth]{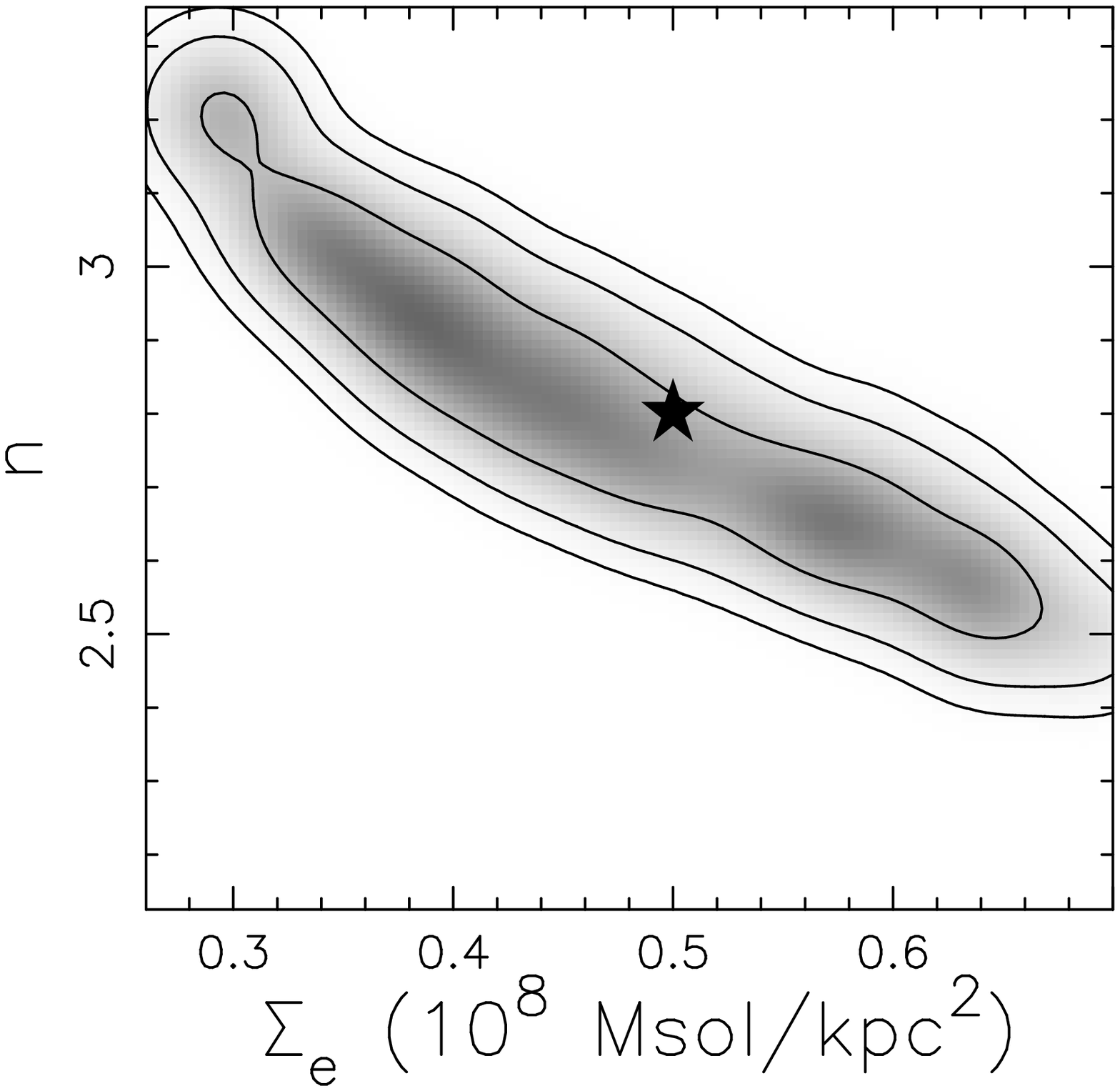} \\
\includegraphics[width=0.28\linewidth]{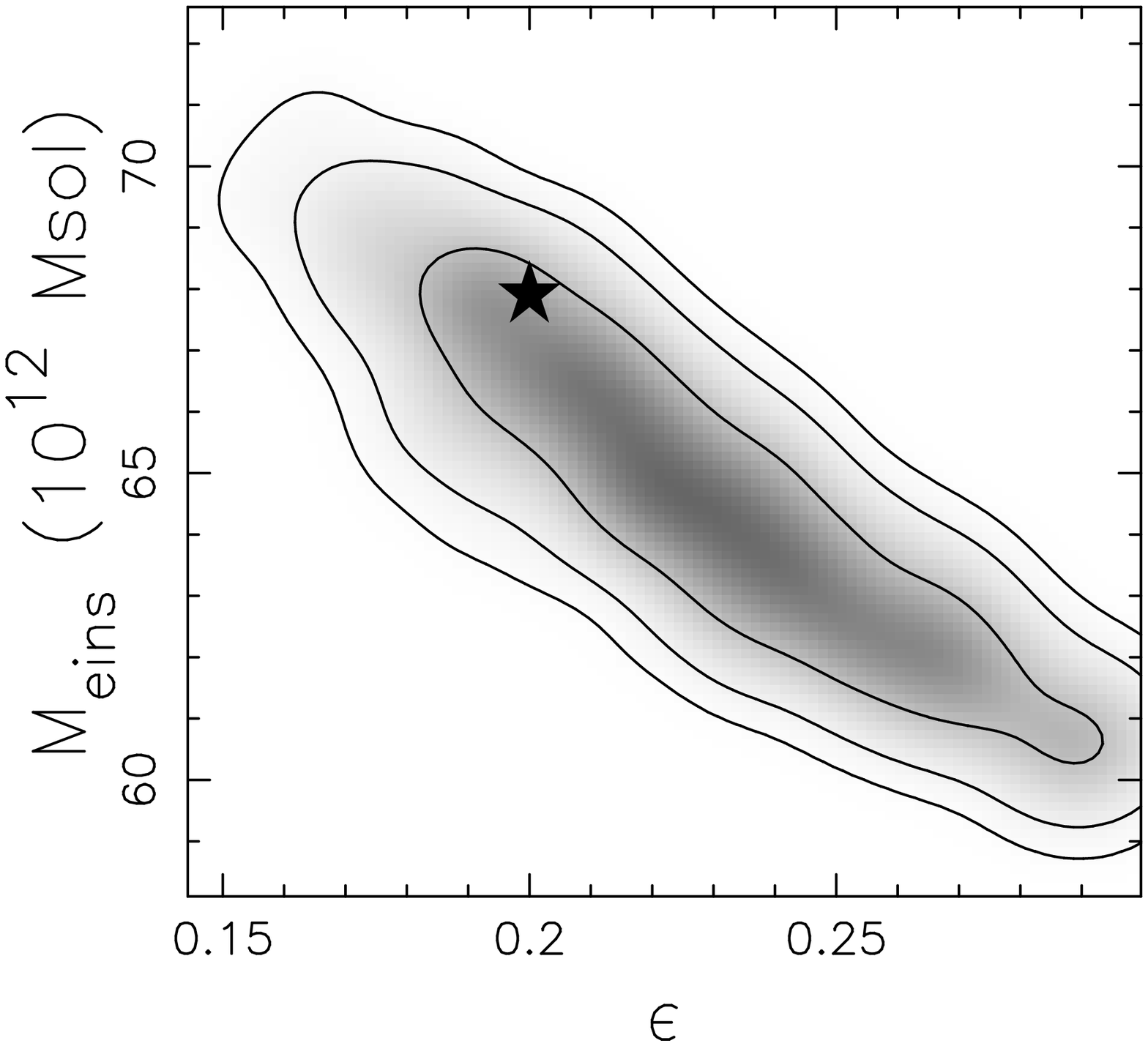} &
\includegraphics[width=0.28\linewidth]{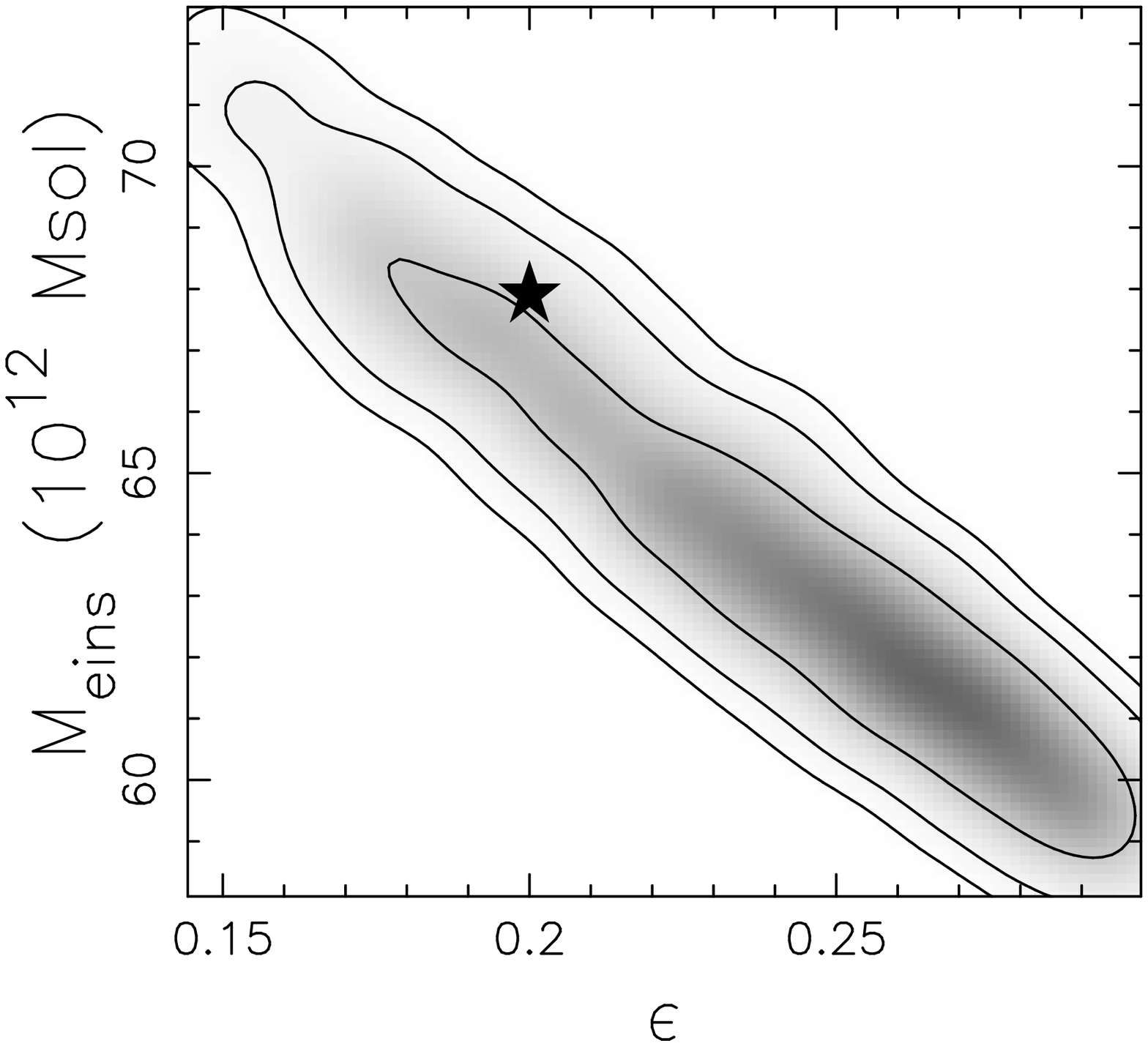} &
\includegraphics[width=0.28\linewidth]{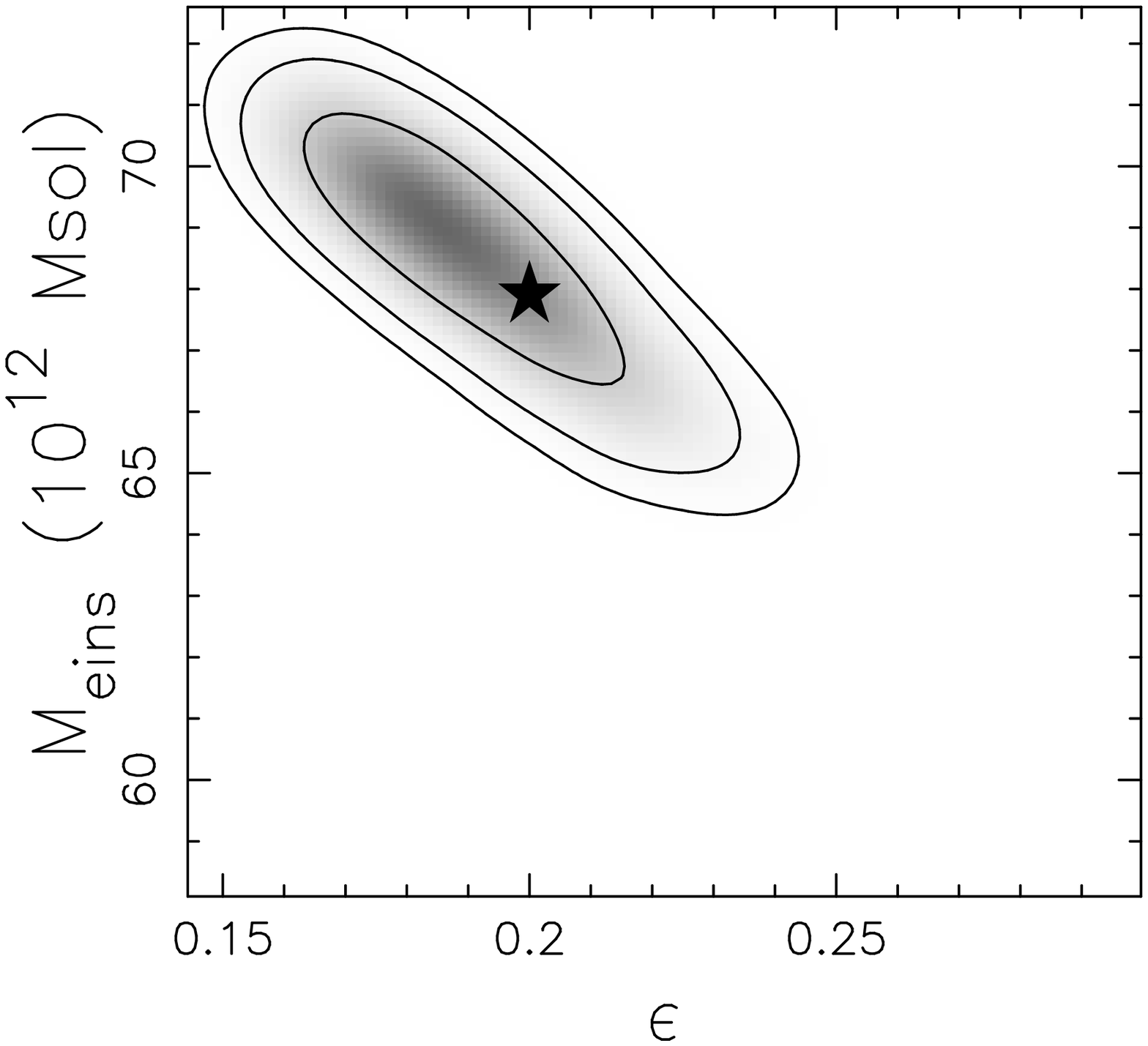} \\
\includegraphics[width=0.28\linewidth]{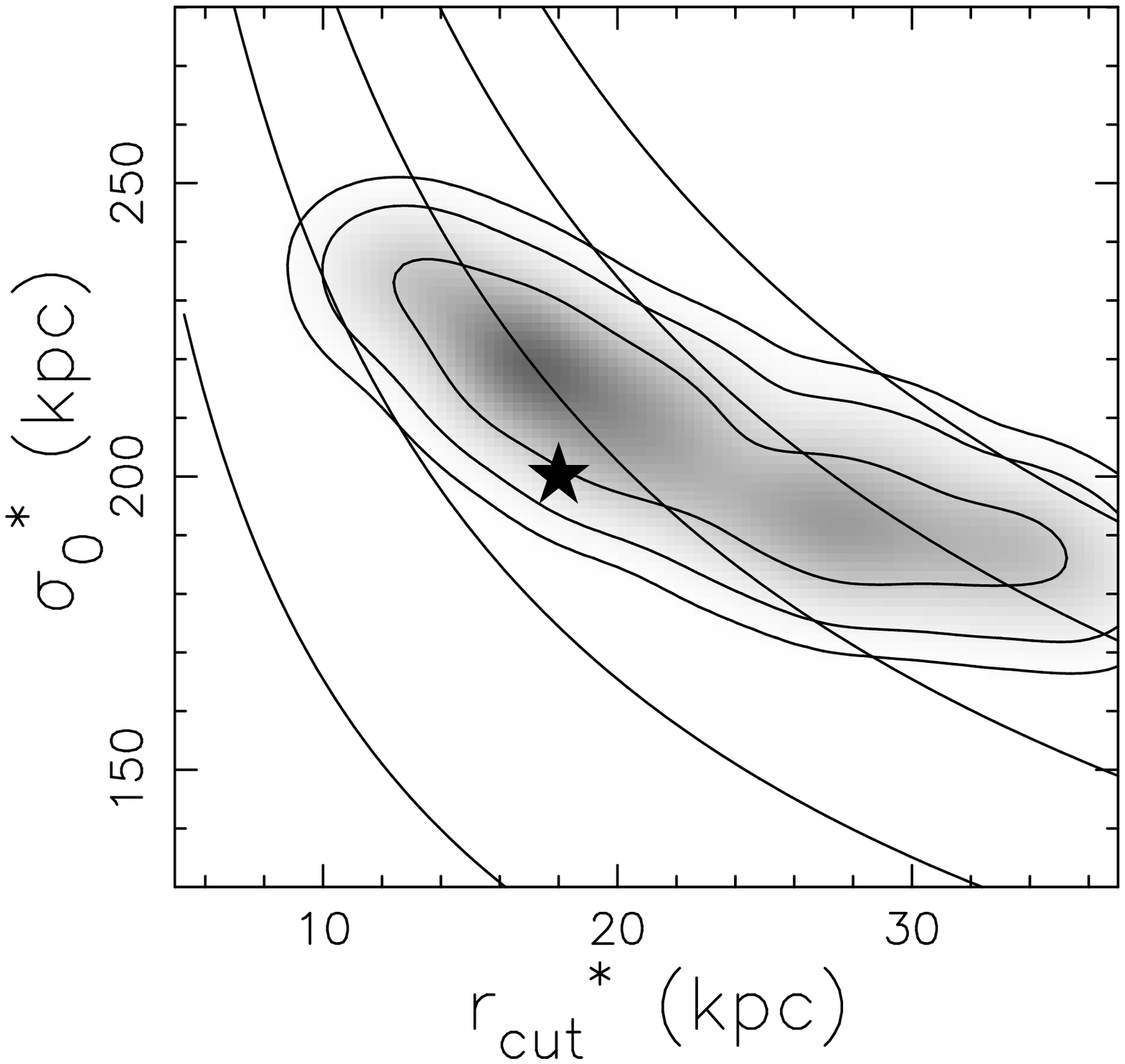} &
\includegraphics[width=0.28\linewidth]{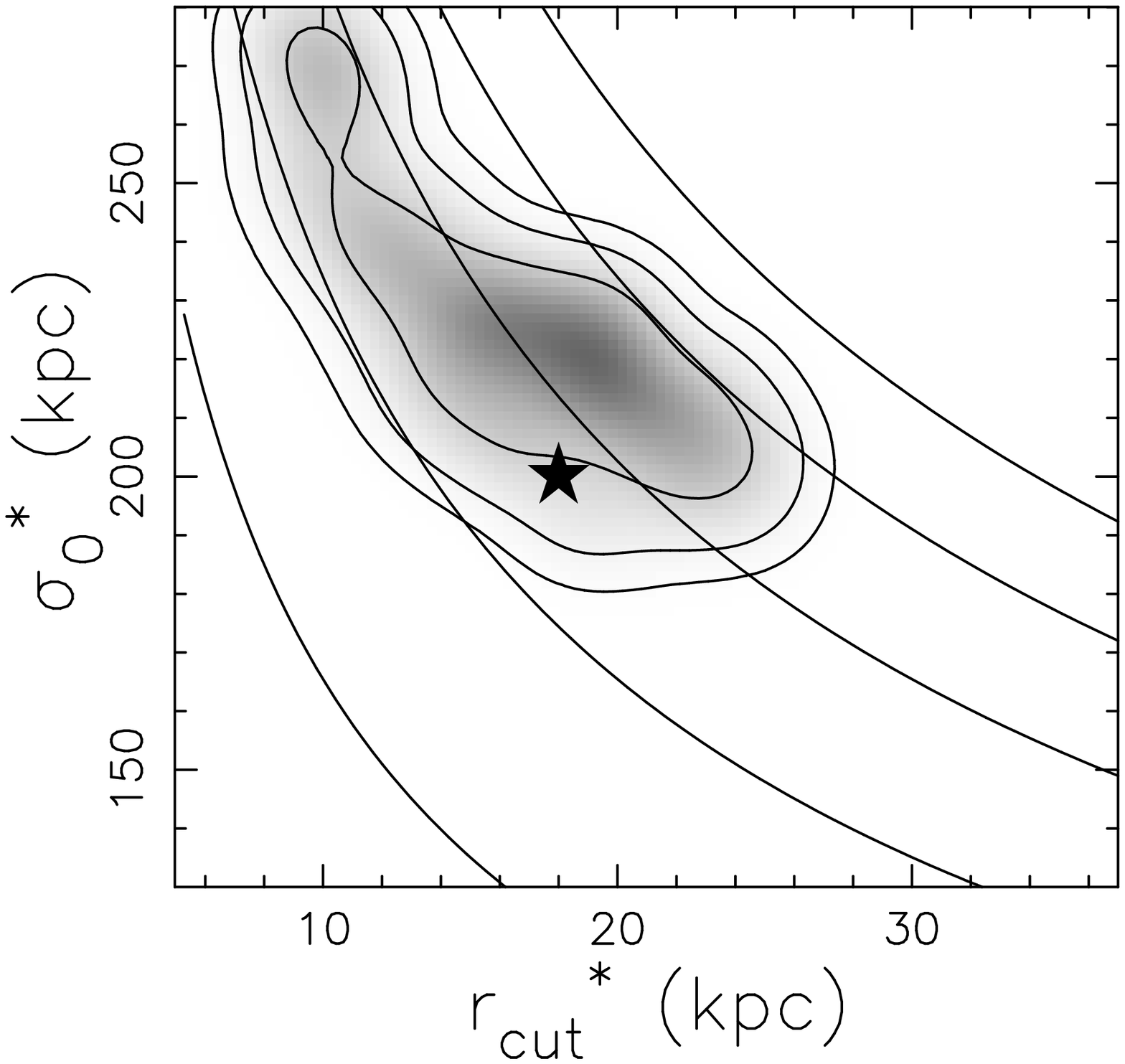} &
\includegraphics[width=0.28\linewidth]{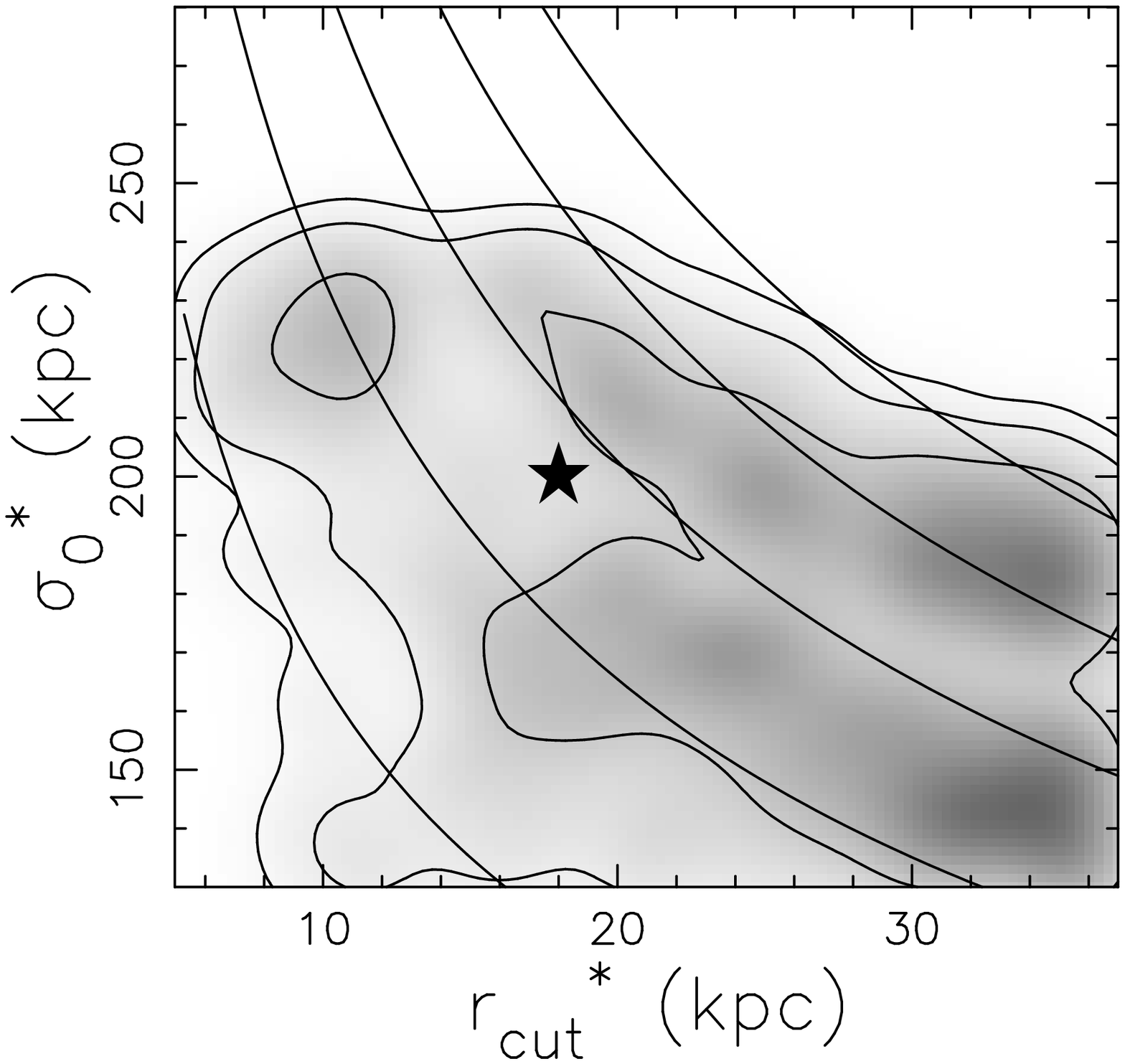} \\
\includegraphics[width=0.28\linewidth]{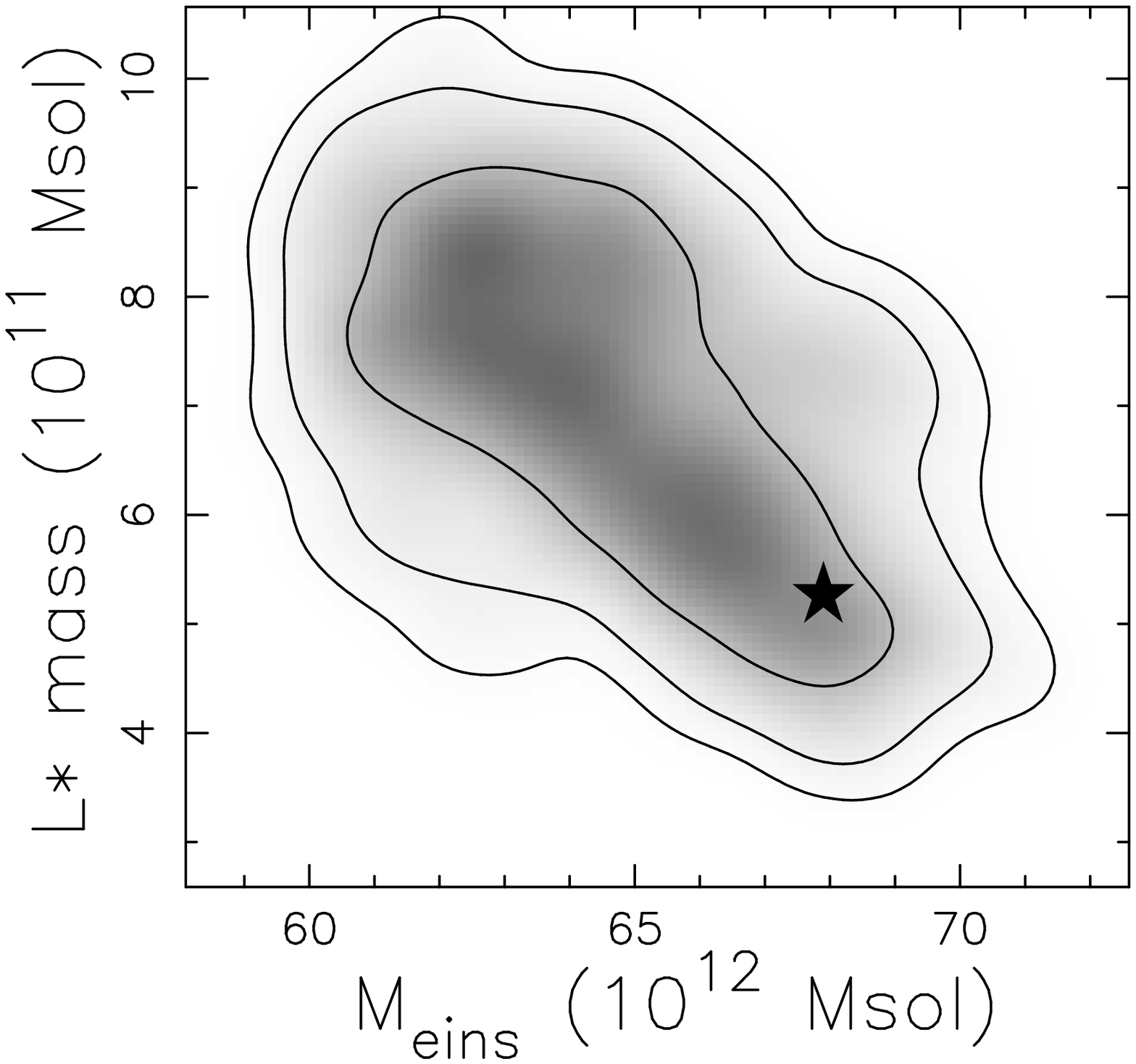} &
\includegraphics[width=0.28\linewidth]{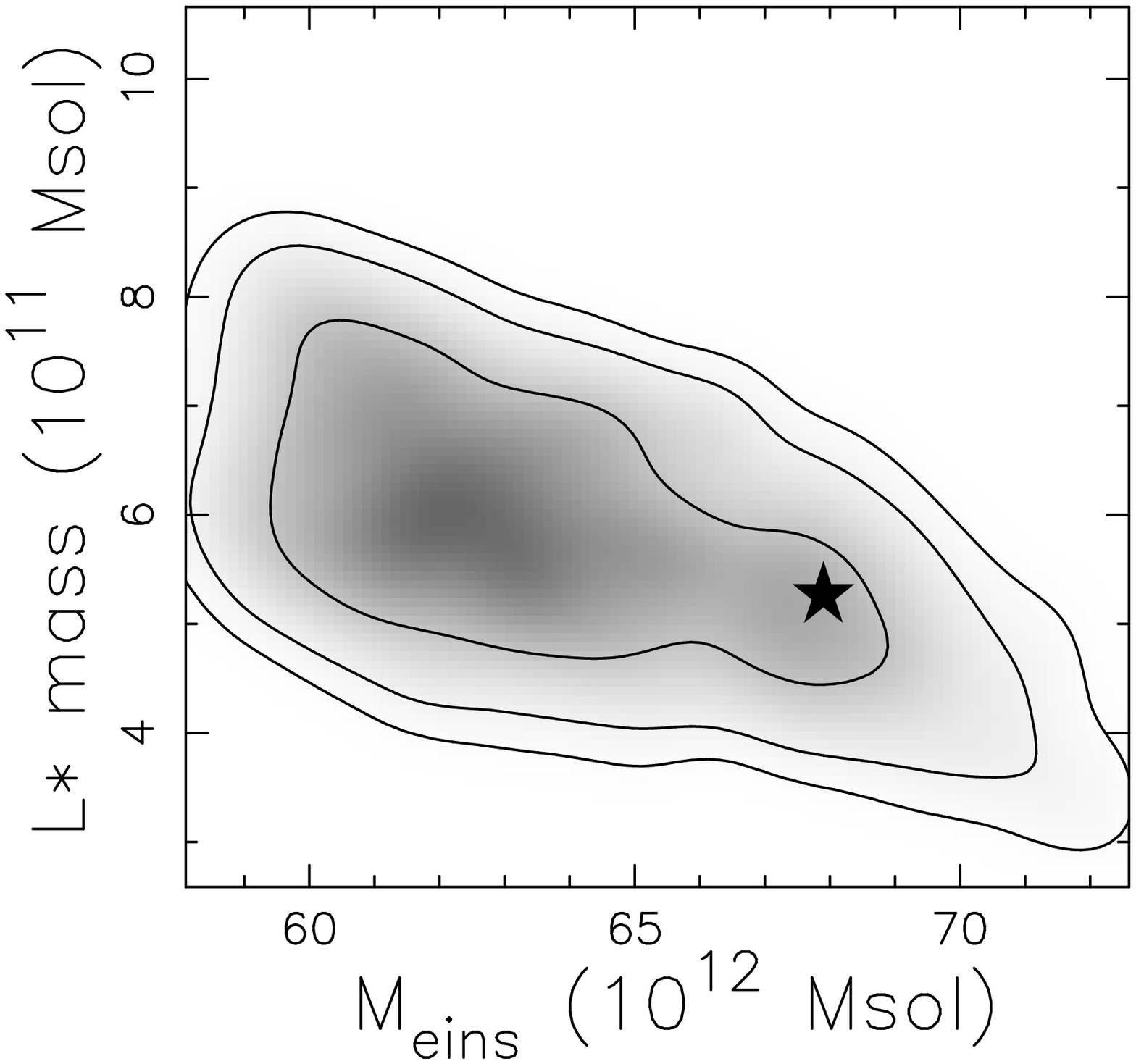} &
\includegraphics[width=0.28\linewidth]{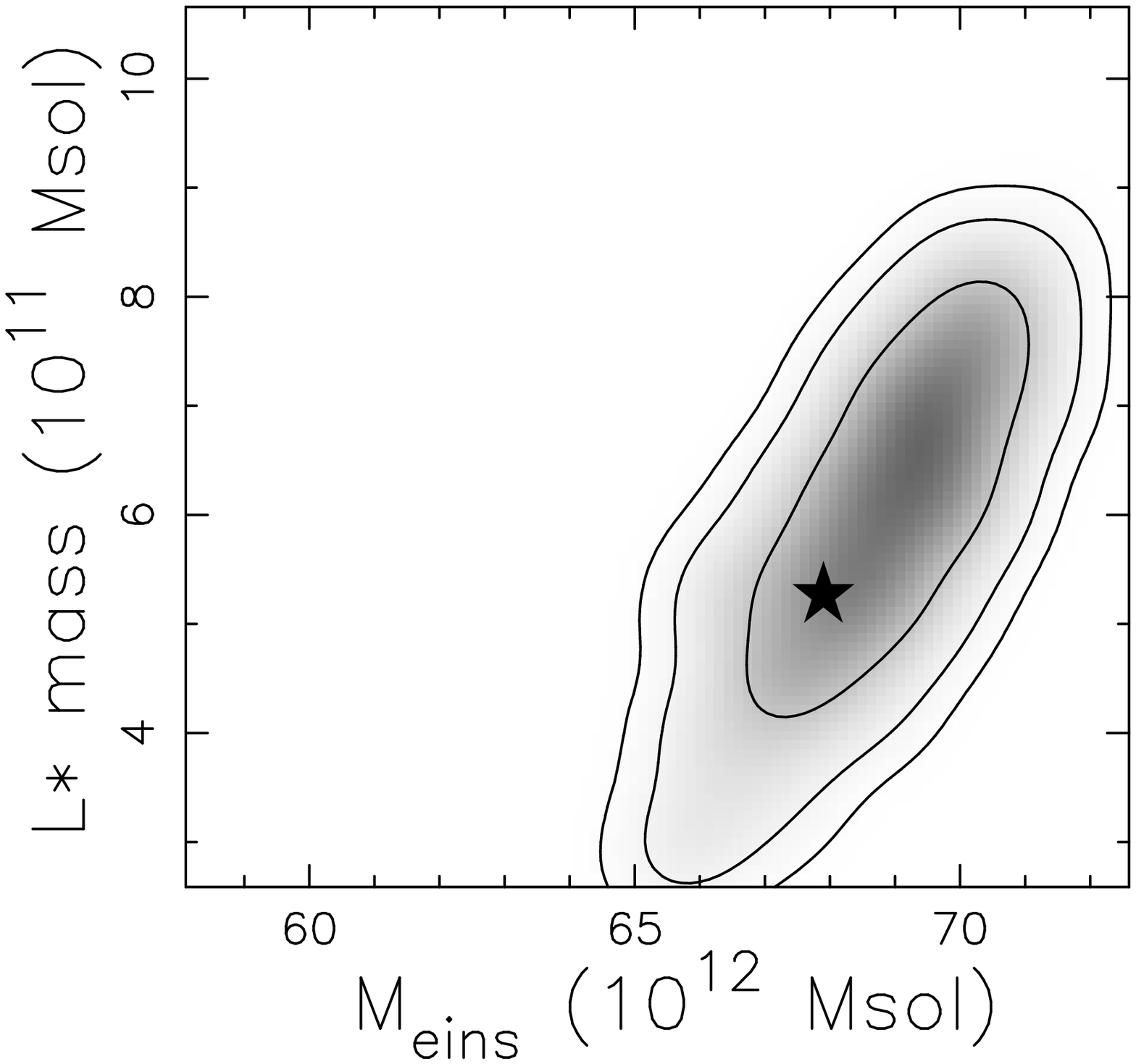} \\
\end{tabular} 
\end{indented} 
\caption{\label{figsersic}
 2D marginalized posterior PDF of the parameters of the cluster-scale
 halo modelled with an S\'ersic potential obtained from left to right
 with Config.~1, 2 and 3 of multiple images respectively. The 3
 contours stand for the 68\%, 95\% and 99\% CL.  The fiducial values
 are marked by the stars. The mass of a $L^\star$ galaxy is the total
 mass for a circular profile. The plotted contours in the
 $r_{cut}^\star$--$\sigma_0^\star$ plot are the iso-mass contours.
 The cluster mass $M_{eins}$ is the total enclosed mass (i.e. galaxy
 subhalos and cluster-scale halo) in the Einstein radius (30'').  }

\end{figure*}

 The obtained posterior PDF is marginalized and plotted in
 Figure~\ref{figsersic}. The estimated parameters are given in
 Table~\ref{tabdegesersic}. 

 First, we note that for the same lensing configuration, the parameters
 of a cluster-scale halo modelled by a S\'ersic potential are more
 difficult to constrain than those of a PIEMD or a NFW
 potential. We understand this to be a result of the effective 
 radius $R_e$ and
 index parameter $n$ mainly impacting the outer region of the mass
 distribution, which is not probed by strong lensing. 

 Second, the ellipticity, the PA, the $M_{eins}$ and the $L^\star$
 mass parameters are degenerate in the same manner as in the previous
 sections, confirming that these degeneracies are
 dependent on the
 lensing configuration alone. 

 Finally, in Table~\ref{tabdegesersic}, we note that the $L^\star$
 cut-off radius is recovered with nearly the same accuracy as in the case where
 the
 cluster-scale halo is modelled with the NFW potential. We suggest
 therefore that the scaling parameters $r_{cut}^\star$ and
 $\sigma_0^\star$ accuracies cannot be lower than about 20\% and 7\%
 respectively. This result is independent of both the model and the lensing
 configuration. 

\begin{table*}
\caption{\label{tabdegesersic} 
 Parameter recovery results for a cluster-scale halo modelled by a
 S\'ersic potential and recovered in 3 different strong lensing
 configurations.  The errors are given at 68\% CL. The $L^\star$
 masses are given for a circular mass component with identical
 dynamical parameters. }

\begin{indented} 
\item[]\begin{tabular*}{\linewidth}[c]{@{}l@{}rr@{ }rr@{ }rr@{ }r} 
\br 
& Input & \multicolumn{2}{c}{Config.1} & \multicolumn{2}{c}{Config.2} & \multicolumn{2}{c}{Config.3} \\
\mr
$\epsilon$ & 0.2 &  0.23 & $\pm$0.03 &  0.24 & $\pm$0.04 &  0.19 &
$\pm$0.01 \\
$PA$ (deg) & 127. &  128.0 & $\pm$0.8 &  121.9 & $\pm$2.3 &  127.5 & $\pm$1.0 \\
$R_e$ (kpc) & 1500. &  1195.7 & $\pm$345.5 &  1630.8 & $\pm$372.4 &  1698.5 & $\pm$319.3 \\
$\Sigma_e$ ($10^8\ M_\odot$) & 0.5 &  0.5 & $\pm$0.1 &  0.5 & $\pm$0.1 &  0.5 & $\pm$0.1 \\
$n$ & 2.8 &  2.9 & $\pm$0.2 &  2.6 & $\pm$0.2 &  2.8 & $\pm$0.2 \\
$r_{cut}^\star$ (kpc) & 18. &  21.0 & $\pm$3.3 &  16.8 & $\pm$4.2 &
25.4 & $\pm$8.2 \\
$\sigma_0^\star$ (km/s) & 200. &  206.6 & $\pm$15.5 &  223.6 &
$\pm$20.2 &  178.0 & $\pm$29.2 \\
$M_{L^\star}$ ($10^{11}\ M_\odot$) & 5.26 &  6.9 & $\pm$1.3 &  5.9 & $\pm$0.9 &  6.1 & $\pm$1.2 \\
$M_{eins}$ ($10^{12}\ M_\odot$) & 67.9 &  64.6 & $\pm$2.0 &  65.5 &
$\pm$3.4 &  68.8 & $\pm$1.3 \\
\br
\end{tabular*}
\end{indented} 
\end{table*} 

\begin{figure}
\begin{indented}
\item[] \includegraphics[width=\linewidth]{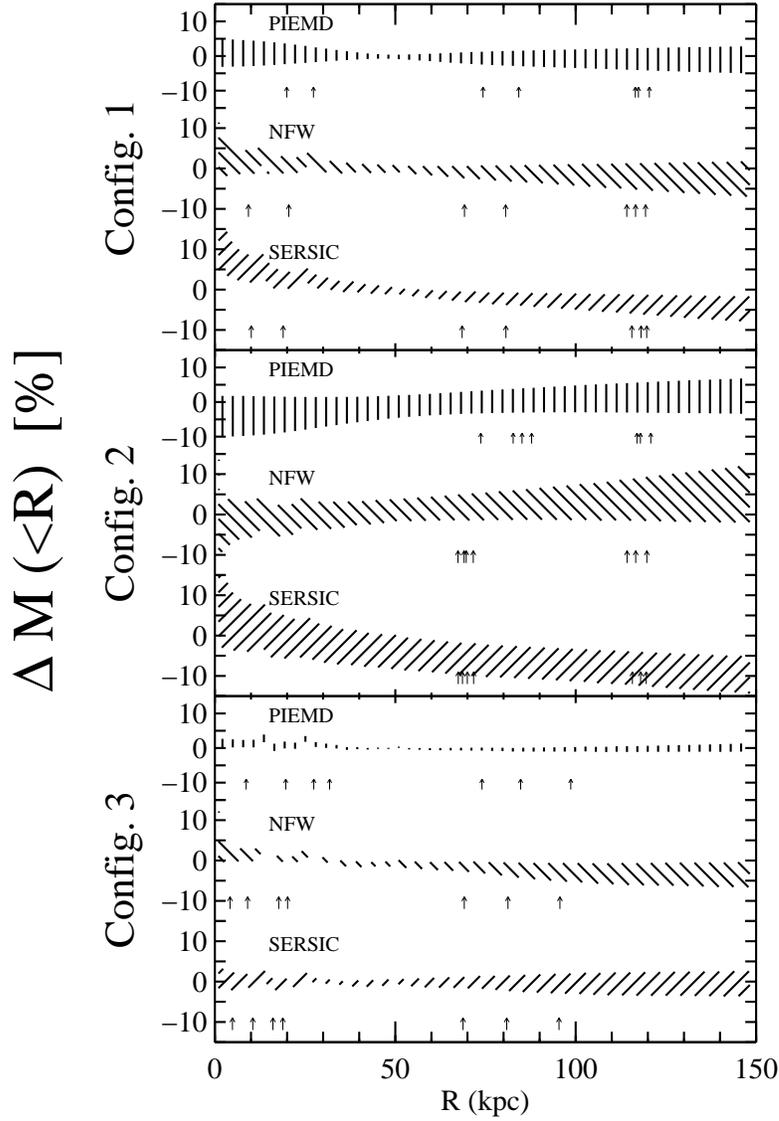} 
\caption{ \label{fig:sumup}
 Relative mass profile recovery in the three configurations for the
 three potentials PIEMD({\sl vertically hatched region}), NFW({\sl
 $-45^\circ$ hatched region}) and S\'ersic({\sl $45^\circ$ hatched
 region}). The arrows below each plot mark the positions of the
 multiple images used as constraints. The error bars are given at
 68\% CL. }

\end{indented}
\end{figure}

 Figure~\ref{fig:sumup} sums up the results found in this section
 concerning the accuracy obtained on the mass profile in each
 configuration for each potential. Although the accuracy depends on
 the lensing configuration it is usually better than 5\% in the region of
 multiple images with no obvious bias.  The accuracy is model
 independent, and is just the noise on the image positions (0.1 arcsec)
 translated into 
 the uncertainty on the parameters.  

\section{Model inference}
\label{sec:evid}

 In this section, we use the Bayesian evidence to rank  models. As an
 example, we consider the controversial inner slope of the density
 profile in clusters of galaxies. In 2004, \citeauthor{sand2004} have
 used a sample of 6 galaxy clusters to show that the slope of the
 central density profile was shallower than $r^{-1}$ as predicted by
 CDM simulations.  In their modelling they were using axisymmetric
 potentials. The same year, \citeauthor{bartelmann2004} reconsider
 these results and conclude that an NFW profile with a $r^{-1}$ inner
 slope could not be ruled out by strong lensing once effects of
 asymmetry and shear were taken into account. 

 In order to illustrate the model inference with the Bayesian
 evidence, we assume here that galaxy clusters actually present an
 inner slope shallower than $r^{-1}$. Then, we show that even when
 accounting for asymmetry and shear, the Bayesian evidence is still
 able to rank models and eventually rule them out.

 To do so, as an input model, we use the PIEMD model from section
 \ref{sec:piemd} i.e. the inner slope is shallower than $r^{-1}$. In
 order to observe the limits of Bayesian inference with the evidence,
 we simulate 6 models in which we change the size of the cluster-scale
 halo core radius. We scale the velocity dispersion accordingly so
 that the enclosed mass at the Einstein radius is maintained. 

 The 3 background galaxies of the previous section are lensed through
 each model. We have to slightly move the sources in the source plane
 so that in every simulation, we always end up with 1 tangential
 system, 1 radial system and 1 singly imaged system. For models with
 $r_{core} < 30$ kpc, we remove the images predicted at the very
 centre of the galaxy cluster because their lensing amplification is
 lower than 1 and in practice they are never observed (either too
 faint or blended in the cD galaxy flux). In contrast, for models with
 $r_{core} \ge 30$ kpc, we keep all the predicted images because their
 lensing amplification is always greater than 1. We add a Gaussian
 noise of FWHM 0.1'' to each image position. 

 Then, we successively fit a SIE, a NFW and a S\'ersic potential to
 the simulated systems of multiple images and report the computed
 evidences in Table~\ref{tab:evid}. As a reference, the last column
 reports the evidence computed when we fit the simulated PIEMD models
 by themselves.  We assume no prior knowledge (in practice, we use
 uniform distributions and adjust the limits so that the posterior PDF
 is not bounded). We also consider the scaling relation parameters
 $r_{cut}^\star$ and $\sigma_0^\star$ as free parameters.

\begin{figure}
\begin{indented}
\item[] \includegraphics[width=\linewidth]{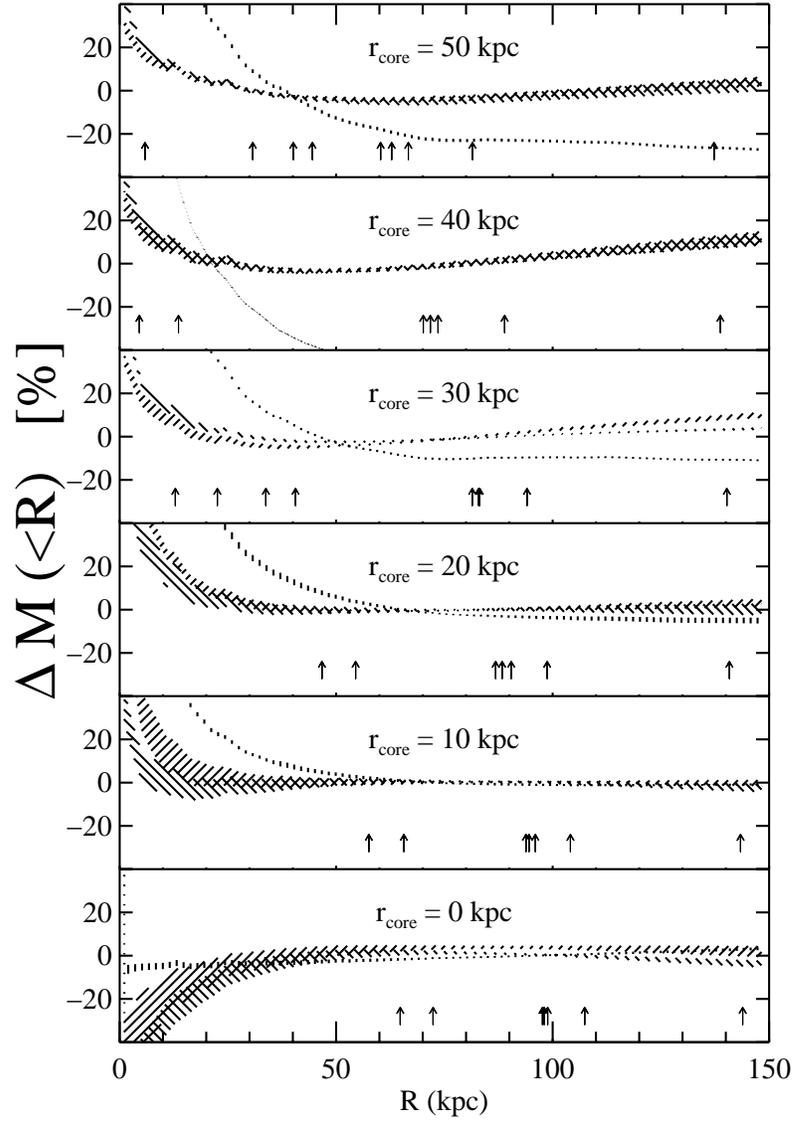} 
\caption{\label{masscumul} 
Aperture mass profile errors relative to the input PIEMD mass profile
for the fitted potentials SIE ({\sl vertically hatched region}), NFW
({\sl $-45^\circ$ hatched region}) and S\'ersic ({\sl $45^\circ$
hatched region}) as a function of the aperture radius. The hatched
width represents the 3$\sigma$ error estimated from the posterior PDF.
The arrows mark the positions of the multiple images used as
constraints.  }

\end{indented} 
\end{figure}

 Figure~\ref{masscumul} shows the aperture mass errors relative to the
 input PIEMD mass profile for the SIE, the NFW and the S\'ersic
 potentials. 

 First, we note that excluding the inner region and when $r_{core} \le
 20$ kpc, the input mass profile is well recovered by all the models.
 Note that in the case $r_{core}=0$ kpc, the SIE aperture mass error
 is smaller than 10\% on the full range of radius.  This ascertain the
 consistency of our SIE and PIEMD models.  Conversely, the SIE
 aperture mass error increases rapidly in the inner region as soon as
 we increase the core radius. In the inner region, the large errors
 are due to the intrinsic slope of each model (see
 Figure~\ref{fig:sdens}).

 The evidences reported in Table~\ref{tab:evid} correctly summarize
 these observations. In particular, the SIE evidence at $r_{core}=0$
 kpc is close to the evidences of the other models.  According to
 \citet{jeffreys1961}, the difference between two models is
 substantial if $1 < \Delta \ln E < 2.5$, strong if $2.5 < \Delta \ln
 E < 5$ and decisive if $\Delta \ln E > 5$.  Following this criteria,
 for $r_{core} \le 20$ kpc, the NFW, the S\'ersic and the SIE models
 are equivalent at fitting the data within the evidence error
 established in section~\ref{mcmc}.  However, when the core radius
 increases, the SIE model can be confidently rejected.

 Now, excluding the SIE models, we can use the evidences to classify
 the models in 2 categories : (i) when $r_{core} \le 20$ kpc, the NFW
 and the S\'ersic models evidences are equivalent to the reference
 PIEMD evidence within the evidence error. The evidences cannot
 confidently rank models. (ii) when $r_{core} > 20$ kpc, the evidences
 drop significantly and the NFW and S\'ersic models are confidently
 ruled out. This corresponds to the appearance of bright images inside
 the core radius (see Figure~\ref{masscumul}) as expected from flat
 core models.  Here, the S\'ersic model evidences are generally better
 than the NFW model evidences although the S\'ersic model contains an
 additional free parameter.  In the $r_{core}=30$ kpc case, the NFW
 and the S\'ersic models evidences are very low because of the
 stringent constraints imposed by the distribution of multiple images
 (a triplet of tangential images at $R=81$ kpc and a set of uniformly
 distributed images below 40 kpc).

 Finally, we conclude that the Bayesian evidence can effectively rank
 strong lensing models even when accounting for asymmetry and shear.
 However, this result strongly depends of the presence of images in the
 cluster centre.

 As we are submitting this paper, some of us are already using
 \lenstool and the evidence inference to study the inner slope of the
 dark matter profile with real data. Their results will be published
 in a forthcoming paper.

\begin{table*}
\caption{\label{tab:evid} Comparison of the log(Evidence) produced
 by the fit of the NFW, SIE and S\'ersic potentials to a core radius
 varying PIEMD potential. The values come from fits performed with
 sets of multiple images described in the text and a Rate equal to 0.1
 }
\begin{indented}
\item[]\begin{tabular}[c]{@{}crrrr}
\br
Core radius (kpc) & $E_{NFW}$ & $E_{Sersic}$ & $E_{SIE}$ & $E_{PIEMD}$ \\
\mr
0 & -27. & -25. & -28. & -20. \\ 
10 & -25. & -23. & -33. & -19. \\ 
20 & -27. & -24. & -146. & -19. \\ 
30 & -198. & -204. & -1391. & -25. \\ 
40 & -81. & -70. & -2795. & -19. \\ 
50 & -86. & -73. & -3260. & -22. \\
\br
\end{tabular}
\end{indented}
\end{table*}

\section{Conclusion} 

 In this study, we have described how to build  a gravitational lensing
 model of galaxy clusters and a set of constraints with multiply and
 singly imaged systems. Then, we have presented a new Bayesian method
 for efficiently exploring its parameter space without falling into
 local maxima of the likelihood PDF. The Bayesian method also gives an
 estimate of the errors and includes prior knowledge. We have
 illustrated the Bayesian posterior PDF analysis by studying the
 degeneracies in the PIEMD, the NFW and the S\'ersic potentials in 3
 different configurations of multiple images. We draw the following
 conclusions.

 (i) Strong degeneracies appear in both the PIEMD, the NFW and the
 S\'ersic potentials. The parameters are clearly dependent and
 compensate in order to produce a constant enclosed mass at the images
 location. The degeneracies are either due to the mathematical
 definitions of the potentials ($\sigma_0$--$r_{core}$,
 $\sigma_0$--$r_{cut}$ for PIEMD, $c$--$r_s$ for NFW,
 $R_e$--$\Sigma_e$, $R_e$--$n$ and $\Sigma_e$--$n$ for S\'ersic) or to
 the configuration of multiple images ($\epsilon$--$PA$,
 $\epsilon$--$L^\star$ galaxy mass, $M_{eins}$--$L^\star$ galaxy
 mass). The latter degeneracies are easily identified by looking at
 the degeneracies between the shape and the dynamical parameters. They
 are model independent. In every case, the enclosed mass in the
 Einstein radius decreases with the model ellipticity.

 (ii) Radial systems of multiple images combined to tangential arcs
 provide unique constraints on the slope of  the mass profile. It is
 therefore important  to identify radial (or central) images in
 the cluster cores.

 (iii) The PIEMD cut-off radius, the S\'ersic effective radius and the
 NFW scale radius are poorly constrained by strong lensing only.
 Hopefully, future parametric methods combining weak and strong
 lensing will provide tighter constraints.

 (iv) Galaxy-scale subhalos degenerate with the cluster-scale halo.
 The best constraints were obtained in lensing configurations
 combining radial and tangential multiple images systems. In this
 case, we barely manage a 20\% accuracy on the cut-off radius of
 subhalos scaled with scaling relations.  As shown by
 \citep{natarajan1998, natarajan2006} weak and strong lensing
 combination can improve this result.

 We have also illustrated how to rank models with the Bayesian
 evidence. We fit a NFW, a S\'ersic and a SIE potential to 6 PIEMD
 simulated clusters with different core radius. We have shown that the
 NFW and the S\'ersic potentials can actually fit systems of multiple
 images produced by clusters with core radius \textit{provided no
 image lie inside the core radius}.  For large core radius, 
 central images appear at the very centre of the cluster and provide
 enough constraints to disentangle PIEMD, NFW or S\'ersic based
 models.

 Although strong lensing is a wonderful tool to infer surface
 densities, it becomes rapidly limited by the models and the observed
 lensing configuration.  For instance, it is not possible to constrain
 the central density slope without radial images. Actually, the
 presence of radial images strongly suggests the presence of a flat
 core. 

 In a forthcoming paper, we will expand this method to constrain
 cosmological parameters with strong lensing. With a large number of
 multiple images with known redshift, one should be able to compare
 the strong lensing cosmography constraints (similarly to the early
 work of \citet{golse2002a} and \citet{soucail2004}) with other
 methods such as the CMB/WMAP results, or Supernovae or cosmic shear
 results.

\ack 
 We acknowledge Bernard Fort and Alain Smette for their support and
 their useful comments. We acknowledge the referees for their useful
 comments. We thank John Skilling for allowing us to use his
 \textsc{bayesys mcmc} sampler.  JPK acknowledge support from CNRS and
 CNES.  This work was supported in part by the European Community's
 Sixth Framework Marie Curie Research Training Network Programme,
 Contract No.  MRTN-CT-2004-505183 "ANGLES". The Dark Cosmology Centre
 is funded by the Danish National Research Foundation.

\bibliographystyle{/home/ejullo/paper/Tex/Njp/bibtex/harvard/jphysicsB}  
\bibliography{jullo}   

\appendix 

\section{Critical lines computation with Marching squares}

 A multiscale marching squares technique has been implemented in
 \textsc{lenstool} to compute the critical lines.  Marching squares is
 a computer graphics algorithm that generates contour lines for a 2D
 scalar field. It is similar to the marching cubes algorithm
 \citep{lorensen1987}. The algorithm proceeds through a scalar field
 taking four neighbour locations at a time (thus forming an imaginary
 square), then determining the line needed to represent the part of
 the contour that passes through this square. The individual lines are
 then fused into the desired contour. 

\begin{figure}
\begin{indented}
\item[]\includegraphics[width=0.8\linewidth]{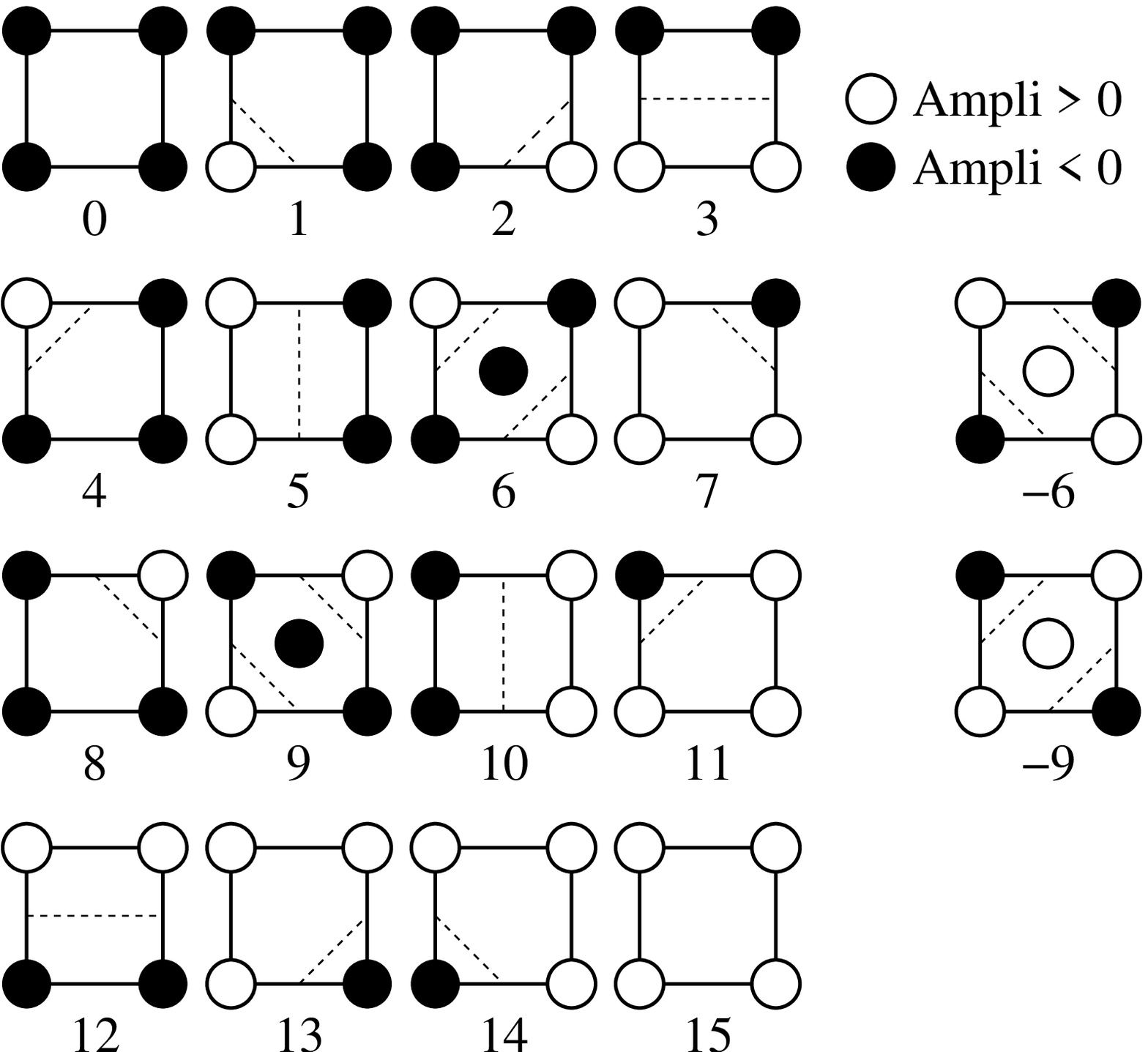}
\end{indented}
\caption{\label{figMS} 
16 square configurations. The empty and filled circles are points with
positive and negative amplification respectively. The dashed lines are
the infered critical lines. }
\end{figure}

 This is done by creating an index to a precalculated array of 16
 ($2^4= 16$) possible line configurations within the square (see
 Figure~\ref{figMS}), by treating each of the 4 scalar values as a bit
 in an 4-bit integer. If the scalar's value is higher than the
 iso-value (i.e. it is inside the contour) then the appropriate bit is
 set to one, while if it is lower (outside), it is set to zero. The
 final value after all 4 scalars are checked, is the actual index to
 the line configuration array. 

 In the critical lines case, the scalar field is not known a-priori.
 Therefore, we adopt a multiscale algorithm to focus towards the
 critical lines. The field is split in two recursively until we reach
 a higher limit for the size of a rectangle. Then, if a critical line
 is detected in a rectangle, it is split further. The rectangles with
 no critical line detected are left aside. Once the size of the
 rectangle has reached a lower limit, a line is kept in memory for
 this rectangle according to the marching squares configurations. The
 individual lines are then fused into the critical lines contour.

 The previous technique was a line following algorithm called
 \textsc{snake}. It starts from the center of a clump and picks
 amplification samples along its way outwards. When an amplification
 sign change is encountered, it precises the infinite amplification
 position and circles the clump until it comes back to its starting
 point along the critical line. \\

 In complex environment, the \textsc{snake} algorithm sometimes gets
 lost and produces incomplete critical lines. Conversely, the
 multiscale marching square algorithm never gets lost and identifies
 all the critical lines in the field. However, it can miss a part of
 critical line if the higher limit is too large. 

\begin{figure}
\begin{indented}
\item[]\includegraphics[width=\linewidth]{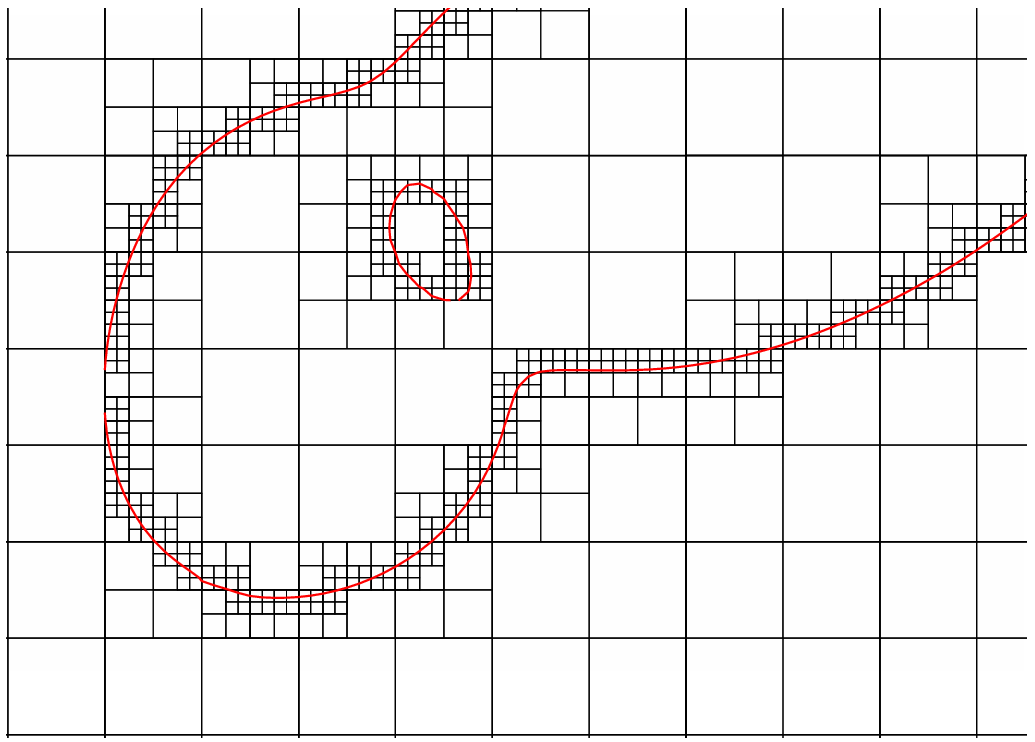}
\end{indented}
\caption{\label{mapms} 
Multiscale marching square field splitting. The boxes represent the
splitting squares and the red lines, the critical curve contour. The
imposed upper and lower limits for the boxes sizes are 10'' and 1''
respectively. The 1'' boxes are not plotted for clarity. }
\end{figure}

\section{Pseudo-elliptical S\'ersic potential}
\label{sec:pseudoser}

 As another addition to \lenstool, we have incorporated the
 S\'ersic density profile \citep{sersic1968} as an alternative
 description of the matter density.  The motivation for including it
 is that as the S\'ersic profile describes the 2D luminosity profile
 of elliptical galaxies \citep{sersic1968, ciotti1991, caon1993}, it
 can be used to separately model the baryonic matter component (which
 should be traced by the light) and the dark matter (DM) component,
 given enough lensing constraints.  In addition,
 \citet{merritt2005,merritt2006}, find that a deprojected S\'ersic
 profile gives a better fit than an NFW profile to the 3D density
 profile of DM halos from simulations. \citet{eliasdottir2007} found
 that given that the surface density distribution is indeed given by a
 S\'ersic profile, but fitted by an NFW using lensing constraints, it
 can lead to unrealistic estimates of the parameters (e.g.\ the
 predicted weak lensing signal and the concentration parameter), making
 the S\'ersic profile an interesting alternative for modelling the DM
 halos themselves.  Finally, the special case of the S\'ersic index
 $n=1$, corresponds to an exponential disk, making it useful for
 modelling spiral galaxies.  Spiral lenses are comparatively rare to
 date, but dedicated efforts are being made to find such lenses, and
 with the inclusion of the S\'ersic profile to Lenstool, it can now be
 used to study and model such lenses. 

 The S\'ersic 2D density profile has three free parameters
 ($n,R_e,\Sigma_e$) and is given by:

\begin{equation}
\label{eq:sersic}
\Sigma_{\mathrm{ser}} =\Sigma_e\exp\left[
-b_n\left(\left(\frac{R}{R_e}\right)^{1/n}-1\right)\right]\;,
\end{equation}

 \noindent where $R$ is the projected radius, $n$ is the S\'ersic
 index, $b_n$ is a constant chosen such that $R_e$ is the radius
 containing one-half of the projected mass and $\Sigma_e$ is the
 density at $R_e$.  The S\'ersic profile reduces to the de Vaucouleurs
 profile for $n=4$, and to the exponential disk for $n=1$.  The other
 parameters of the S\'ersic profile in \lenstool are its position on
 the sky, its position angle and its ellipticity. 

 The elliptic version of the S\'ersic profile is calculated using the
 pseudo-elliptical approximation developed by \citet{golse2002a}. It
 is introduced in the expression of the circular S\'ersic potential by
 substituting $R$ by $R_\epsilon$, using the following elliptical
 coordinate system:

\begin{equation}
\left\{
\begin{array}{l}
x_\epsilon = \sqrt{(1-\epsilon)}x \\
y_\epsilon = \sqrt{(1+\epsilon)}y \\
R_\epsilon=\sqrt{x_\epsilon^2 + y_\epsilon^2} \\
\phi = \arctan(y_\epsilon/x_\epsilon)\;.
\end{array}
\right.
\end{equation}

 In this definition, $\epsilon = (a^2-b^2)/(a^2+b^2)$ where $a$ and
 $b$ are respectively the semi-major and the semi-minor axis of the
 elliptical potential. From the elliptical lens potential
 $\varphi_\epsilon(r) \equiv \varphi(r_\epsilon)$,
 \citeauthor{golse2002a} propose generic expressions to compute the
 elliptical deviation angle $\balpha_\epsilon(\mathbf{r})$, the
 convergence $\kappa_\epsilon(\mathbf{r})$, the shear
 $\gamma_\epsilon(\mathbf{r})$ and the projected mass density
 $\Sigma_\epsilon(\mathbf{r})$:

\begin{equation}
\Sigma_\epsilon(\mathbf{r}) = \Sigma(r_\epsilon) + 
\epsilon \cos 2 \phi_\epsilon (\bar{\Sigma}(r_\epsilon) -
\Sigma(r_\epsilon))\;.
\end{equation}

 The pseudo-elliptical developments are limited to small
 ellipticities. For instance for the NFW, when $\epsilon > 0.25$, the
 surface iso-densities become increasingly boxy/peanut. Similarly for
 the S\'ersic potential, we have found that when $\epsilon > 0.25$,
 the goodness of fit (defined in \citeauthor{golse2002a}) measured at
 $R_\epsilon = R_e$ becomes larger than 10\%. We also fit the relation
 between $\epsilon_\Sigma$ and $\epsilon$ and found $\epsilon_\Sigma =
 3.55 \epsilon - 3.42 \epsilon^2$ with a $\chi^2 = 10^{-5}$.

 The ellipticities of the potentials used in this paper and of the
 projected mass densities $\epsilon_\Sigma$ are linearly proportional
 through multiplicative factors (reported in Table~\ref{potentials}).

 The range of valid surface density axis ratio $q = b/a$ provided by
 the pseudo-elliptical approximation for the SIE, the NFW and the
 S\'ersic potentials are $q_{SIE} > 0.65$, $q_{NFW} > 0.53$ and
 $q_{Sersic} > 0.44$ respectively. From N-body simulations
 \citet{oguri2003} found that the most probable projected axis ratio
 is $q = 0.6$. The pseudo-elliptical technique is therefore able to
 model most of the triaxial halos.

 In case of highly elliptical mass distributions, the PIEMD model
 \citep{kk1993} produces elliptical iso-densities because the
 ellipticity has been introduced directly in the projected mass
 distribution and not at the level of the potential.

\end{document}